\pdfoutput=1
\documentclass[11pt,twoside,a4paper,cmspaper,final,collab]{cms-tdr}
\def\svnVersion{f02ae5e}\def\svnDate{2023/07/19}\def\cmsCernNoTag{CERN-EP-2022-113}\def\cmsCernDate{\today}\def\cmsMessage{Published in the Journal of High Energy Physics as \href{http://dx.doi.org/10.1007/JHEP07(2023)095}{\doi{10.1007/JHEP07(2023)095}.}}
\begin{document}\cmsNoteHeader{HIG-21-002}

\newcommand{\lep}{\ensuremath{\ell}\xspace}
\newcommand{\leplep}{\ensuremath{\ell\ell}\xspace}
\newcommand{\Plepton}{\ensuremath{\ell}\xspace}
\newcommand{\Pnu}{\ensuremath{\PGn}\xspace}
\newcommand{\Pggx}{\ensuremath{\PGg^{\ast}}\xspace}
\newcommand{\Pbottom}{\ensuremath{\PQb}\xspace}
\newcommand{\APbottom}{\ensuremath{\PAQb}\xspace}
\newcommand{\Ptop}{\ensuremath{\PQt}\xspace}
\newcommand{\APtop}{\ensuremath{\PAQt}\xspace}
\newcommand{\Pquark}{\ensuremath{\PQq}\xspace}
\newcommand{\APquark}{\ensuremath{\PAQq}\xspace}
\newcommand{\Pgluon}{\ensuremath{\Pg}\xspace}
\newcommand{\HH}{\ensuremath{\PH\PH}\xspace}
\newcommand{\ggHH}{\ensuremath{\Pgluon\Pgluon\PH\PH}\xspace}
\newcommand{\qqHH}{\ensuremath{\Pquark\Pquark\PH\PH}\xspace}
\newcommand{\mHH}{\ensuremath{m_{\PH\PH}}\xspace}
\newcommand{\V}{\ensuremath{\PV}\xspace}
\newcommand{\tH}{\ensuremath{\Ptop\PH}\xspace}
\newcommand{\tHq}{\ensuremath{\Ptop\PH\PQq}\xspace}
\newcommand{\tHW}{\ensuremath{\Ptop\PH\PW}\xspace}
\newcommand{\ttH}{\ensuremath{\Ptop\APtop\PH}\xspace}
\newcommand{\tZ}{\ensuremath{\Ptop\PZ}\xspace}
\newcommand{\tW}{\ensuremath{\Ptop\PW}\xspace}
\newcommand{\ttZ}{\ensuremath{\Ptop\APtop\PZ}\xspace}
\newcommand{\ttW}{\ensuremath{\Ptop\APtop\PW}\xspace}
\newcommand{\X}{\ensuremath{\textrm{X}}\xspace}
\newcommand{\PJpsi}{\ensuremath{\JPsi}\xspace}
\newcommand{\PUpsilon}{\ensuremath{\Upsilon}\xspace}

\newcommand{\fb}{\ensuremath{\,\text{fb}}\xspace}
\newcommand{\pb}{\ensuremath{\,\text{pb}}\xspace}

\newcommand{\PWst}{{\HepParticle{\PW}{}{\ast}}\Xspace}
\newcommand{\WWWW}{\ensuremath{\PW\PWst\PW\PWst}\xspace}
\newcommand{\WWtt}{\ensuremath{\PW\PWst\Pgt\Pgt}\xspace}
\newcommand{\WWZZ}{\ensuremath{\PW\PWst\PZ\PZst}\xspace}
\newcommand{\ZZZZ}{\ensuremath{\PZ\PZst\PZ\PZst}\xspace}
\newcommand{\ZZtt}{\ensuremath{\PZ\PZst\Pgt\Pgt}\xspace}
\newcommand{\tttt}{\ensuremath{\Pgt\Pgt\Pgt\Pgt}\xspace}

\renewcommand{\ss}{\ensuremath{\text{ss}}\xspace}
\newcommand{\jet}{\ensuremath{\text{j}}\xspace}

\newcommand{\zeroLeptonFourTau}{\ensuremath{4\tauh}\xspace}
\newcommand{\oneLeptonThreeTau}{\ensuremath{1\Plepton + 3\tauh}\xspace}
\newcommand{\twoLeptonssZeroTau}{\ensuremath{2\Plepton\ss}\xspace}
\newcommand{\twoLeptonTwoTau}{\ensuremath{2\Plepton + 2\tauh}\xspace}
\newcommand{\threeLeptonZeroTau}{\ensuremath{3\Plepton}\xspace}
\newcommand{\threeLeptonOneTau}{\ensuremath{3\Plepton + 1\tauh}\xspace}
\newcommand{\fourLeptonZeroTau}{\ensuremath{4\Plepton}\xspace}
\newcommand{\threeLeptonCR}{\ensuremath{3\Plepton~\PW\PZ}\xspace}
\newcommand{\fourLeptonCR}{\ensuremath{4\Plepton~\PZ\PZ}\xspace}

\newcommand{\llss}{\twoLeptonssZeroTau}
\newcommand{\lllnot}{\threeLeptonZeroTau}
\newcommand{\llll}{\fourLeptonZeroTau}
\newcommand{\lllt}{\threeLeptonOneTau}
\newcommand{\lltt}{\twoLeptonTwoTau}
\newcommand{\lttt}{\oneLeptonThreeTau}
\newcommand{\noltttt}{\zeroLeptonFourTau}

\newcommand{\Zll}{\ensuremath{\PZ \to \Plepton\Plepton}\xspace}
\newcommand{\Zee}{\ensuremath{\PZ \to \Pe\Pe}\xspace}
\newcommand{\Zmm}{\ensuremath{\PZ \to \PGm\PGm}\xspace}

\newcommand{\metHT}{\ensuremath{\HT^\text{miss}}\xspace}
\newcommand{\metLD}{\ensuremath{\pt^\text{miss,LD}}\xspace}

\newcommand{\kappal}{\ensuremath{\kappa_{\lambda}}\xspace}
\newcommand{\yt}{\ensuremath{\text{y}_{\Ptop}}\xspace}
\newcommand{\kappat}{\ensuremath{\kappa_{\Ptop}}\xspace}
\newcommand{\kappaV}{\ensuremath{\kappa_{\V}}\xspace}
\newcommand{\kappaVV}{\ensuremath{\kappa_{2\V}}\xspace}
\newcommand{\cg}{\ensuremath{\text{c}_{\Pgluon}}\xspace}
\newcommand{\cgg}{\ensuremath{\text{c}_{2\Pgluon}}\xspace}
\newcommand{\ctwo}{\ensuremath{\text{c}_{2}}\xspace}

\newlength\cmsTabSkip\setlength{\cmsTabSkip}{1ex}
\providecommand{\cmsCommonTable}[1]{\resizebox{0.90\textwidth}{!}{#1}}
\providecommand{\cmsTableThreeCol}[1]{\resizebox{\textwidth}{!}{#1}}
\providecommand{\cmsTableTwoCol}[1]{\resizebox{\textwidth}{!}{#1}}

\newcolumntype{C}[1]{>{\centering\let\newline\\\arraybackslash\hspace{0pt}}m{#1}}
\newcolumntype{L}[1]{>{\raggedright\let\newline\\\arraybackslash\hspace{0pt}}m{#1}}
\newcolumntype{R}[1]{>{\raggedleft\let\newline\\\arraybackslash\hspace{0pt}}m{#1}}

\newlength\cmsFigWidth
\ifthenelse{\boolean{cms@external}}{\setlength\cmsFigWidth{0.49\columnwidth}}{\setlength\cmsFigWidth{0.49\textwidth}}
\ifthenelse{\boolean{cms@external}}{\providecommand{\cmsLeft}{left\xspace}}{\providecommand{\cmsLeft}{left\xspace}}
\ifthenelse{\boolean{cms@external}}{\providecommand{\cmsRight}{right\xspace}}{\providecommand{\cmsRight}{right\xspace}}
\ifthenelse{\boolean{cms@external}}{\providecommand{\cmsMid}{middle\xspace}}{\providecommand{\cmsMid}{middle\xspace}}
\ifthenelse{\boolean{cms@external}}{\providecommand{\cmsBottom}{lower\xspace}}{\providecommand{\cmsBottom}{lower\xspace}}
\ifthenelse{\boolean{cms@external}}{\providecommand{\cmsTop}{upper\xspace}}{\providecommand{\cmsTop}{upper\xspace}}
\ifthenelse{\boolean{cms@external}}{\providecommand{\cmsCenter}{center\xspace}}{\providecommand{\cmsCenter}{center\xspace}}      

\cmsNoteHeader{HIG-21-002}

\title{Search for Higgs boson pairs decaying to \texorpdfstring{$\WWWW$}{WWWW}, \texorpdfstring{$\WWtt$}{WWtautau}, and \texorpdfstring{$\tttt$}{tautautautau} in proton-proton collisions at \texorpdfstring{$\sqrt{s} = 13\TeV$}{sqrt(s) = 13 TeV}}

\date{\today}

\abstract{The results of a search for Higgs boson pair (\HH) production in the \WWWW, \WWtt, and \tttt decay modes are presented. The search uses 138\fbinv of proton-proton collision data recorded by the CMS experiment at the LHC at a center-of-mass energy of 13\TeV from 2016 to 2018. Analyzed events contain two, three, or four reconstructed leptons, including electrons, muons, and hadronically decaying tau leptons. No evidence for a signal is found in the data. Upper limits are set on the cross section for nonresonant \HH production, as well as resonant production in which a new heavy particle decays to a pair of Higgs bosons. For nonresonant production, the observed (expected) upper limit on the cross section at 95\% confidence level (\CL) is 21.3 (19.4) times the standard model (SM) prediction. The observed (expected) ratio of the trilinear Higgs boson self-coupling to its value in the SM is constrained to be within the interval $-6.9$ to 11.1 ($-6.9$ to 11.7) at 95\% \CL, and limits are set on a variety of new-physics models using an effective field theory approach. The observed (expected) limits on the cross section for resonant \HH production range from 0.18 to 0.90 (0.08 to 1.06)\pb at 95\% \CL for new heavy-particle masses in the range 250--1000\GeV.}

\hypersetup{
    pdfauthor={CMS Collaboration},
    pdftitle={Search for Higgs boson pairs decaying to WWWW, WWtautau, and tautautautau in proton-proton collisions at sqrt(s) = 13 TeV},
    pdfsubject={CMS},
    pdfkeywords={CMS, Hadron-hadron scattering, Higgs physics}
}

\maketitle

\section{Introduction}
\label{sec:introduction}

Since the discovery of the Higgs (\PH) boson~\cite{Higgs-Discovery_ATLAS,Higgs-Discovery_CMS,Higgs-Discovery_CMS_long}, many of its properties have already been measured with high precision~\cite{CMS-HIG-17-031,ATLAS-HIGG-2018-57,CMS-HIG-21-013}.
One important property that remains largely unknown is the \PH boson self-coupling.
A precise measurement of this coupling is necessary to determine the shape of the Higgs potential, and thus verify that the mechanism breaking the electroweak gauge symmetry is indeed the Higgs mechanism~\cite{Englert:1964et,Higgs:1964ia,Higgs:1964pj,Guralnik:1964eu,Higgs:1966ev,Kibble:1967sv} of the standard model (SM)~\cite{Glashow:1961tr,Weinberg:1967tq,Salam:1968rm}.
The SM predicts the existence of both trilinear and quartic \PH boson self-couplings.
Due to the very low predicted cross section for triple \PH boson production, the SM quartic self-coupling will not be experimentally accessible at the CERN LHC, even with the full integrated luminosity of 3000\fbinv scheduled to be delivered after the high-luminosity LHC upgrade~\cite{deFlorian:2019app,HL-LHC-TDR}.
The strength of the trilinear self-coupling, however, can be determined using measurements of \PH boson pair (\HH) production.

\begin{figure}[h!]
    \centering\includegraphics*[width=0.46\textwidth]{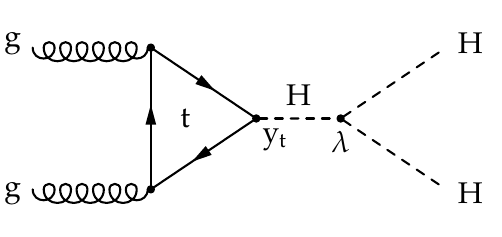}\hfill
    \centering\includegraphics*[width=0.46\textwidth]{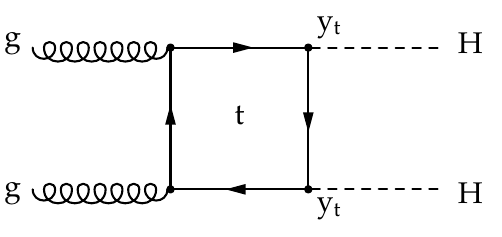}
    \caption{
        Leading order Feynman diagrams for SM nonresonant \HH production via gluon fusion, including the ``triangle'' diagram (left) and the ``box'' diagram (right).
    }
    \label{fig:Feynman_ggHH_and_qqHH_sm}
\end{figure}

In the SM, most \HH pairs are produced in two types of processes.
The Feynman diagrams for the dominant ``gluon fusion'' (\ggHH) process at leading order (LO) in perturbative quantum chromodynamics (QCD) are shown in Fig.~\ref{fig:Feynman_ggHH_and_qqHH_sm}.
The left ``triangle'' diagram amplitude varies proportionally to the \PH boson self-coupling ($\lambda$) and the Yukawa coupling of the top quark (\yt), while the right ``box'' diagram amplitude is insensitive to $\lambda$ and varies as $\yt^{2}$.
The triangle and box diagrams interfere destructively, so the \ggHH cross section exhibits a strong dependence on both $\lambda$ and \yt.
The \ggHH cross section in the SM has been computed to be $31.1^{+2.1}_{-7.2}$\fb at next-to-next-to-LO (NNLO) accuracy in QCD using the FT$_\text{approx}$ scheme, in which the true top quark mass is used for the real radiation matrix elements, while the virtual part is computed using an infinite top quark mass~\cite{Grazzini:2018hh}.
The predicted SM cross section for the subdominant ``vector boson fusion'' (\qqHH) process is $1.73 \pm 0.04$\fb at next-to-NNLO accuracy in QCD~\cite{Dreyer:2018qbw}.

Deviations of the coupling strength modifiers \mbox{\kappal$=\lambda/\lambda^\text{SM}$} and \mbox{\kappat$=\yt/\yt^\text{SM}$} from unity would affect both the rate of \HH production and kinematic distributions of the \HH signal.
The \HH invariant mass (\mHH) is particularly sensitive to changes in \kappal and \kappat, as these couplings affect the triangle and box diagram amplitudes differently.
Because SM \ggHH and \qqHH production do not include a heavy resonant particle, and typically result in a broad \mHH distribution, they are referred to as ``nonresonant''.
Changes in \kappal and \kappat also influence the rate of single Higgs boson production and the Higgs boson decay branching fractions~\cite{Degrassi:2016wml,Maltoni:2017ims}.

\begin{figure}[h!]
    \centering\includegraphics[width=0.325\textwidth]{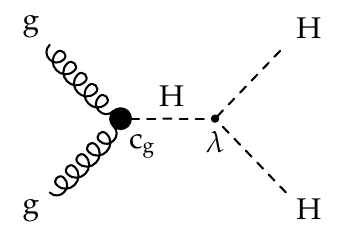}\hfill
    \centering\includegraphics[width=0.255\textwidth]{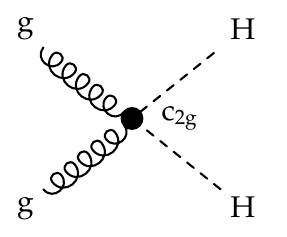}\hfill
    \centering\includegraphics[width=0.355\textwidth]{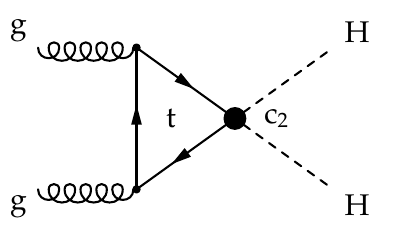}
    \caption{
        Leading order Feynman diagrams for nonresonant \HH production via gluon fusion in an EFT approach, where loop-mediated contact interactions between (\cmsLeft) two gluons and one \PH boson, 
        (\cmsMid) two gluons and two \PH bosons, and (\cmsRight) two top quarks and two \PH bosons are parametrized by three effective couplings: \cg, \cgg, and \ctwo.
    }
    \label{fig:Feynman_ggHH_eft}
\end{figure}

The presence of undiscovered particles or interactions, predicted by a variety of theoretical models beyond the SM, may alter the \HH production rate as well as observable kinematic distributions.
Such particles could give rise to loop diagrams similar to the one shown on the left of Fig.~\ref{fig:Feynman_ggHH_and_qqHH_sm}.
These diagrams may significantly enhance the \HH production rate, as they occur at the same loop level as \HH production in the SM.
Since no particles beyond those predicted by the SM have been observed so far, their mass may be at the \TeVns scale or higher, well above the scale of electroweak symmetry breaking.
Loop contributions of such heavy particles can be approximated as contact interactions with the \PH boson using an effective field theory (EFT) approach~\cite{Buchmuller:1985jz,Grzadkowski:2010es}.
Following Ref.~\cite{Carvalho:2015ttv}, the contact interactions relevant for \HH production are parametrized by the couplings \cg, \cgg, and \ctwo, referring to the interactions between two gluons and one \PH boson, two gluons and two \PH bosons, and two top quarks and two \PH bosons, respectively.
The corresponding Feynman diagrams for \ggHH production are shown in Fig.~\ref{fig:Feynman_ggHH_eft}.
The LO diagrams for \qqHH production contain no gluons or top quarks, so the impacts of \cg, \cgg, and \ctwo are only considered in the \ggHH signal in this publication.

\begin{figure}[h!]
    \centering\includegraphics*[width=0.34\textwidth]{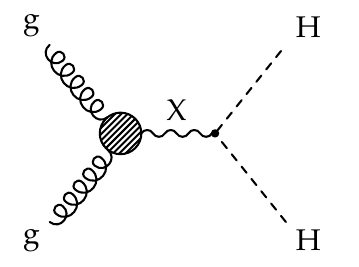}
    \caption{
        Leading order Feynman diagram for resonant \HH production.
    }
    \label{fig:Feynman_ggHH_resonant}
\end{figure}

An excess of \HH signal events may also result from decays of new heavy particles, denoted as \X, into pairs of \PH bosons.
Various theoretical models of new physics postulate such decays, in particular two-Higgs-doublet models~\cite{Craig:2013hca,Nhung:2013lpa}, composite-Higgs models~\cite{Grober:2010yv,Contino:2010mh}, Higgs portal models~\cite{Englert:2011yb,No:2013wsa}, and models inspired by warped extra dimensions~\cite{Randall:1999ee}.
In the last class of models, the new heavy particles may have spin 0 (``radions'') or spin 2 (``gravitons'')~\cite{Cheung:2000rw}.
In this paper, the resulting ``resonant'' \HH production is sought for mass values of \X from 250 to 1000\GeV, and the width of \X is assumed to be negligible compared to the experimental resolution in \mHH.
This would create a peak in the reconstructed \mHH distribution around the mass $m_{\X}$ of the resonance.
The Feynman diagram for this process is shown in Fig.~\ref{fig:Feynman_ggHH_resonant}.
For resonance masses above 1\TeV the strongest constraints are given by searches for \HH production targeting \PH boson decays to bottom quarks~\cite{ATLAS:2022hwc,CMS:2021roc,ATLAS-HDBS-2018-40}, as the selection and reconstruction efficiency for hadronic decays increases, in particular in the trigger, and relevant backgrounds decrease with energy. For leptonic decay modes, the selection and reconstruction efficiency in general is high and as such do not increase notably for high masses above 1 \TeV.

Phenomenological studies of the prospects for discovering \HH signal in the \WWWW decay mode are documented in Refs.~\cite{Baur:2002rb,Baur:2002qd,Li:2015yia,Adhikary:2017jtu,Ren:2017jbg}, where the symbol $\ast$ denotes virtual particles.
The ATLAS Collaboration published results of a search for nonresonant and resonant \HH pairs decaying to \WWWW based on 36\fbinv of proton-proton ($\Pp\Pp$) collision data recorded at $\sqrt{s} = 13$\TeV~\cite{Aaboud:2018ksn}, placing an upper limit of 160 times the SM predicted cross section for nonresonant \HH production at 95\% confidence level (\CL).
Searches for \HH production in $\Pp\Pp$ collisions at $\sqrt{s} = 7$, 8, and 13\TeV have previously been performed by the CMS and ATLAS Collaborations in the decay modes $\Pbottom\Pbottom\PGg\PGg$~\cite{CMS:2020tkr,ATLAS-HDBS-2018-34}, $\Pbottom\Pbottom\Pbottom\Pbottom$~\cite{ATLAS:2020jgy,ATLAS:2022hwc,CMS:2022cpr,CMS-B2G-22-003,Sirunyan:2018zkk}, $\Pbottom\Pbottom\PGt\PGt$~\cite{CMS-HIG-20-010,ATLAS-HDBS-2018-40,ATLAS:2020azv}, $\Pbottom\Pbottom\PW\PWst$~\cite{Sirunyan:2017guj,CMS:2021roc,ATLAS-HDBS-2018-33,ATLAS:2018fpd}, and $\PW\PWst\PGg\PGg$~\cite{Aaboud:2018ewm}.
Limits on \HH production obtained from a combination of some of these analyses have been published by the CMS and ATLAS Collaborations~\cite{Sirunyan:2018ayu,2020135103}.

Searches targeting the $\Pbottom\Pbottom\PGt\PGt$~\cite{CMS-HIG-20-010}, $\Pbottom\Pbottom\Pbottom\Pbottom$~\cite{CMS:2022cpr,CMS-B2G-22-003}, and $\Pbottom\Pbottom\PGg\PGg$~\cite{CMS:2020tkr} final states in CMS, and $\Pbottom\Pbottom\PGt\PGt$~\cite{ATLAS-HDBS-2018-40} and $\Pbottom\Pbottom\PGg\PGg$~\cite{ATLAS-HDBS-2018-34} in ATLAS, provide the strongest constraints on nonresonant \HH production to date, with observed (expected) 95\% \CL upper limits ranging from 3.3 to 9.9 (3.9 to 7.8) times the SM predicted cross section.
The corresponding lower bounds on \kappal vary from $-1.5$ to $-3.3$ ($-2.4$ to $-5.0$ expected), with upper bounds between 6.7 and 9.4 (7.7 to 12.0 expected).
The ATLAS $\Pbottom\Pbottom\PGg\PGg$ analysis places a 95\% \CL upper limit of 0.64\pb on resonant \HH production with a mass around 250\GeV (where 0.39\pb was expected)~\cite{ATLAS-HDBS-2018-34}, while the ATLAS resonant $\Pbottom\Pbottom\Pbottom\Pbottom$ search constrains higher mass hypotheses most strongly, with observed and expected limits around 0.01\pb at 1\TeV~\cite{ATLAS:2022hwc}.
The ATLAS $\Pbottom\Pbottom\PGt\PGt$ performs best for many mass points in between~\cite{ATLAS-HDBS-2018-40}.
The only published \HH search using an EFT approach comes from CMS in the $\Pbottom\Pbottom\PGg\PGg$ final state, with 95\% \CL upper limits on the \HH production cross section ranging from 0.1 to 0.6\pb, depending on the EFT scenario~\cite{CMS:2020tkr}.

This paper presents the first search for \PH boson pairs decaying to \WWWW, \WWtt, and \tttt.
Both nonresonant and resonant \HH production in final states with multiple reconstructed leptons, \ie, electrons (\Pe), muons (\PGm), or hadronically decaying tau leptons (\tauh) are covered.
The search is based on LHC $\Pp\Pp$ collision data recorded by the CMS experiment at a center-of-mass energy of 13\TeV, corresponding to an integrated luminosity of 138\fbinv.
Signal candidate events are subdivided into seven mutually exclusive ``search categories'' based on \lep (\Pe, \PGm) and \tauh multiplicity: two same-sign \lep with fewer than two \tauh (\llss), three \lep with no \tauh (\lllnot), four \lep (\llll), three \lep with one additional \tauh (\lllt), two \lep with two \tauh (\lltt), one \lep with three \tauh (\lttt), or four \tauh with no \lep (\noltttt).
In final states with a total of four \lep and \tauh, the charge sum of all \lep and \tauh candidates is required to be zero.
The seven search categories target \HH signal events in which the \PH boson pair decays into \WWWW, \WWtt, or \tttt.
Multivariate analysis (MVA) methods are used to distinguish the \HH signal from backgrounds.

The paper is structured as follows.
A brief overview of the CMS detector is given in Section~\ref{sec:detector}.
Section~\ref{sec:datasets} lists the data sets and simulation samples used.
The reconstruction of \Pe, \PGm, \tauh, and jets, along with various kinematic observables, is detailed in Section~\ref{sec:eventReconstruction}.
This is followed by a description, in Section~\ref{sec:eventSelection}, of the event selection criteria defining the seven search categories.
The multivariate methods used to distinguish the \HH signal from backgrounds are detailed in Section~\ref{sec:analysisStrategy}.
The estimation of these backgrounds is described in Section~\ref{sec:backgroundEstimation}, followed by an outline of the relevant systematic uncertainties in Section~\ref{sec:systematicUncertainties}.
The statistical procedure used to extract limits on the \HH production rate in the SM, as well as constraints on SM coupling strengths, EFT benchmark scenarios, and resonant \HH production rates are presented in Section~\ref{sec:results}.
The paper concludes with a summary in Section~\ref{sec:summary}.

\section{The CMS detector}
\label{sec:detector}

The central feature of the CMS apparatus is a superconducting solenoid of 6\unit{m} internal diameter, providing a magnetic field of 3.8\unit{T}.
Within the solenoid volume are a silicon pixel and strip tracker, a lead tungstate crystal electromagnetic calorimeter (ECAL), and a brass and scintillator hadron calorimeter (HCAL), each composed of one barrel and two endcap sections.
The silicon tracker measures charged particles within the pseudorapidity range $\abs{\eta} < 2.5$ for data recorded in 2016, and within the range $\abs{\eta} < 3.0$ for data recorded in 2017 and 2018.
The ECAL is a fine-grained hermetic calorimeter with quasi-projective geometry, and is divided into a barrel region covering $\abs{\eta} < 1.5$, and two endcaps that extend to $\abs{\eta} = 3.0$.
The HCAL barrel and endcaps similarly cover the region $\abs{\eta} < 3.0$.
Forward calorimeters extend beyond these endcaps to $\abs{\eta} = 5.0$.
Muons are detected within the range $\abs{\eta} < 2.4$ by gas-ionization detectors embedded in the steel flux-return yoke outside the solenoid.
Collision events of interest are selected using a two-tiered trigger system.
The level-1 trigger, composed of custom hardware processors, uses information from the calorimeters and muon detectors to select less than 100\unit{kHz} of events from a 40\unit{MHz} base event rate, within a fixed latency of 4\mus~\cite{Sirunyan:2020zal}.
The second tier, known as the high-level trigger, is a processor farm which runs a version of the full event reconstruction software optimized for fast processing, and reduces the event rate to around 1\unit{kHz} before data storage~\cite{Khachatryan:2016bia}.
A more detailed description of the CMS detector, together with a definition of the coordinate system used and the most relevant kinematic variables, can be found in Ref.~\cite{Chatrchyan:2008zzk}.

\section{Data samples and Monte Carlo simulation}
\label{sec:datasets}

The analyzed $\Pp\Pp$ collision data correspond to an integrated luminosity of 138\fbinv, collected by the CMS detector over three years: 36\fbinv in 2016, 42\fbinv in 2017, and 60\fbinv in 2018~\cite{CMS-LUM-17-003,CMS-PAS-LUM-17-004,CMS-PAS-LUM-18-002}.
This analysis uses triggers requiring one or more reconstructed \Pe, \PGm, or \tauh candidates to be associated with the same collision vertex.
The exact triggers and their thresholds varied slightly from year to year because of changes in luminosity and detector conditions, as well as improvements to the trigger algorithms.
The transverse momentum ($\pt$) thresholds imposed by the trigger on the ``leading'' (highest \pt), ``subleading'' (second-highest \pt), and third \Pe, \PGm, or \tauh, and the corresponding $\eta$ requirements for each year are shown in Table~\ref{tab:triggers}.
All triggers include identification and isolation requirements on the \Pe, \PGm, and \tauh candidates~\cite{Khachatryan:2016bia}.
When combined, the triggers achieve an efficiency of 95--100\% for simulated SM \HH signal events in each of the seven search categories.

\begin{table}[h!]
    \topcaption{
        Selection requirements on \pt and $\eta$ of reconstructed electrons (\Pe), muons (\PGm), and hadronically decaying tau leptons (\tauh) applied by the triggers used in this analysis.
        The trigger \pt thresholds for leading, subleading, and third \Pe, \PGm, or \tauh are separated by commas.
        For trigger thresholds that varied over time, the range of variation is indicated.
    }
    \centering
    \begin{tabular}{l|l}
        Trigger        & Selection requirements for reconstructed \Pe, \PGm, and \tauh objects                \\
        \hline
        Single \Pe     & $\pt(\Pe) > 27$--35\GeV                                                              \\
        Single \PGm    & $\pt(\PGm) > 22$--27\GeV                                                             \\
        [\cmsTabSkip]
        Double \Pe     & $\pt(\Pe) > 23$, 12\GeV                                                              \\
        \Pe + \PGm     & $\pt(\Pe) > 23$\GeV, $\pt(\PGm) > 8$\GeV                                             \\
        \PGm + \Pe     & $\pt(\PGm) > 23$\GeV, $\pt(\Pe) > 8$--12\GeV                                         \\
        Double \PGm    & $\pt(\PGm) > 17$, 8\GeV                                                              \\
        \Pe + \tauh    & $\pt(\Pe) > 24$\GeV, $\pt(\tauh) > 20$--30\GeV, $\abs{\eta(\Pe, \tauh)} < 2.1$       \\
        \PGm + \tauh   & $\pt(\PGm) > 19$--20\GeV, $\pt(\tauh) > 20$--27\GeV, $\abs{\eta(\PGm, \tauh)} < 2.1$ \\
        Double \tauh   & $\pt(\tauh) > 35$--40\GeV, $\abs{\eta(\tauh)} < 2.1$                                 \\
        [\cmsTabSkip]
        Triple \Pe     & $\pt(\Pe) > 16$, 12, 8\GeV                                                           \\
        Two \Pe + \PGm & $\pt(\Pe) > 12$, 12\GeV, $\pt(\PGm) > 8$\GeV                                         \\
        Two \PGm + \Pe & $\pt(\PGm) > 9$, 9\GeV, $\pt(\Pe) > 9$\GeV                                           \\
        Triple \PGm    & $\pt(\PGm) > 12$, 10, 5\GeV                                                          \\
    \end{tabular}
    \label{tab:triggers}
\end{table}

Monte Carlo (MC) simulated samples are used to model \HH signal events and a wide range of SM background processes that produce final states with \Pe, \PGm, or \tauh.
Background MC samples include processes producing a single \PW or \PZ boson, two bosons ($\PW\PW$, $\PW\PZ$, $\PZ\PZ$, $\PW\Pgg$, and $\PZ\Pgg$), three bosons ($\PW\PW\PW$, $\PW\PW\PZ$, $\PW\PZ\PZ$, $\PZ\PZ\PZ$, and $\PW\PZ\Pgg$), a single \PH boson (via gluon fusion, vector boson fusion, or associated production with a $\PW$ or $\PZ$ boson), a single top quark, a top quark-antiquark pair (\ttbar), and top quarks associated with one or more bosons (\ttW, \ttZ, \ttH, \tHq, and \tHW).
All MC samples were generated using either \MGvATNLO v2~\cite{Alwall:2014hca,Kalogeropoulos:2018cke}, \POWHEG v2~\cite{Nason:2004rx,Frixione:2007vw,Alioli:2010xd}, \MCFM v7~\cite{MCFM1,MCFM2,MCFM3}, or \PYTHIA v8.2~\cite{Sjostrand:2014zea}.
All samples that include a \PH boson were produced for a \PH boson mass of 125\GeV.
Specific details of the simulated processes are summarized in Table~\ref{tab:MC_datasets}.

\begin{table}[h!]
    \topcaption{
        The MC generators that are used to simulate \HH signal and background processes.
        The order of MC simulation and cross section calculation both refer to the perturbative expansion in QCD.
        Additional higher order electroweak (EW) corrections, if present, are indicated separately.
    }
    \centering
    \cmsCommonTable{
        \begin{tabular}{lcc}
            Process   & MC generator (order) & Cross section order     \\
            \hline
            \ggHH                                                         & \MGvATNLO v2 (LO)~\cite{Hespel:2014sla,Oliveira:2014kla} & NNLO FT$_\text{approx}$  \\
                                                                          & \POWHEG v2 (NLO)~\cite{POWHEGHH1,POWHEGHH2,POWHEGHH3}    &                          \\
            \qqHH                                                         & \MGvATNLO v2 (LO)                                        & N3LO                     \\
            [\cmsTabSkip]
            Single \PH boson production                                   &                                                          &                          \\
            \hspace*{1.0cm} (via gluon fusion)                            & \POWHEG v2 (NLO)~\cite{POWHEGGGFH}                       & N3LO QCD, NLO EW         \\
            \hspace*{1.0cm} (via vector boson fusion)                     & \POWHEG v2 (NLO)~\cite{POWHEGVBFH}                       & NNLO QCD, NLO EW         \\
            \hspace*{1.0cm} (with a \PW or a \PZ boson)                   & \POWHEG v2 (NLO)~\cite{POWHEGVH}                         & NNLO QCD, NLO EW         \\
            \hspace*{1.0cm} (with a pair of top quarks)                   & \MGvATNLO v2 (NLO)                                       & NLO QCD, NLO EW          \\
            \hspace*{1.0cm} (with a single top quark)                     & \MGvATNLO v2 (LO)                                        & NLO                      \\
            [\cmsTabSkip]
            \PW                                                           & \MGvATNLO v2 (LO)                                        & NNLO                     \\
            \PZ                                                           & \MGvATNLO v2 (LO)                                        & NNLO QCD, NLO EW         \\
            [\cmsTabSkip]
            $\PW\PW$ (double-parton interaction)                          & \PYTHIA v8.2 (LO)                                        & LO                       \\
            \phantom{$\PW\PW$} (same-sign pair)                           & \MGvATNLO v2 (LO)                                        & LO                       \\
            \phantom{$\PW\PW$} (opposite-sign pair)                       & \POWHEG v2 (NLO)~\cite{POWHEGVV1,POWHEGVV2}              & NNLO                     \\
            $\PW\PZ$                                                      & \MGvATNLO v2 (NLO)                                       & NNLO                     \\
            $\PZ\PZ$ (quark-initiated)                                    & \POWHEG v2 (NLO)~\cite{POWHEGVV1,POWHEGVV2}              & NNLO                     \\
            \phantom{$\PZ\PZ$} (gluon-initiated)                          & \MCFM v7 (LO)~\cite{MCFMGGZZ}                            & NLO                      \\
            [\cmsTabSkip]
            $\PW\Pgg$, $\PZ\Pgg$, $\PW\PZ\Pgg$, $\cPqt\Pgg$, $\ttbar\Pgg$ & \MGvATNLO v2 (NLO   )                                    & NLO                      \\
            [\cmsTabSkip]
            $\PW\PW\PW$, $\PW\PW\PZ$, $\PW\PZ\PZ$, $\PZ\PZ\PZ$            & \MGvATNLO v2 (NLO)                                       & NLO                      \\
            [\cmsTabSkip]
            Single top                                                    & \POWHEG v2 (NLO)~\cite{POWHEGST2}                        & NLO                      \\
            \hspace*{1.0cm} (with a $\PW$ boson)                          & \POWHEG v2 (NLO)~\cite{POWHEGST1}                        & NLO                      \\
            \hspace*{1.0cm} (with a $\PZ$ boson)                          & \MGvATNLO v2 (NLO)                                       & NLO                      \\
            $\ttbar$                                                      & \POWHEG v2 (NLO)~\cite{POWHEGTTBAR}                      & NNLO                     \\
            $\ttbar\ttbar$                                                & \MGvATNLO v2 (NLO)                                       & NLO                      \\
            [\cmsTabSkip]
            $\ttbar\PW$                                                   & \MGvATNLO v2                                             & NLO QCD, NLO EW          \\
                                                                          & (NLO QCD, NLO EW   )                                     &                          \\
            $\ttbar\PZ$                                                   & \MGvATNLO v2 (NLO)                                       & NLO QCD, NLO EW          \\
            $\ttbar\PW\PW$, $\ttbar\PW\PZ$, $\ttbar\PZ\PZ$                & \MGvATNLO v2 (LO)                                        & LO                       \\
        \end{tabular}
    }
    \label{tab:MC_datasets}
\end{table}

The parton distribution functions (PDFs) of the proton are modeled using the \textrm{NNPDF3.0} and \textrm{NNPDF3.1} PDF sets~\cite{NNPDF:2014otw,NNPDF:2017mvq,Butterworth:2015oua,Rojo:2015acz,Accardi:2016ndt}.
Parton shower, hadronization processes, and \PGt decays are modeled by \PYTHIA, using the tunes \textrm{CP5}, \textrm{CUETP8M1}, \textrm{CUETP8M2}, or \textrm{CUETP8M2T4}~\cite{PYTHIA_CUETP8M1tune_CMS,Sirunyan:2019dfx,PYTHIA_MonashTune}, depending on the process and the data-taking period that is being modeled.
The matching of matrix elements to parton showers is performed using the \textrm{MLM} scheme~\cite{Alwall:2007fs} for the LO samples and the \textrm{FxFx} scheme~\cite{Frederix:2012ps} for the NLO samples.
The interactions of particles with the CMS detector material was simulated in detail using \GEANTfour~\cite{Agostinelli:2002hh}.
Simulated events were reconstructed using the same procedure as in data.
The response of the trigger is included in the simulation.
Additional $\Pp\Pp$ interactions (pileup) were generated with \PYTHIA and overlaid on all MC events, with event weights used to match the collision multiplicity to the distribution inferred from data.
Residual differences between data and simulation are rectified by applying corrections to simulated events.

A variety of \HH signal samples were generated at LO and NLO accuracy in QCD to simulate nonresonant \HH production, covering the \ggHH and \qqHH production processes,
with the \PH bosons decaying to either $\PW\PWst$, $\PZ\PZst$, or $\PGt\PGt$.
The NLO samples are used to extract the rate of the \HH signal from the data, while LO samples with a larger number of simulated events are used to train machine learning algorithms.
Separate \ggHH samples are produced for SM \HH production and for a total of twelve EFT benchmark (BM) scenarios in the Higgs Effective Field Theory (HEFT) approach~\cite{Carvalho:2015ttv}.
These benchmarks, along with the seven benchmarks from Ref.~\cite{Capozi:2019xsi}, represent different combinations of \kappal, \kappat, \cg, \cgg, and \ctwo HEFT parameter values, and are chosen to probe distinct classes of \HH kinematic configurations.
These benchmarks are referred to as JHEP04 BM1-12, and JHEP03 BM1-7, respectively.
The benchmark JHEP04 BM8 is complemented by a modified version of this benchmark, published in Ref.~\cite{Buchalla:2018yce}, denoted as JHEP04 BM8a.
The parameter values of these twenty BM scenarios are shown in Table~\ref{tab:HH_benchmarks}.
The values of the \cg and \cgg couplings published in Ref.~\cite{Capozi:2019xsi} have been scaled by factors of 1.5 and $-3$, respectively, to convert them to the convention introduced for these couplings in Ref.~\cite{Carvalho:2015ttv}.
In order to increase the number of simulated events and to model kinematic configurations not explicitly generated, such as JHEP03 BM1-7, the \ggHH samples are merged and the events in the merged samples are reweighted, using the procedure documented in Ref.~\cite{Carvalho:2017vnu}, to match the distributions in \mHH and $\abs{\cos\theta^{\ast}}$ computed at NLO accuracy and published in Ref.~\cite{Buchalla:2018yce}.
This procedure is applied to the LO and NLO \ggHH samples separately.
The symbol $\cos\theta^{\ast}$ denotes the cosine of the polar angle of one \PH with respect to the beam axis in the \HH rest frame.
The \qqHH samples are produced only for SM \HH production.

\begin{table}[h!]
    \topcaption{
        Parameter values for \kappal, \kappat, \ctwo, \cg, and
        \cgg in MC samples modeling twenty benchmark scenarios in the EFT approach, plus SM
        \HH production.
    }
    \centering
    \begin{tabular}{l|ccccc}
        Benchmark   & \kappal & \kappat & \ctwo  & \cg     & \cgg   \\
        \hline
        JHEP04 BM1  & 7.5     & 1.0     & $-$1.0 & 0.0     & 0.0    \\
        JHEP04 BM2  & 1.0     & 1.0     & 0.5    & $-$0.8  & 0.6    \\
        JHEP04 BM3  & 1.0     & 1.0     & $-$1.5 & 0.0     & $-$0.8 \\
        JHEP04 BM4  & $-$3.5  & 1.5     & $-$3.0 & 0.0     & 0.0    \\
        JHEP04 BM5  & 1.0     & 1.0     & 0.0    & 0.8     & $-$1.0 \\
        JHEP04 BM6  & 2.4     & 1.0     & 0.0    & 0.2     & $-$0.2 \\
        JHEP04 BM7  & 5.0     & 1.0     & 0.0    & 0.2     & $-$0.2 \\
        JHEP04 BM8  & 15.0    & 1.0     & 0.0    & $-$1.0  & 1.0    \\
        JHEP04 BM8a & 1.0     & 1.0     & 0.5    & 4/15    & 0.0    \\
        JHEP04 BM9  & 1.0     & 1.0     & 1.0    & $-$0.6  & 0.6    \\
        JHEP04 BM10 & 10.0    & 1.5     & $-$1.0 & 0.0     & 0.0    \\
        JHEP04 BM11 & 2.4     & 1.0     & 0.0    & 1.0     & $-$1.0 \\
        JHEP04 BM12 & 15.0    & 1.0     & 1.0    & 0.0     & 0.0    \\
        [\cmsTabSkip]
        JHEP03 BM1  & 3.94    & 0.94    & $-$1/3 & 0.75    & $-$1   \\
        JHEP03 BM2  & 6.84    & 0.61    & 1/3    & 0       & 1      \\
        JHEP03 BM3  & 2.21    & 1.05    & $-$1/3 & 0.75    & $-$1.5 \\
        JHEP03 BM4  & 2.79    & 0.61    & 1/3    & $-$0.75 & $-$0.5 \\
        JHEP03 BM5  & 3.95    & 1.17    & $-$1/3 & 0.25    & 1.5    \\
        JHEP03 BM6  & 5.68    & 0.83    & 1/3    & $-$0.75 & $-$1   \\
        JHEP03 BM7  & $-$0.10 & 0.94    & 1      & 0.25    & 0.5    \\
        [\cmsTabSkip]
        SM          & 1.0     & 1.0     & 0.0    & 0.0     & 0.0    \\
    \end{tabular}
    \label{tab:HH_benchmarks}
\end{table}

Resonant \HH production was simulated at LO for both spin-0 (radion) and spin-2 (graviton) scenarios with $m_{\X}$ values of 250, 260, 270, 280, 300, 320, 350, 400, 450, 500, 550, 600, 650, 700, 750, 800, 850, 900, and 1000\GeV.

\section{Event reconstruction}
\label{sec:eventReconstruction}

The CMS particle-flow (PF) algorithm~\cite{Sirunyan:2017ulk} aims to reconstruct and identify each individual particle in an event, using an optimized combination of information from the various elements of the CMS detector.
The particles are subsequently classified into five mutually exclusive types: electrons, muons, photons, and charged and neutral hadrons.
These particles are then combined to reconstruct hadronic \PGt decays, jets, and the missing transverse momentum in the event.

The candidate vertex with the largest value of summed physics-object $\pt^2$ is taken to be the primary $\Pp\Pp$ interaction vertex.
The physics objects used for this determination are the jets, clustered using the infrared and collinear safe anti-\kt algorithm~\cite{Cacciari:2008gp, Cacciari:2011ma}, with the tracks assigned to candidate vertices as inputs, and the associated missing transverse momentum, taken as the negative vector sum of the \pt of those jets.

Electrons are reconstructed within the geometric acceptance of the tracking detectors ($\abs{\eta} < 2.5$) by combining information from the tracker and the ECAL~\cite{CMS:2020uim}.
They are initially identified using an MVA classifier which distinguishes real electrons from hadrons, along with requirements that the track be associated with the collision vertex, and limits on hadronic energy deposits separated by $\Delta R < 0.4$ from the electrons (their ``isolation'').
The angular separation between two particles is defined as $\Delta R = \sqrt{(\eta_{1} - \eta_{2})^{2} + (\phi_{1} - \phi_{2})^{2}}$, where the symbol $\phi$ refers to the azimuthal angle of the particle.
Electrons passing this initial selection are referred to as ``loose''.
In this analysis, events with electrons originating from hadron decays (``nonprompt''), or with hadrons misidentified as electrons, constitute the largest source of background.
This motivates the use of an additional MVA classifier, which is trained to select ``prompt'' electrons from \PW, \PZ, and \PGt lepton decays, and to reject nonprompt or misidentified electrons.
This MVA classifier was previously used for measurements of \ttH production in events with multiple leptons~\cite{Sirunyan:2020icl}.
It combines observables comparing measurements of the electron energy and direction in the tracker and the ECAL, the compactness of the electron cluster, the bremsstrahlung emitted along the electron trajectory, and the electron isolation.
Two levels of thresholds on the output of this MVA classifier are used in the analysis, referred to as the ``tight'' and ``medium'' electron selections for the more and less restrictive thresholds, respectively.
The tight selection has an average efficiency of 60\% for electrons from SM \HH decays.
Only the electrons passing the tight selections are used to reconstruct signal candidate events, while data events with electrons passing the medium selections and failing the tight selections are used to estimate the contribution of misidentified- and nonprompt-electron backgrounds in each search category.
Compared to Ref.~\cite{Sirunyan:2020icl}, this analysis uses lower thresholds on the MVA classifier output for the medium and tight electron selections, in order to increase the efficiency in particular for low-\pt electrons, which frequently appear in the \HH signal events studied in this analysis.
Electrons from photon conversions in the tracker are suppressed by requiring that the track is missing no hits in the innermost layers of the silicon tracker, and is not matched to a reconstructed conversion vertex.
In the \twoLeptonssZeroTau category, further electron selection criteria are applied, which require agreement among three independent measurements of the electron charge, including the Gaussian sum filter and Kalman filter track curvatures, as well as the ECAL supercluster position~\cite{Khachatryan:2015hwa}.
The remaining charge misidentification rate is measured to be less than 0.1\% for $\abs{\eta} < 1.479$, and under 0.4\% for $\abs{\eta} > 1.479$.
The charge quality requirement reduces the electron identification efficiency by about $4\%$.

Muons are reconstructed by extrapolating tracks in the silicon tracker to hits in the gas-ionization muon detectors embedded in the steel flux-return yoke outside the solenoid~\cite{Sirunyan:2018}.
To pass the initial loose identification requirement for this analysis, muons must satisfy criteria related to isolation and track proximity to the primary interaction vertex, as well as track quality observables and matching between the tracker and muon chambers.
Additional requirements on the prompt vs.\ nonprompt muon identification MVA classifier from Ref.~\cite{Sirunyan:2020icl} serve to select muons passing a tight selection for signal candidate events, and a medium selection for nonprompt background estimation.
Inputs to this MVA classifier include energy deposits close to the muon in the ECAL and HCAL, the hits and track segments reconstructed in the muon detectors located outside the solenoid, the quality of the spatial matching between the track segments reconstructed in the silicon tracker and in the muon detectors, and the isolation of the muon with respect to other particles.
Again, lower selection thresholds on the MVA classifier output compared to Ref.~\cite{Sirunyan:2020icl} bring higher efficiency for the \HH signal, amounting to 80\% per muon in simulated SM \HH events for the tight selection.
In the \twoLeptonssZeroTau channel, the uncertainty in the curvature of the muon track is required to be less than 20\% to ensure a high-quality charge measurement~\cite{Sirunyan:2020icl}.
This requirement reduces the muon identification efficiency by about 2\%.

Hadronic decays of tau leptons are identified using the ``hadrons-plus-strips'' algorithm~\cite{Sirunyan:2018pgf}.
This algorithm classifies individual hadronic decay modes of the \PGt by combining charged hadrons from the PF reconstruction with neutral pions.
The latter are reconstructed by clustering electrons and photons into rectangular strips, which are narrow in $\eta$ but wide in the $\phi$ direction.
The spread in $\phi$ accounts for photons originating from neutral pion decays that convert into electron-positron ($\Pem\Pep$) pairs while traversing the silicon tracker.
The $\Pem$ and $\Pep$ are bent in opposite directions in $\phi$ by the magnetic field, and may further emit bremsstrahlung photons before reaching the ECAL.
The decay modes considered in this analysis produce one charged pion or kaon plus up to two neutral pions (collectively referred to as ``one-prong'' \tauh), or three charged pions or kaons plus zero or one neutral pion (referred to as ``three-prong'' \tauh).
The \textsc{DeepTau} algorithm~\cite{CMS:2022prd} distinguishes true \tauh objects from quark and gluon jets, electrons, and muons using a convolutional artificial neural network (NN)~\cite{lecun1989} with 42 high-level observables as input, together with low-level information obtained from the silicon tracker, ECAL, HCAL, and the muon detectors.
The former include the \pt, $\eta$, $\phi$, and mass of the \tauh candidate, the reconstructed \tauh decay mode, its isolation with respect to charged and neutral particles, and the estimated distance that the \PGt lepton traverses between its production and decay.
For three-prong \tauh candidates, this distance is determined by reconstructing the decay vertex, while for one-prong \tauh candidates, the transverse impact parameter of the charged pion track with respect to the primary $\Pp\Pp$ interaction vertex is used as an estimate of the distance.
The low-level information quantifies the particle activity within two $\eta \times \phi$ grids, centered on the direction of the \tauh candidate: an inner grid of size $0.2 \times 0.2$, filled with $0.02 \times 0.02$ cells, and an outer grid of size $0.5 \times 0.5$ (partially overlapping with the inner grid), with $0.05 \times 0.05$ cells.
Selected \tauh candidates in this analysis must have $\pt > 20$\GeV and $\abs{\eta} < 2.3$, and are subjected to two levels of thresholds on the NN output that separates \tauh from quark and gluon jets, referred to as the tight and medium \tauh selections, respectively.

Hadronic jets (\jet) are reconstructed with the anti-\kt algorithm using the particles reconstructed with the PF algorithm as input, and serve to identify $\PH \to \PW\PWst \to \jet\jet\Plepton\Pnu$ decays in this analysis.
Jets reconstructed with size parameters of 0.4 (``small-radius jets'') and 0.8 (``large-radius jets'') are both used: two small-radius jets to reconstruct the two quarks from low-\pt \PW boson decays, or a single large-radius jet to reconstruct high-\pt \PW boson decays, where the quarks are collimated.
Overlap between small-radius jets and electrons, muons, and $\tauh$ is resolved by discarding those small-radius jets 
that contain one or more PF particles matched to an electron, a muon, or a constituent of a \tauh passing the medium selection criteria.
In case of large-radius jets, electrons and muons passing the loose selection are removed from the collection of PF particles used as input to the jet reconstruction, 
so that leptons produced in $\PH \to \PW\PWst \to \jet\jet\Plepton\Pnu$ decays of Lorentz-boosted \PH bosons are not clustered into those jets.

The effect of pileup on the reconstruction of large-radius jets is mitigated by applying the pileup per particle identification algorithm (PUPPI)~\cite{Sirunyan:2020foa,Bertolini:2014bba} to the collection of particles used as input to the jet reconstruction.
For small-radius jets, the effect of pileup is reduced by removing charged particles identified with pileup vertices from the jet reconstruction, 
and applying corrections to the jet energy to account for neutral particles from pileup.

After calibration, the jet energy resolution at the central rapidities amounts to 15--20\% at 30\GeV, 10\% at 100\GeV, and 5\% at 1\TeV~\cite{Khachatryan:2016kdb}.
This analysis only considers jets reconstructed in the region $\abs{\eta} < 2.4$.
Small-radius jets must have $\pt > 25$\GeV, while large-radius jets must have $\pt > 170$\GeV.
Additional criteria requiring that each large-radius jet contain exactly two identifiable, energetic subjets are applied to specifically select those from boosted hadronic \PW boson decays~\cite{Sirunyan:2019quj}.

Events containing small-radius jets identified with the hadronization of bottom quarks ($\Pbottom$ jets) are vetoed in this analysis.
The \textsc{DeepJet} algorithm~\cite{Bols_2020} exploits observables related to the long lifetime of $\Pbottom$ hadrons and the higher particle multiplicity and mass of $\Pbottom$ jets compared to light quark and gluon jets.
Both ``loose'' and ``medium'' $\Pbottom$ jet selections on the \textsc{DeepJet} output are employed in this analysis, corresponding to $\Pbottom$ jet selection efficiencies of 84 and 70\%, while the misidentification rates for light-quark or gluon jets are 11 and 1.1\%, respectively.

The missing transverse momentum vector $\ptvecmiss$ is computed as the negative vector $\pt$ sum of all the particles reconstructed by the PF algorithm in an event, and its magnitude is denoted as $\ptmiss$~\cite{Sirunyan:2019kia}.
The $\ptvecmiss$ is modified to account for corrections to the energy scale of the reconstructed jets in the event.
A linear discriminant, denoted as \metLD, is employed to remove background events in which the reconstructed \ptmiss arises from resolution effects.
The discriminant is defined by the relation $\metLD = 0.6 \ptmiss + 0.4 \metHT$, where \metHT corresponds to the magnitude of the vector \pt sum of \Pe, \PGm, and \tauh passing the medium selection criteria, and small-radius jets satisfying the criteria detailed above~\cite{Sirunyan:2018shy}.

\section{Event selection}
\label{sec:eventSelection}

Events are selected with the aim of maximizing the acceptance for \HH decays to \WWWW, \WWtt, and \tttt, while simultaneously rejecting the large backgrounds from multijet production, single and pair production of \PW and \PZ bosons, and \ttbar production.
To achieve this, each event must contain multiple reconstructed \lep or \tauh associated with the primary interaction vertex.
The \lep and \tauh may originate from the decay of a $\PW$ boson or a $\PGt$ lepton.
Seven mutually exclusive search categories, distinguished by the number of reconstructed \lep and \tauh candidates, are included in the analysis: \llss, \lllnot, \llll, \lllt, \lltt, \lttt, and \noltttt.
Here ``\ss'' indicates a same-sign \leplep pair, with two leptons of identical electric charge.
The \lep and \tauh candidates selected in any of the seven search categories must pass the tight selection criteria described in Section~\ref{sec:eventReconstruction}.
In addition, they are required to pass category-specific \pt thresholds motivated by the trigger selection.
Further requirements are placed on the sum of \lep and \tauh charges, and, in two categories, on the discriminant \metLD and the multiplicity of jets.

The leading and subleading leptons in the \llss category must pass \pt selection thresholds of 25 and 15\GeV, respectively.
Events in this category are required to contain two or more small-radius jets, or at least one large-radius jet, targeting hadronic \PW boson decays.
Dielectron events must have $\metLD > 30$\GeV and $m(\leplep) < 81$\GeV or $m(\leplep) > 101$\GeV, in order to suppress
charge-misidentified \Zee background.
If the event contains a \tauh, the charge of the \tauh must be opposite to the charge of the leptons.
After this selection, the main backgrounds in the \llss category arise from $\PW\PZ$ production, from $\PW\Pgg$ events in which the photon converts into an $\Pem\Pep$ pair and either the $\Pem$ or the $\Pep$ is not reconstructed, and from events in which one or both reconstructed leptons are due to a nonprompt \lep or a misidentified hadron, as shown in Table~\ref{tab:event_yields}.
The ``other'' background given in the table is dominated by same-sign \PW boson pairs and $\PW\PW\PW$ production.
The \WWWW decay mode accounts for roughly 70\% of SM \HH signal events selected in the \llss category, with \WWtt events accounting for the other 30\%.

In the \lllnot category, the leading, subleading, and third \lep are required to have \pt values greater than 25, 15, and 10\GeV, respectively, and the sum of their charges must be either $+1$ or $-1$.
At least one small- or large-radius jet must be present, and the \metLD quantity must be greater than 30\GeV, or 45\GeV if there is at least one same-flavor opposite-sign (SFOS) \leplep pair in the event.
Again, backgrounds are dominated by $\PW\PZ$ production and events with misidentified \lep.
Notable contributions to the ``other'' background arise from $\PW\PW\PW$ and $\PW\PW\PZ$ production.
The signal composition is similar to the \llss category.

The \llll category has identical lepton selection criteria to the \lllnot category, except that the third \lep must have $\pt > 15$\GeV, and a fourth \lep with $\pt > 10$\GeV is required, and the sum of the four lepton charges is required to be equal to zero.
In this category and all the remaining categories, there are no selection requirements on jets or \metLD.
Almost $70\%$ of signal events come from the \WWWW decay mode, and about 30\% from \WWtt, while $\PZ\PZ$ production accounts for 85\% of the background.

Events in the \lllt category are required to satisfy the \lllnot criteria on the \lep objects, except that an additional \tauh with $\pt > 20$\GeV and charge opposite to the sum of the \lep charges is required.
Background events in which the reconstructed \tauh fails a loose selection on the NN output of the \textsc{DeepTau} algorithm that separates \tauh from electrons, or falls near the ECAL barrel-endcap transition region in $1.460 < \abs{\eta} < 1.558$ are removed.
About 70\% of signal events come from the $\WWtt$ decay mode, while $\PZ\PZ$ production and events with at least one misidentified \lep or \tauh dominate the background.

In the \lltt category, the leading and subleading \lep are required to pass \pt thresholds of 25 and 15\GeV, while the two \tauh must have $\pt > 20$\GeV.
The sum of \lep plus \tauh charges is required to be zero.
Signal contributions are mostly from the \WWtt (60\%) and \tttt (40\%) decay modes, while background contributions arise from $\PZ\PZ$ production and events with a misidentified \lep or \tauh candidate.

In the \lttt category, the \lep is required to satisfy the conditions $\abs{\eta} < 2.1$ and $\pt > 20$ (15)\GeV if it is an electron (muon).
The leading, subleading, and third \tauh must have $\pt > 40$, 30, and 20\GeV, respectively, and the sum of \tauh and \lep charges is required to be zero.
Background events containing a \Zee decay where one electron is misidentified as a \tauh are vetoed by discarding events containing an \Pe-\tauh pair of opposite charge and mass $71 < m(\Pe\tauh) < 101$\GeV, and in which the \tauh either fails a loose selection on the discriminant that separates \tauh from electrons, or falls into the region $1.460 < \abs{\eta} < 1.558$.
Around 80\% of \HH signal events selected in the \lttt category are from \tttt and 20\% from the \WWtt decay mode, while the majority of background events stem from $\PZ\PZ$ production or contain a misidentified \lep or \tauh.

The \noltttt category requires the leading and subleading \tauh to pass \pt thresholds of 40 and 30\GeV, respectively, and the third and fourth \tauh to have $\pt > 20$\GeV.
Given the extremely low backgrounds in this category, no charge sum criterion or \Zee veto is applied.
Almost all signal events come from the \tttt decay mode, while 55\% of the background events contain at least one misidentified \tauh candidate, and the remainder arises from $\PZ\PZ$ (30\%) and single Higgs boson ($15\%$) production.

In all seven search categories, the background contamination from processes with top quarks is reduced by discarding events with at least one selected small-radius jet passing the medium \Pbottom jet identification, or at least two passing the loose \Pbottom jet identification.
Leptons originating from low-mass Drell--Yan production, decays of \PJpsi and \PUpsilon mesons, cascade decays of bottom quarks, and photon conversions are removed by vetoing events containing any pair of loose \lep with mass $m(\leplep) < 12$\GeV.
To eliminate overlap with events selected in the ongoing search for \HH production in the $\Pbottom\APbottom\PZ\PZ$, $\PZ\PZ \to 4\lep$ decay mode, no event in the \llss, \lllnot, and \llll categories may contain two SFOS loose \leplep pairs with a mass of the four-$\lep$ system of less than 140\GeV.
In addition, to reduce the \Zll background, these three categories along with \lltt and \lllt exclude events where any SFOS loose \leplep pair has an invariant mass of 81--101\GeV (\PZ boson veto).

A summary of the event selection criteria applied in the different categories is given in Table~\ref{tab:event_selection0}.
Criteria that are common to all seven search categories are given in Table~\ref{tab:event_selection1}.

Two control regions (CRs) are used to validate the modeling of the $\PW\PZ$ and $\PZ\PZ$ backgrounds.
These CRs match the signal regions of the \threeLeptonZeroTau and \fourLeptonZeroTau categories, but with the \PZ boson veto inverted,
and are referred to as the ``\threeLeptonCR'' CR and ``\fourLeptonCR'' CR, respectively.

The number of events selected in the signal regions of each of the seven search categories and in the \threeLeptonCR and \fourLeptonCR CRs are given in Table~\ref{tab:event_yields}.
The contribution expected from nonresonant \HH production with event kinematics as predicted by the SM, but 30 times the SM cross section, is given separately for \HH decays into \WWWW, \WWtt, and \tttt in the upper three rows of each table.
The event yields given in the rows labeled \WWWW include a small contribution from \HH decays into \WWZZ and \ZZZZ, and, similarly, the numbers quoted in the rows labeled \WWtt include a small contribution from \HH decays into \ZZtt.

\begin{table}[!ht]
    {\centering
        \topcaption{
            Event selection criteria applied in the seven search categories.
            The \pt thresholds for \lep and \tauh with the highest, second-, third-, and fourth-highest \pt are separated by slashes.
            The symbol ``\NA'' indicates that no requirement is applied.
        }
        \label{tab:event_selection0}
        \cmsTableThreeCol{
            \begin{tabular}{lC{4.2cm}C{4.2cm}C{4.2cm}}
                Category                & \llss                                               & \lllnot                                              & \llll                                                \\
                \hline
                Targeted \HH decays     & \WWWW                                               & \WWWW                                                & \WWWW                                                \\
                [\cmsTabSkip]
                Trigger                 & Single- and                                         & Single-, double-                                     & Single-, double-                                     \\
                                        & double-lepton                                       & and triple-lepton                                    & and triple-lepton                                    \\
                [\cmsTabSkip]
                Lepton \pt              & $>$25 / 15\GeV                                      & $>$25 / 15 / 10\GeV                                  & $>$25 / 15 / 15 / 10\GeV                             \\
                Lepton charge sum       & $\pm 2$, with charge quality                        & $\pm 1$                                              & 0                                                    \\
                                        & requirements applied                                &                                                      &                                                      \\
                Dilepton invariant mass & $\abs{m_{\lep\lep} - m_{\PZ}} > 10$\GeV$^{\dagger}$ & $\abs{m_{\lep\lep} - m_{\PZ}} > 10$\GeV$^{\ddagger}$ & $\abs{m_{\lep\lep} - m_{\PZ}} > 10$\GeV$^{\ddagger}$ \\
                [\cmsTabSkip]
                Jets                    & $\geq$2 small-radius jets or                        & $\geq$1 small-radius jet or                          & \NA                                                  \\
                                        & $\geq$1 large-radius jet                            & $\geq$1 large-radius jet                             &                                                      \\
                [\cmsTabSkip]
                Missing $\pt$           & $\metLD > 30$\GeV$^{\mathsection}$                  & $\metLD > 30$\GeV$^{\|}$                             & \NA                                                  \\
            \end{tabular}}

        \vspace*{0.4 cm}

        \cmsTableTwoCol{
            \begin{tabular}{lC{6cm}C{6cm}}
                Category                & \lllt                                                & \lltt                                                \\
                \hline
                Targeted \HH decays     & \WWtt                                                & \WWtt, \tttt                                         \\
                [\cmsTabSkip]
                Trigger                 & Single-, double-,                                    & Single- and                                          \\
                                        & and triple-lepton                                    & double-lepton                                        \\
                [\cmsTabSkip]
                Lepton \pt              & $>$25 / 15 / 10\GeV                                  & $>$25 / 15\GeV                                       \\
                \tauh \pt               & $>$20\GeV                                            & $>$20\GeV                                            \\
                Lepton and \tauh charge & \lep and \tauh charges sum to 0                      & \lep and \tauh charges sum to 0                      \\
                Dilepton invariant mass & $\abs{m_{\lep\lep} - m_{\PZ}} > 10$\GeV$^{\ddagger}$ & $\abs{m_{\lep\lep} - m_{\PZ}} > 10$\GeV$^{\ddagger}$ \\
            \end{tabular}
        }

        \vspace*{0.4 cm}

        \cmsTableTwoCol{
            \begin{tabular}{lC{6cm}C{6cm}}
                Category                & \lttt                                                     & \noltttt                 \\
                \hline
                Targeted \HH decays     & \tttt                                                     & \tttt                    \\
                [\cmsTabSkip]
                Trigger                 & Single-lepton, lepton+\tauh                               & Double-\tauh             \\
                                        & and double-\tauh                                          &                          \\
                [\cmsTabSkip]
                Lepton $\eta$           & $\abs{\eta} < 2.1$                                        & \NA                      \\
                Lepton \pt              & $>$20\GeV (\Pe) or $>$15\GeV (\PGm)                       & \NA                      \\
                \tauh \pt               & $>$40 / 30 / 20\GeV                                       & $>$40 / 30 / 20 / 20\GeV \\
                Lepton and \tauh charge & \lep and \tauh charges sum to 0                           & \tauh charges sum to 0   \\
                [\cmsTabSkip]
                \Zee veto               & $\abs{m_{\Pe\tauh} - 86\GeV} > 15$\GeV$^{\mathparagraph}$ & \NA                      \\
            \end{tabular}
            \par}
    }

    \vspace*{0.2 cm}

    $^{\dagger}$ Applied to all SFOS \leplep pairs and electron pairs with the same charge. \\
    $^{\ddagger}$ Applied to all SFOS \leplep pairs. \\
    $^{\mathsection}$ Only applied to events containing two electrons. \\
    $^{\|}$ Tightened to $\metLD > 45$\GeV if event contains a SFOS \leplep pair. \\
    $^{\mathparagraph}$ For \tauh classified as electrons by the \textsc{DeepTau} algorithm or with $1.460 < \abs{\eta} < 1.558$. \\
\end{table}

\begin{table}[!ht]
    \topcaption{
        Reconstructed object and event selection requirements in all seven search categories.
        Electrons or muons in the \leplep pairs include any leptons passing the loose selection criteria.
    }
    \label{tab:event_selection1}
    \centering
    \cmsCommonTable{
        \begin{tabular}{lC{9cm}}
            Object and event properties     & Selection criteria                                                                    \\
            \hline
            Lepton and \tauh pseudorapidity & $\abs{\eta} < 2.5$ for \Pe, $\abs{\eta} < 2.4$ for \PGm, $\abs{\eta} < 2.3$ for \tauh \\
            Dilepton invariant mass         & $m_{\lep\lep} > 12$\GeV (all \leplep pairs)                                           \\
            Four-lepton invariant mass      & $m_{4\lep} > 140$\GeV (any two SFOS \leplep pairs)                                    \\
            \Pbottom jet veto               & 0 medium and $\leq 1$ loose \Pbottom-tagged small-radius jet                          \\
        \end{tabular}
    }
    \vspace*{0.2 cm}
\end{table}

\begin{table}[!ht]
    \topcaption{
        The number of expected and observed events in each of the seven search categories, and in two CRs, which validate the modeling of the $\PW\PZ$ and $\PZ\PZ$ backgrounds.
        The symbol ``\NA'' indicates that the background is not relevant for the category.
        The \HH signal represents the sum of the \ggHH and \qqHH production processes and is normalized to 30 times the event yield expected in the SM, corresponding to a cross section of about 1\pb.
        The event yields are obtained by performing the event selection and applying appropriate corrections to the simulated events.
        Quoted uncertainties represent the sum of statistical and systematic components.
        Uncertainties that are smaller than half the value of the least significant digit have been rounded to zero.
    }
    \label{tab:event_yields}
    \centering
    \begin{tabular}{lr@{ $\pm$ }lr@{ $\pm$ }lr@{ $\pm$ }l}
        Process                          & \multicolumn{2}{c}{\llss}   & \multicolumn{2}{c}{\lllnot} & \multicolumn{2}{c}{\llll}                                           \\
        \hline
        SM $\HH\to\WWWW$  ($\times\,30$) & 73                          & 6                           & 33                        & 3                          & 2.2  & 0.2 \\
        SM $\HH\to\WWtt$  ($\times\,30$) & 31                          & 3                           & 12                        & 1                          & 0.9  & 0.1 \\
        SM $\HH\to\tttt$  ($\times\,30$) & 3                           & 0                           & 1                         & 0                          & 0.1  & 0.0 \\
        [\cmsTabSkip]
        $\PW\PZ$                         & 1999                        & 122                         & 1318                      & 78                         & 0.4  & 0.1 \\
        $\PZ\PZ$                         & 121                         & 3                           & 109                       & 3                          & 53.9 & 3.1 \\
        Misidentified \lep               & 4842                        & 1327                        & 510                       & 94                         & 2.2  & 1.1 \\
        Conversion electrons             & 804                         & 174                         & 117                       & 24                         & 0.7  & 0.3 \\
        Electron charge misid.           & 394                         & 61                          & \multicolumn{2}{c}{\,\, \NA}  & \multicolumn{2}{c}{\, \NA}              \\
        Single Higgs boson               & 214                         & 6                           & 61                        & 1                          & 2.4  & 0.3 \\
        Other backgrounds                & 2740                        & 338                         & 289                       & 29                         & 4.0  & 0.5 \\
        [\cmsTabSkip]
        Total expected background        & 11\,114                     & 1387                        & 2404                      & 128                         & 63.7 & 3.3 \\
        \hline
        Data                             & \multicolumn{2}{c}{10\,344} & \multicolumn{2}{c}{2621}    & \multicolumn{2}{c}{62}                                              \\
    \end{tabular}
    \begin{tabular}{lr@{ $\pm$ }lr@{ $\pm$ }lr@{ $\pm$ }lr@{ $\pm$ }l}
        Process                          & \multicolumn{2}{c}{\lllt} & \multicolumn{2}{c}{\lltt} & \multicolumn{2}{c}{\lttt}  & \multicolumn{2}{c}{\noltttt}                                                                       \\
        \hline
        SM $\HH\to\WWWW$  ($\times\,30$) & 0.9                       & 0.1                       & 0.2                        & 0.0                          & 0.2                        & 0.0                        & 0.3 & 0.0 \\
        SM $\HH\to\WWtt$  ($\times\,30$) & 4.1                       & 0.3                       & 3.9                        & 0.4                          & 0.6                        & 0.1                        & 0.1 & 0.0 \\
        SM $\HH\to\tttt$  ($\times\,30$) & 0.9                       & 0.1                       & 2.3                        & 0.3                          & 2.6                        & 0.4                        & 1.3 & 0.2 \\
        [\cmsTabSkip]
        $\PW\PZ$                         & 0.2                       & 0.0                       & \multicolumn{2}{c}{$<$0.1} & \multicolumn{2}{c}{$<$0.1}   & \multicolumn{2}{c}{$<$0.1}                                          \\
        $\PZ\PZ$                         & 24.1                      & 1.4                       & 18.4                       & 1.3                          & 1.9                        & 0.2                        & 0.7 & 0.1 \\
        Misidentified \lep and \tauh     & 23.9                      & 6.6                       & 31.9                       & 10.1                         & 2.2                        & 2.1                        & 2.2 & 1.6 \\
        Conversion electrons             & 0.1                       & 0.0                       & 0.1                        & 0.1                          & \multicolumn{2}{c}{$<$0.1} & \multicolumn{2}{c}{$<$0.1}             \\
        Single Higgs boson               & 3.8                       & 0.4                       & 2.8                        & 0.7                          & 0.8                        & 0.4                        & 0.4 & 0.3 \\
        Other backgrounds                & 2.8                       & 0.4                       & 2.2                        & 0.8                          & 0.1 & 0.1  & \multicolumn{2}{c}{$<$0.1}             \\
        [\cmsTabSkip]
        Total expected background        & 54.9                      & 6.8                       & 55.4                       & 10.3                          & 5.0                        & 2.2                        & 3.4 & 1.6 \\
        \hline
        Data                             & \multicolumn{2}{c}{55}    & \multicolumn{2}{c}{55}    & \multicolumn{2}{c}{$6$}    & \multicolumn{2}{c}{$1$}                                                                            \\
    \end{tabular}
    \begin{tabular}{lr@{ $\pm$ }lr@{ $\pm$ }l}
        Process                   & \multicolumn{2}{c}{\threeLeptonCR CR} & \multicolumn{2}{c}{\fourLeptonCR CR}                                 \\
        \hline
        $\PW\PZ$                  & 12\,565                               & 705                                  & \multicolumn{2}{c}{$<$1}      \\
        $\PZ\PZ$                  & 765                                   & 47                                   & 2000                     & 108 \\
        Misidentified \lep        & 804                                   & 211                                  & 13                       & 4  \\
        Conversion electrons      & 106                                   & 21                                   & 2                        & 0  \\
        Other backgrounds         & 625                                   & 76                                   & 60                       & 8  \\
        [\cmsTabSkip]
        Total expected background & 14\,866                               & 742                                  & 2074                     & 108 \\
        \hline
        Data                      & \multicolumn{2}{c}{14\,994}           & \multicolumn{2}{c}{2096}                                             \\
    \end{tabular}
\end{table}

\section{Analysis strategy}
\label{sec:analysisStrategy}

The rate of the \HH signal is extracted through a binned maximum likelihood (ML) fit to the distributions in the output of boosted decision tree (BDT) classifiers~\cite{Breiman:1984jka}, which are trained to discriminate the \HH signal from backgrounds, along with kinematic distributions from the two CRs above.
The data from each of the three years are fit separately.
Three classifiers are trained for each of the seven search categories using a mix of MC simulation from all three years, targeting nonresonant \HH production and resonant \HH production from the decay of heavy particles of spin 0 and of spin 2.
The binning is chosen with the objective of maximizing the sensitivity for the \HH signal, while maintaining sufficient background events in each bin to keep the statistical uncertainty in the background prediction under control.
In the two categories with high event yields (\llss and \lllnot) the BDT output binning is chosen such that each bin contains a similar number of expected \HH signal events.
The four categories containing events with $\tauh$ (\lllt, \lltt, \lttt, and \noltttt) have low event yields and sizable background contributions arising from the misidentification of \lep and \tauh candidates, which are determined from data and statistically limited.
For these categories, we choose the binning for each BDT output distribution such that a similar number of expected background events is contained in each bin.
In the \llll category, the fact that the background is dominated by $\PZ\PZ$ production, which is modeled by the MC simulation with low statistical uncertainties,
allows one to choose the binning in the same way as for the \llss and \lllnot categories.
The number of bins is determined by the condition that the relative statistical uncertainty in the background prediction in each bin does not exceed 15\%.
Higher bin numbers correspond to a higher BDT output value, and feature a higher signal-to-background ratio.
For the SM \HH signal, the bins with the highest BDT output values feature a signal-to-background ratio up to 10 times higher than the inclusive ratio in each category.

\begin{figure}[!ht]
    \centering\includegraphics[width=0.46\textwidth]{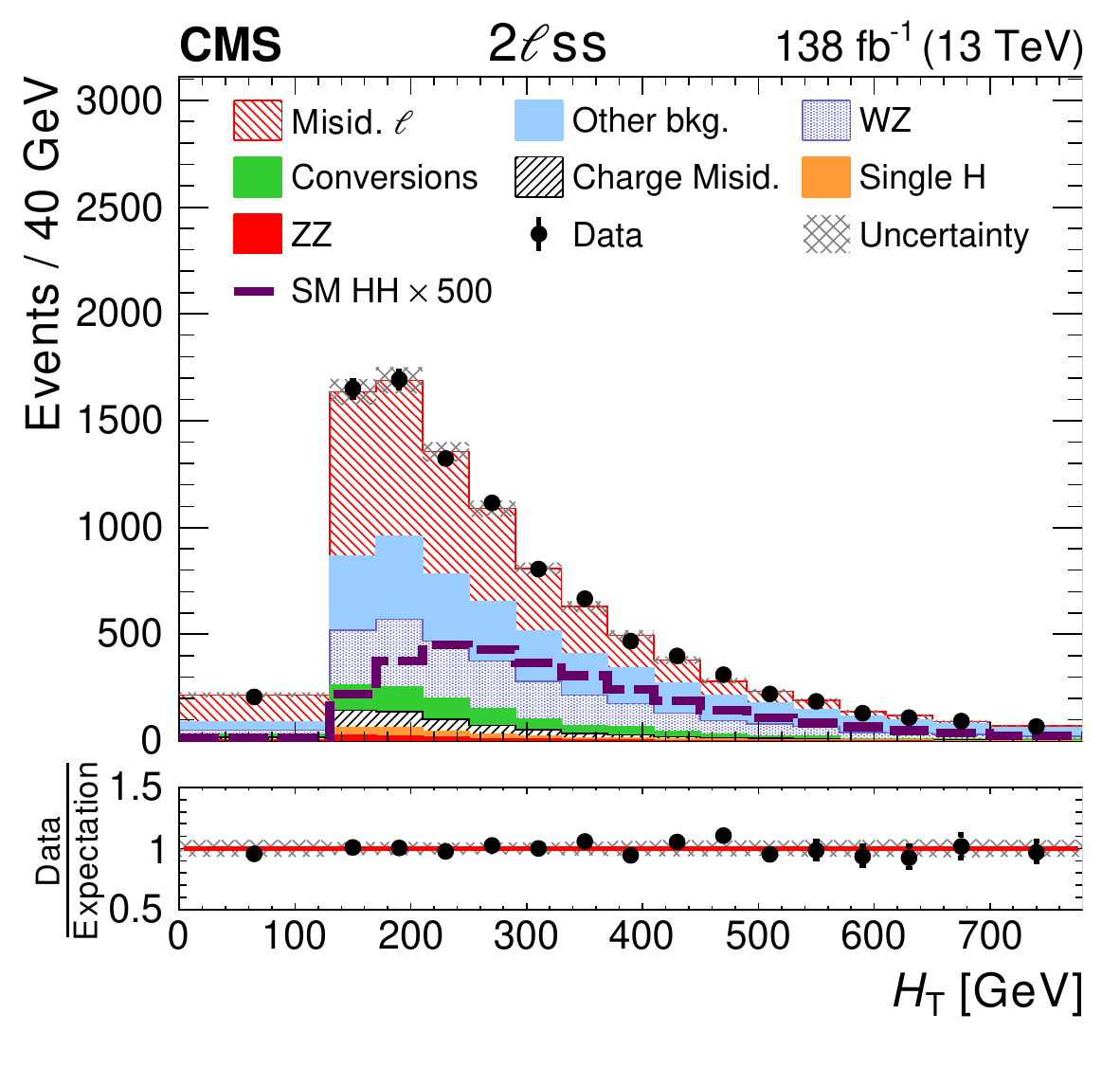}\hfill
    \centering\includegraphics[width=0.46\textwidth]{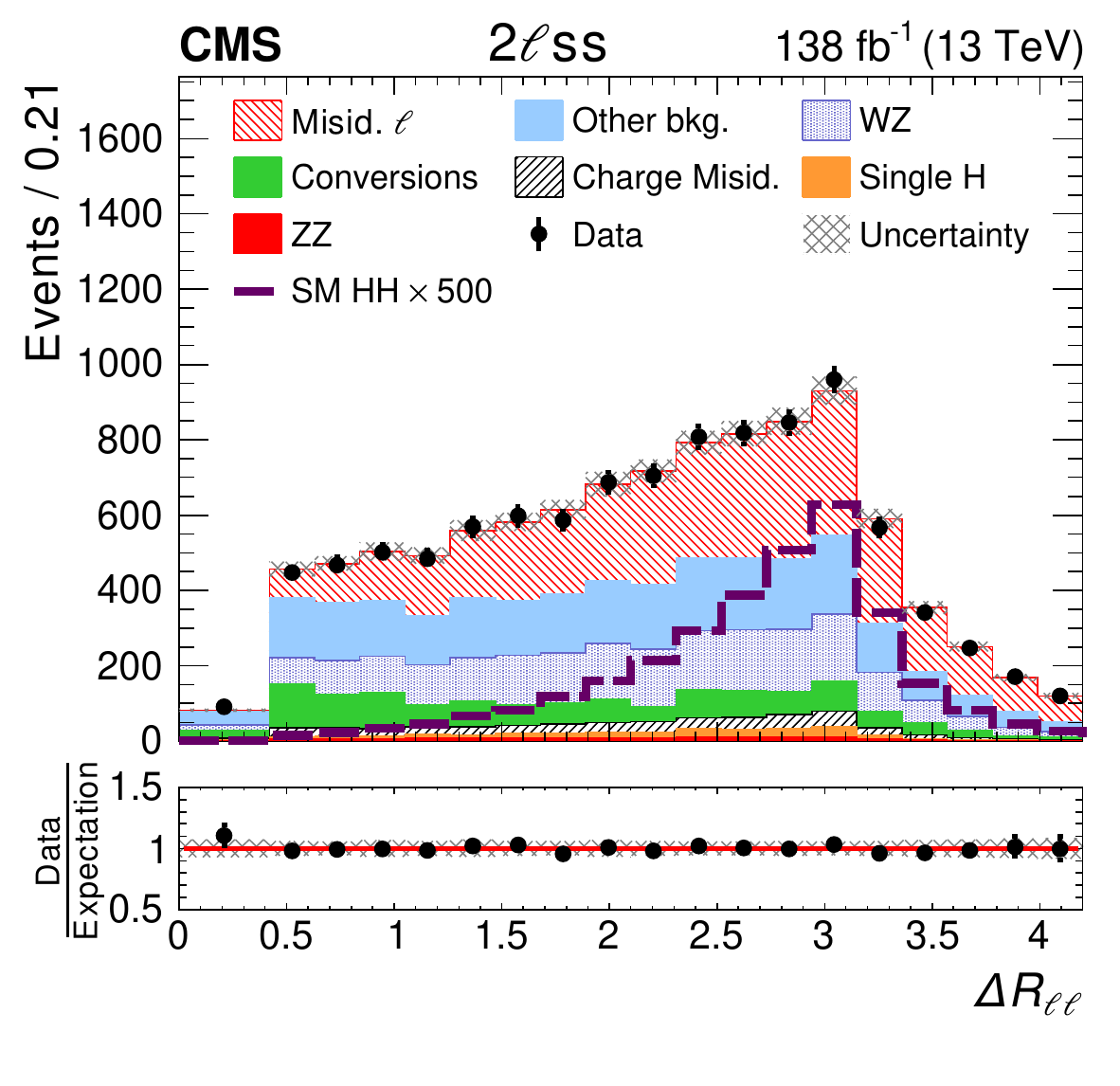}\\
    \centering\includegraphics[width=0.46\textwidth]{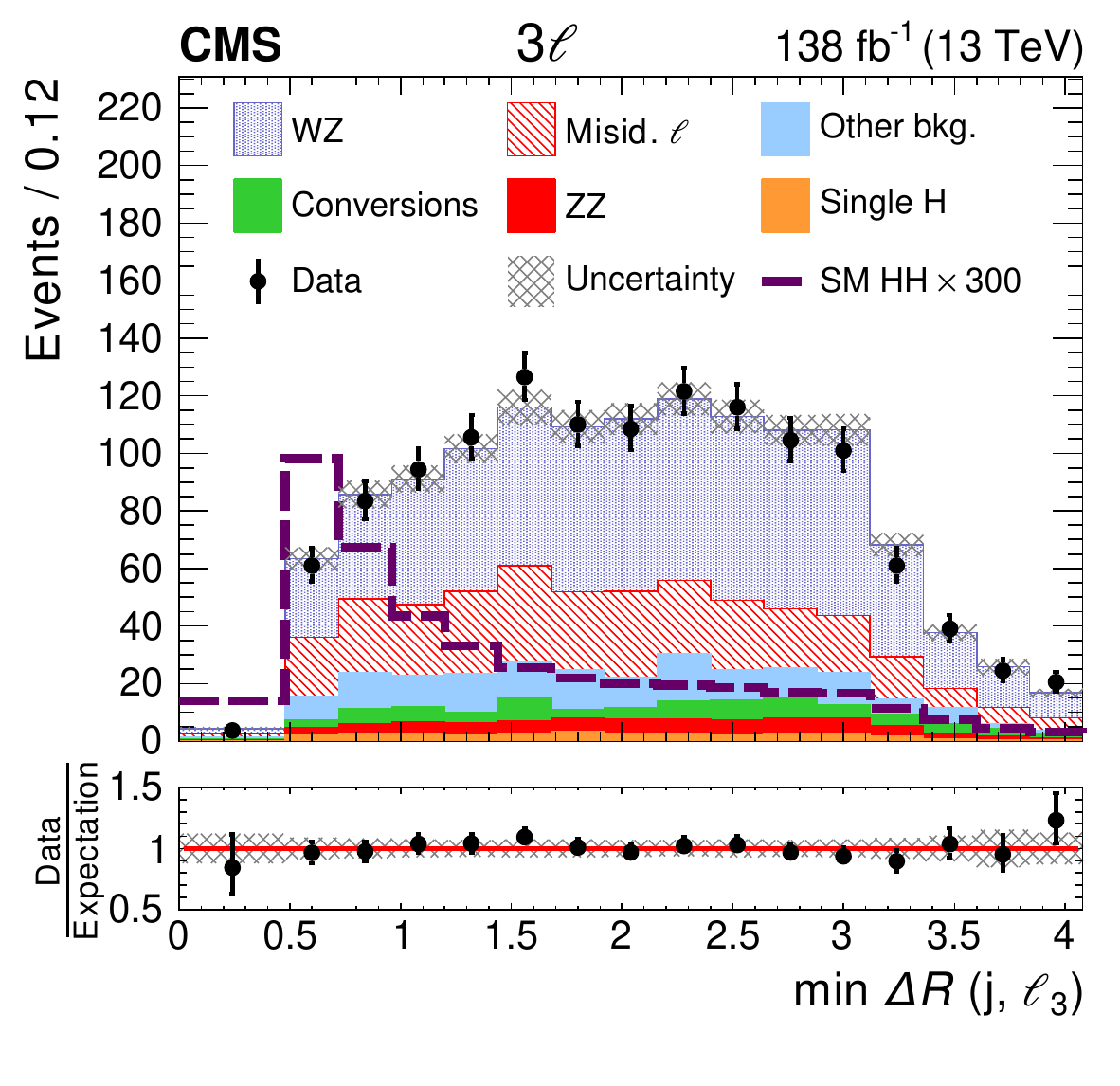}\hfill
    \centering\includegraphics[width=0.46\textwidth]{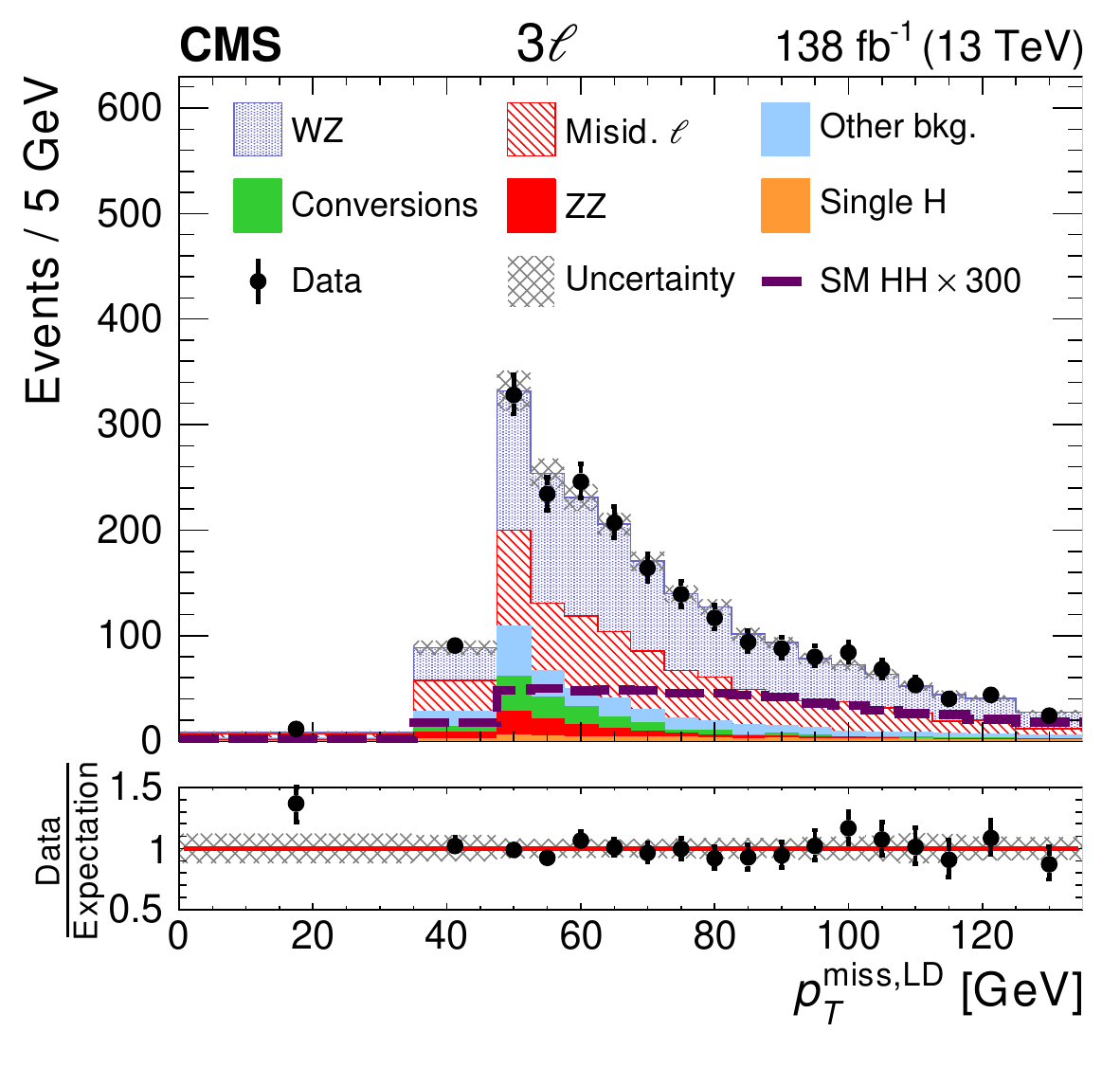}
    \caption{
        Distributions in a few observables used as inputs to the BDT classifiers in the \llss and \lllnot categories: the scalar \pt sum, denoted as \HT, of the two reconstructed \lep and all small-radius jets in the \llss category (\cmsTop \cmsLeft); the angular separation $\Delta R$ between the two \lep in the \llss category (\cmsTop \cmsRight); the angular separation between $\lep_{3}$ and the nearest small-radius jet in the \lllnot category (\cmsBottom \cmsLeft); and \metLD in the \lllnot category (\cmsBottom \cmsRight).
        The $\lep_{3}$ in the \lllnot category is defined as the \lep that is not part of the opposite-sign \leplep pair of lowest mass.
        The normalization and shape of the distributions expected for the different background processes 
        are shown for the values of nuisance parameters obtained from an ML fit in which the \HH signal is constrained to be zero.
        The gray shaded area indicates the sum of statistical and systematic uncertainties on the background prediction obtained from this ML fit.
    }
    \label{fig:bdtinputs}
\end{figure}

The inputs to the BDT classifiers differ by search category and include the \pt and $\eta$ of reconstructed \lep and \tauh; the angular separation $\Delta R$ and invariant mass of \leplep, $\lep\tauh$, and $\tauh\tauh$ pairs; the $\Delta R$ and invariant mass between an \lep or \tauh candidate and the nearest jet(s); the number of jets in the event; the discriminant \metLD; the scalar \pt sum of all reconstructed \Pe, \PGm, \tauh, and jets; the ``visible'' mass of the Higgs boson pair, given by the mass of the system of reconstructed \Pe, \PGm, \tauh, and jets; and where applicable, the ``full'' mass of the \HH system, including neutrinos, reconstructed using the algorithm from Ref.~\cite{Ehataht:2018nql} designed for reconstructing Higgs pair decays into \Pgt leptons.
This algorithm targets \HH signal events decaying to \tttt and thus works best in the \zeroLeptonFourTau and \oneLeptonThreeTau search categories.
Distributions in some of the observables used as inputs to the BDT classifiers in the \llss and \lllnot categories are shown in Fig.~\ref{fig:bdtinputs}.

These observables are complemented by further inputs, which parametrize the BDT as a function of the model parameters: the Higgs boson couplings $\lambda$, \yt, \cg, \cgg, and \ctwo for nonresonant \HH production, and the mass of the heavy particle \X in resonant \HH production.
When training the BDT that targets nonresonant \HH production, the values for the couplings are chosen according to the twelve EFT benchmark scenarios given in Ref.~\cite{Carvalho:2015ttv} and the SM, indicated by thirteen binary inputs to the BDT.
The BDT classifiers used for the analysis of resonant \HH production are trained separately for spin-0 and spin-2 on the full set of resonance masses listed in Section~\ref{sec:datasets}, and the resonance mass is used as an input to the BDT.
Each simulated background event is replicated multiple times in the training sample, with different values assigned to the Higgs boson couplings and the mass of the heavy particle $\X$.

The training is performed using simulated samples of signal and background events.
The signal events used in the training consist of \ggHH events in the \HH decay modes \WWWW, \WWtt, and \tttt.
Background contributions arising from the misidentification of \lep and \tauh candidates and from the mismeasurement of the electron charge are included in the simulation.
The signal and background events used in the training are required to pass the event selection criteria for the respective search category, described in Section~\ref{sec:eventSelection}.
The number of training events is increased by applying the medium \lep and \tauh identification criteria instead of the tight ones.
Weights are applied to background events arising from different sources, such that the relative fractions of different types of backgrounds in the training match the fractions expected in the signal region of the analysis, \ie when the tight \lep and \tauh identification criteria are applied.
The MC samples used for the BDT training overlap with the samples used to model signal and background contributions in the analysis.
To avoid potential biases, the training samples are split into two samples of equal size, based on even and odd event numbers.
The BDTs trained on even events are evaluated on odd events, and vice versa, thereby ensuring that BDTs are not trained and evaluated on the same events.
The training is performed using the $\textsc{XGBoost}$ algorithm~\cite{Chen:2016btl}, interfaced to the $\textsc{Scikit-learn}$ machine learning library~\cite{scikit-learn}.
The parameters of the BDT training (so-called ``hyperparameters'') are optimized using the particle swarm optimization algorithm described in Ref.~\cite{Tani:2020dyi}.

\section{Background estimation}
\label{sec:backgroundEstimation}

Background contributions are classified as either ``reducible'' or ``irreducible''.
In this analysis, three types of reducible backgrounds are considered, arising from misidentified \lep or \tauh, electron charge misidentification, and electrons from photon conversions.
Background events in which all selected \lep and \tauh come from \PW, \PZ, or \PH boson decays, and are reconstructed with the correct charge, are considered ``irreducible''.
The \lep/\tauh misidentification and electron charge misidentification backgrounds are both determined from data,
while electron conversions and irreducible backgrounds are modeled using MC simulation.

The \lep/\tauh misidentification background (which includes nonprompt leptons) is the largest reducible background in all search categories.
Nonprompt \lep are either electrons or muons produced in bottom and charm quark decays, or muons that originate from pion and kaon decays.
Hadronic jets may also be misidentified as electrons or \tauh.
The \lep/\tauh misidentification background estimate is detailed in Section~\ref{sec:backgroundEstimation_fakes}.
The electron charge misidentification background is only relevant for the \twoLeptonssZeroTau search category, and is described in Section~\ref{sec:backgroundEstimation_flips}.
The modeling of photon conversion events by the MC simulation has been validated in data as described in Ref.~\cite{CMS:2017moi,Sirunyan:2020icl}.

The main contribution to the irreducible background arises from $\PW\PZ$ production in the \twoLeptonssZeroTau and \threeLeptonZeroTau categories, and $\PZ\PZ$ production in the remaining five categories.
The production of pairs of bosons (\PGg, \PW, \PZ, or \PH) other than $\PW\PZ$, $\PZ\PZ$ and \HH, and production of bosons with top quarks, including $\PW\PGg$, $\PZ\PGg$, $\PW\PH$, $\PZ\PH$, $\tH$, $\ttH$, \tW, \ttW, \tZ, \ttZ, $\Ptop\PGg$, and $\Ptop\APtop\Pgg$, constitute subdominant additional backgrounds.
The \tZ and \ttZ backgrounds also include contributions from off-shell $\Ptop\APtop\Pggx$ and $\Ptop\Pggx$ production.
Background processes which include at least one top quark are suppressed by the \Pbottom jet veto described in Section~\ref{sec:eventSelection}, but are still sizable compared to the expected \HH signal.
All irreducible backgrounds are modeled using the MC simulation.

\begin{figure}
    \centering\includegraphics[width=0.46\textwidth]{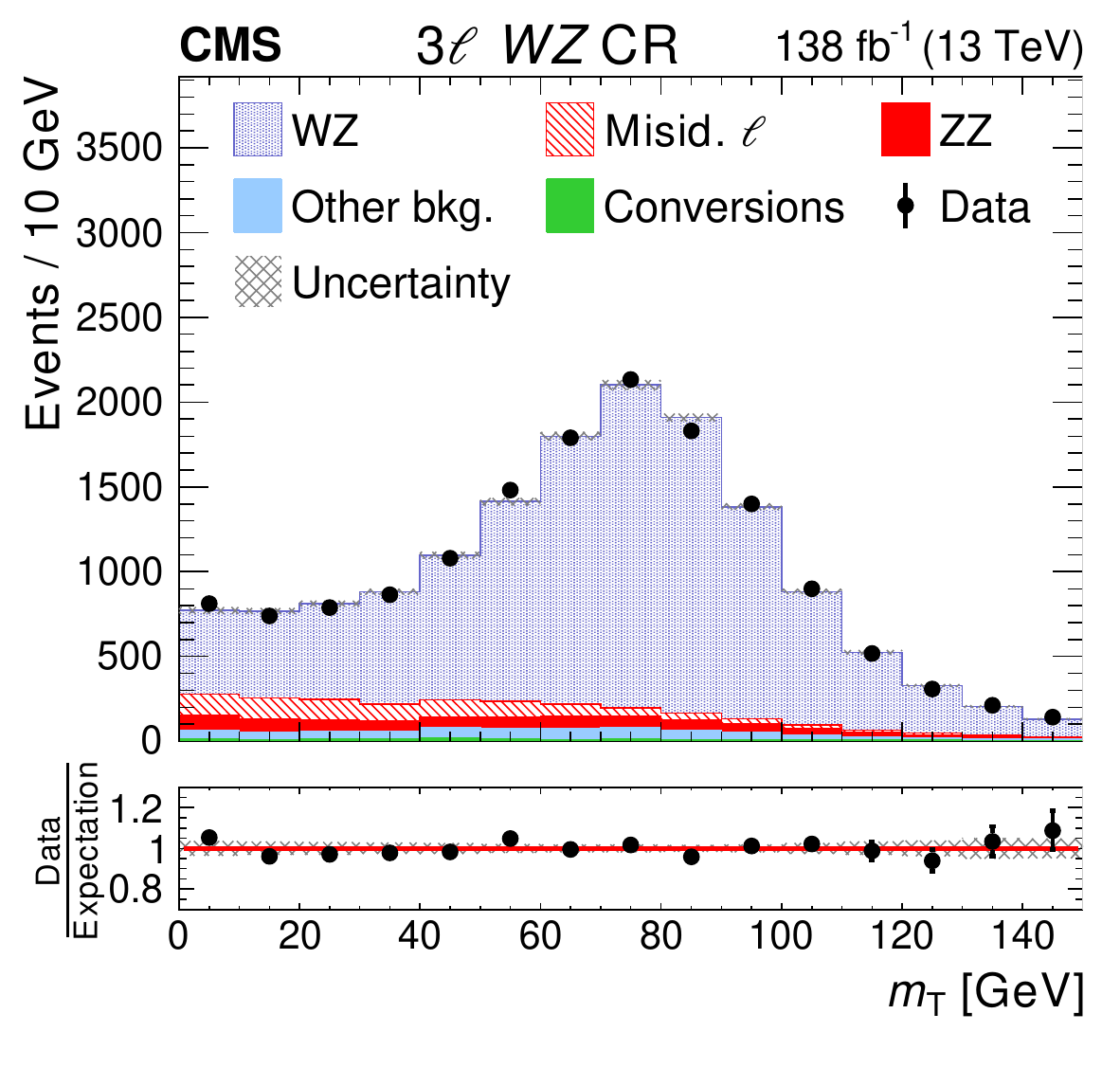}\hfill
    \centering\includegraphics[width=0.46\textwidth]{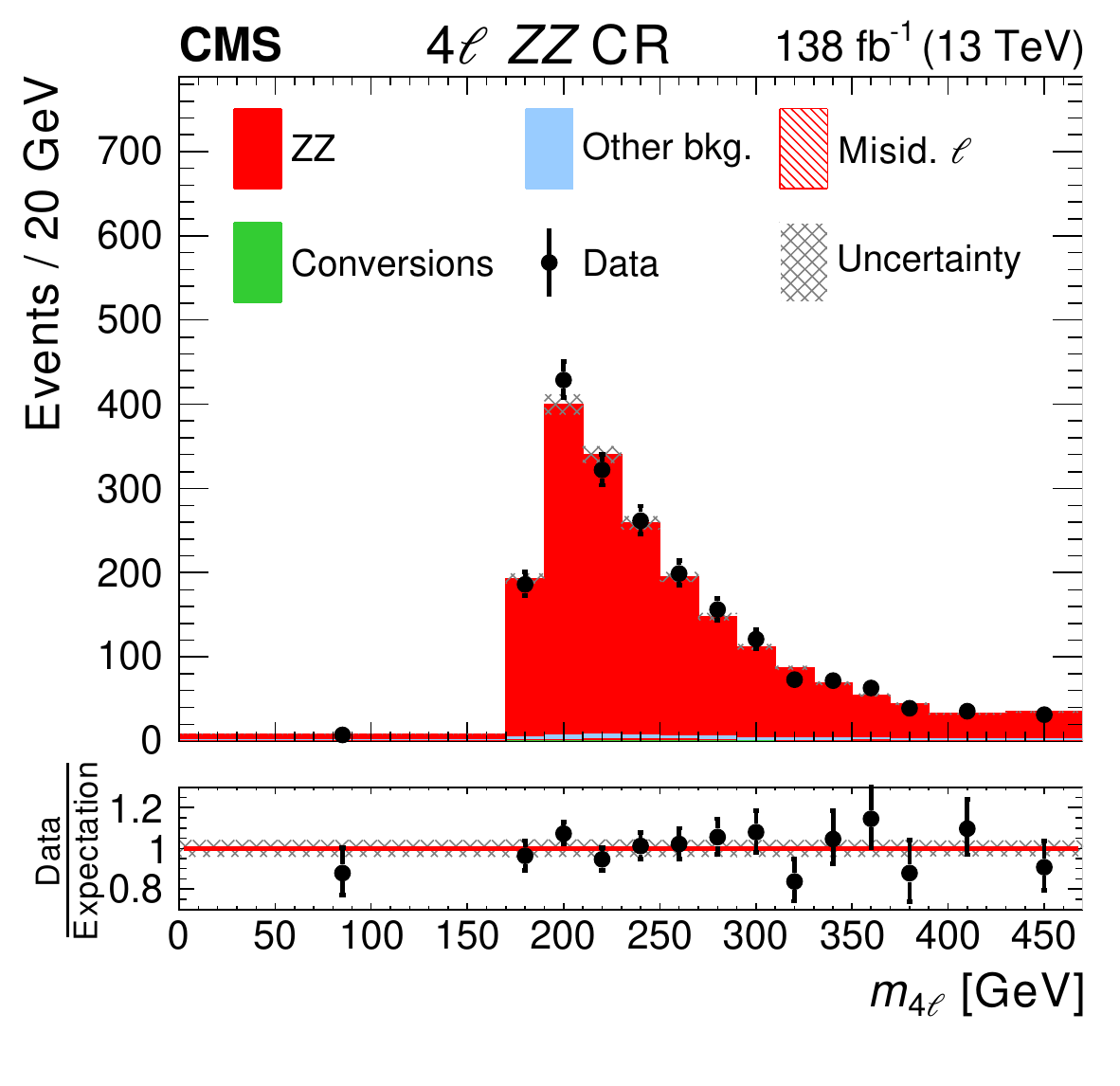}
    \caption{
        Distributions in \mT in the \threeLeptonCR CR (\cmsLeft) and in $m_{4\lep}$ in the \fourLeptonCR CR (\cmsRight).
        The normalization and shape of the distributions expected for $\PW\PZ$, $\PZ\PZ$, and other background processes
        are shown for the values of nuisance parameters obtained from the ML fit described in Section~\ref{sec:results}.
        The gray shaded area indicates the sum of statistical and systematic uncertainties on the background prediction obtained from the ML fit.
    }
    \label{fig:postfitPlotsCR}
\end{figure}

The modeling of the dominant irreducible $\PW\PZ$ and $\PZ\PZ$ backgrounds is validated using the ``\threeLeptonCR'' and ``\fourLeptonCR'' CRs introduced in Section~\ref{sec:eventSelection}.
Distributions in kinematic observables from these CRs (shown in Fig.~\ref{fig:postfitPlotsCR}) are included in the ML fit that is used to extract the $\PH\PH$ signal, described in Section~\ref{sec:results}.
This provides in-situ constraints on the $\PW\PZ$ and $\PZ\PZ$ backgrounds
and on systematic uncertainties related to lepton identification and trigger efficiency.
The transverse mass, $\mT = \sqrt{2 \, \pt^{\lep} \, \ptmiss \left( 1 - \cos\Delta\phi \right)}$, in the \threeLeptonCR CR is computed using the \lep that is not identified as originating from the \PZ boson decay.
The symbol $\Delta\phi$ refers to the angle in the transverse plane between the \lep momentum and the \ptvecmiss.
The observable $m_{4\lep}$ refers to the mass of the $4\lep$ system in the \fourLeptonCR CR.

\begin{figure}
    \centering\includegraphics[width=0.46\textwidth]{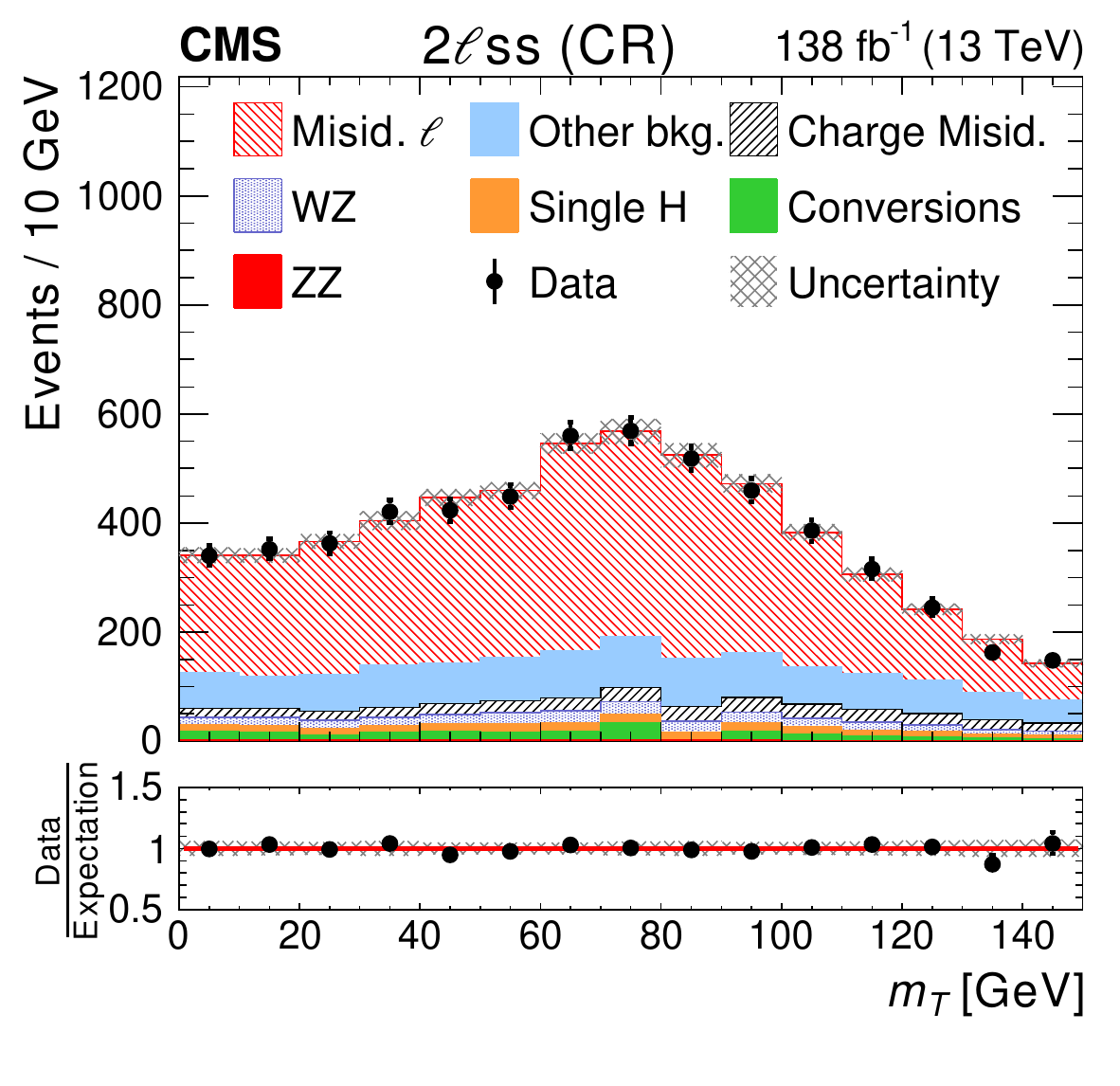}\hfill
    \centering\includegraphics[width=0.46\textwidth]{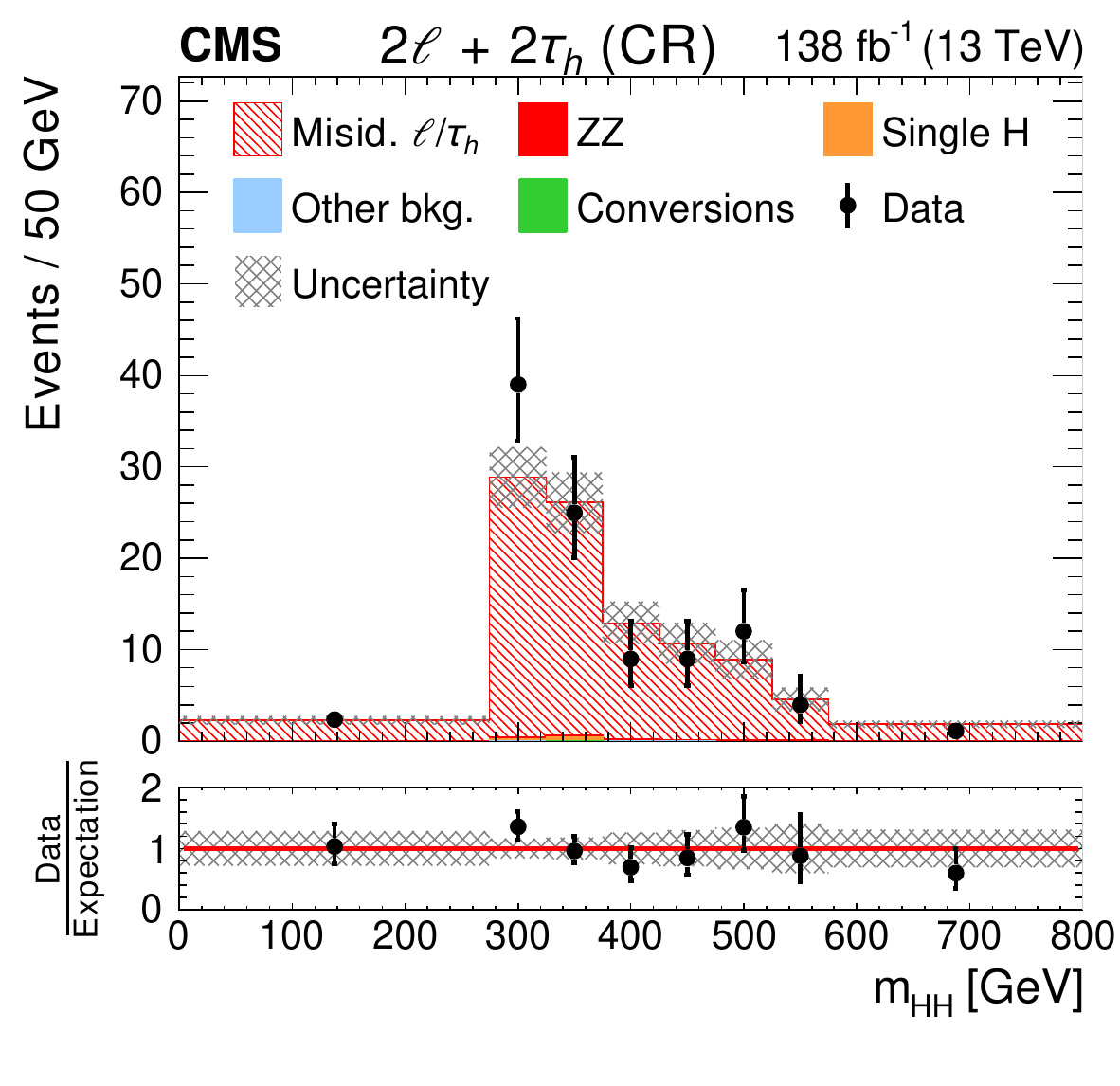}
    \caption{
        Distributions in \mT in the \llss CR (\cmsLeft) and in the mass of the \HH candidate in the \lltt CR (\cmsRight).
        The normalization and shape of the distributions expected for the misidentified \lep/\tauh background and other background processes 
        are shown for the values of nuisance parameters obtained from an ML fit in which the \HH signal is constrained to be zero.
        The gray shaded area indicates the sum of statistical and systematic uncertainties on the background prediction obtained from this ML fit.
    }
    \label{fig:addval}
\end{figure}

The modeling of the reducible \lep/\tauh misidentification background is validated in two further CRs, the ``\llss CR'' and the ``\lltt CR''.
They are based on the signal regions (SRs) of the \llss and \lltt categories.
In the \llss CR, no \Pbottom jet veto is applied, and at least one small-radius jet passing the medium \Pbottom jet identification is required.
The \lltt CR differs from the SR of the \lltt category in that the sum of \lep plus \tauh charges is required to be non-zero, and no \PZ boson veto is applied.
The \llss CR is dominated by events with misidentified \lep, while the \lltt CR is dominated by events with misidentified \tauh.
Distributions in the transverse mass \mT in the \llss CR and in the mass of the \HH candidate in the \lltt CR, reconstructed by the algorithm described in Ref.~\cite{Ehataht:2018nql}, are shown in Fig.~\ref{fig:addval}.
The transverse mass in the \llss CR is computed using the leading \lep.
The data agree well with the background prediction in both CRs.

Simulated events are only considered as irreducible background if every selected \Pe, \PGm, and \tauh candidate matches a prompt MC generator-level counterpart.
Events with at least one selected electron from a photon conversion, and the remaining \lep and \tauh candidates matched to prompt leptons in MC simulation, are classified as conversion background.
Electrons that are misidentified as \tauh, and \tauh that are misidentified as \Pe are also modeled using the MC simulation.
All other simulated events are discarded, as the \lep/\tauh misidentification and charge misidentification backgrounds are estimated from data, as described below.

\subsection{Lepton and \texorpdfstring{\tauh}{hadronic tau} misidentification background}
\label{sec:backgroundEstimation_fakes}

The background from events with misidentified \lep and \tauh candidates is estimated using the ``fake factor'' or ``FF'' method from Ref.~\cite{Sirunyan:2018shy}.
An estimate of this background's contribution to the SR of each search category is obtained by selecting a sample of events that satisfy all selection criteria of the SR for the respective search category, except that the \Pe, \PGm, and \tauh are required to pass the medium selections instead of the tight ones.
The sample of events thus obtained is referred to as the application region (AR) of the FF method.
Events in which every \lep and \tauh satisfies the tight selections are excluded from the AR.

The prediction for misidentification backgrounds in the SR is obtained by applying suitably chosen weights $w$ to the events selected in the AR, where $w$ is given by the expression

\begin{equation}
    w = (-1)^{n+1} \, \prod_{i=1}^{n} \, \frac{f_{i}(\pt, \eta)}{1 - f_{i}(\pt, \eta)}.
    \label{eq:FF_weights}
\end{equation}

The product extends over all \Pe, \PGm, and \tauh that pass the medium, but fail the tight selection criteria, and $n$ refers to the total number of such \lep and \tauh.
The symbol $f_{i}(\pt, \eta)$ corresponds to the probability for a single \Pe, \PGm, or \tauh that passes the medium selection to also pass the tight selection.
These probabilities are measured separately for \Pe, \PGm, and \tauh candidates, parametrized as a function of \pt and $\eta$, and vary between $5$ and $30\%$.
The contributions of irreducible backgrounds to the AR are subtracted based on the MC expectation of such processes.
The alternating sign in Eq.~(\ref{eq:FF_weights}) is necessary to avoid double-counting arising from events with more than one misidentified \lep or \tauh~\cite{Sirunyan:2018shy}.

The probabilities $f_{i}(\pt, \eta)$ for electrons and muons are measured in multijet events, as described in Ref.~\cite{Sirunyan:2020icl}.
The $f_{i}(\pt, \eta)$ for \tauh are measured using $\Zmm$+jets events, where the misidentified \tauh candidates arise from quark or gluon jets.
These events are selected by requiring a muon pair passing the tight selection,
with opposite charge and invariant mass $60 < m_{\PGm\PGm} < 120$\GeV, plus at least one \tauh candidate that passes the medium \tauh selection.
The leading and subleading muons must have $\pt > 25$ and 15\GeV, respectively.
Events must also pass the $\Pbottom$ jet veto described in Section~\ref{sec:eventSelection} to remove \ttbar background.

\subsection{Charge misidentification background}
\label{sec:backgroundEstimation_flips}

The electron charge misidentification background in the $\twoLeptonssZeroTau$ category is estimated using the method described in Ref.~\cite{Sirunyan:2020icl}.
A sample of dielectron events passing all selection criteria of the SR of the $\twoLeptonssZeroTau$ category, except that both electrons are required to have opposite- instead of same-sign charge, is selected and assigned appropriately chosen weights.
The weights are computed by summing the probabilities for the charge of either electron to be mismeasured.
The probability for the mismeasurement of the electron charge is determined using $\Zee$ events, and ranges from under 0.1\% in the barrel up to 0.4\% in the endcap.
The probability for mismeasuring the charge of muons is negligible~\cite{Sirunyan:2020icl}.

\section{Systematic uncertainties}
\label{sec:systematicUncertainties}

Multiple sources of systematic uncertainty affect the predicted event yields, the distributions in the output of the BDT classifiers, or both.
These uncertainties may be theoretical, affecting the predicted cross section or decay kinematics of the collision process, or experimental, accounting for differences in object reconstruction and calibration between data and the MC simulation, or for uncertainties on the estimates of the \lep/\tauh misidentification and electron charge misidentification background obtained from data.
The systematic uncertainties may be correlated or uncorrelated across the three data-taking years, and among the various signal and background processes considered in the analysis.

The SM prediction for the \ggHH production cross section at $\sqrt{s} = 13$\TeV has a relative uncertainty of $+6.7$\%/$-23.2$\%~\cite{Baglio:2020wgt}, while the \qqHH cross section uncertainty is $\pm 2.1\%$~\cite{Dreyer:2018qbw}.
The predicted \PH boson decay branching fractions to $\PW\PWst$, $\PGt\PGt$, and $\PZ\PZst$ have relative uncertainties of 1.54\%, 1.65\%, and 1.54\%, respectively~\cite{LHCHiggsCrossSectionWorkingGroup:2016ypw}.
Correlations between these uncertainties have a negligible effect.
Alternate \HH predictions are generated with the renormalization and factorization scales varied up and down by a factor of 2.
Variations that increase the factorization scale and decrease the renormalization scale (and vice versa) are excluded, following the recommendation of Ref.~\cite{LHCHiggsCrossSectionWorkingGroup:2016ypw}.
All theoretical uncertainties in the \HH signal model are correlated across all three data-taking years and among the seven search categories.
The uncertainties in the \PH boson decay branching fractions and the effect of renormalization and factorization scale uncertainties in the signal acceptance impact the measurement of cross sections for both nonresonant and resonant \HH production.
Conversely, the uncertainties in the SM prediction for the \ggHH and \qqHH cross sections only affect the measurement of the \HH production rate as a ratio to the SM prediction.

Theoretical uncertainties also affect the irreducible background prediction.
The relative uncertainties in the cross sections of the dominant $\PW\PZ$ and $\PZ\PZ$ backgrounds are 2.1 and 6.3\%, respectively~\cite{GRAZZINI2016179,CASCIOLI2014311,PhysRevD.92.094028}.
The uncertainties in the cross sections for the subdominant single \PH boson backgrounds range from 2 to 9\% for \ggHH, \qqHH, $\PW\PH$, and $\PZ\PH$.
The cross sections for the production of \PW, \PZ, or \PH bosons with one or two top quarks are known with uncertainties of 8--15\%.
The event yields of extremely rare backgrounds not mentioned above (\eg, triple boson or four top quark production) are given a conservative uncertainty of 50\%, since the analysis has little sensitivity to these processes.
Following Ref.~\cite{Sirunyan:2020icl}, background contributions arising from photon conversions are assigned a 30\% yield uncertainty.
The theoretical uncertainties affecting background cross sections are partially correlated among different processes.
Here, contributions arising from uncertainties in the proton PDFs are correlated among processes with a similar initial state.
Processes involving single \PH boson production are an exception.
These uncertainties are uncorrelated from other background processes but correlated among each other depending on the initial state.
Uncertainties arising from the choice of the renormalization and factorization scales are correlated for processes with similar production modes, 
for example among all processes involving diboson production ($\PW\PW$, $\PW\PZ$, $\PZ\PZ$, $\PW\Pgg$, and $\PZ\Pgg$).
Uncertainties in $\alpha_{s}$ are correlated among all background processes.
The theoretical cross section uncertainties for signal processes are uncorrelated with those of background processes, but otherwise follow the same uncertainty scheme for proton PDF, scale, and $\alpha_{s}$ contributions.
All theoretical cross section uncertainties are treated as correlated across the different data-taking years and among all seven search categories.

The rate of the misidentified \lep/\tauh background is assigned a 30\% uncertainty in all search categories,
to account for variations in the misidentification rates between the ARs of the FF method and the multijet ($\Zmm$+jets) event samples used to measure the $f_{i}(\pt, \eta)$ for \Pe and \PGm (\tauh).
In the \lllt and \lttt categories, an additional uncertainty of 30\% (uncorrelated with the other 30\% uncertainty) is assigned to the rate of the misidentified \lep/\tauh background, to account for the extra uncertainty arising from the modified \tauh selection criteria that suppress the misidentification of electrons as \tauh.
The effect of statistical uncertainties in the probabilities $f_{i}(\pt, \eta)$ for electrons and muons is evaluated by varying these probabilities in bins of \pt and $\eta$ and determining the resulting change in the shape of the BDT classifier output distribution obtained for the misidentified \lep/\tauh background.
For \tauh, the effect of statistical uncertainties in $f_{i}(\pt, \eta)$ is evaluated by fitting the probabilities in bins of $\eta$ with functions that are linear in \pt, varying the slope of these functions up and down within the uncertainties obtained from the fit, and determining the resulting change in the shape of the BDT classifier output distribution.

An additional uncertainty in the BDT output shape in each category is evaluated for events with a nonprompt or misidentified \lep or \tauh as follows: Simulated events passing all signal selection criteria are compared to those with at least one \lep or \tauh candidate failing the tight identification criteria, scaled according to the FF method described in
Section~\ref{sec:backgroundEstimation}, but with the probabilities $f_{i}(\pt, \eta)$ taken from the MC simulation instead of from the data.
The ratio of these two shapes is fitted with a linear function, which is convoluted with the misidentified \lep/\tauh background prediction from the data to serve as an uncertainty in the BDT output shape for these events in the SR.
The systematic uncertainties associated with the misidentified \lep/\tauh background prediction and the uncertainty associated with the electron charge misidentification rate are treated as uncorrelated among the different data-taking years.

The rate of the electron charge misidentification background in the \llss category is assigned a 30\% uncertainty.
It covers the uncertainty on the electron charge misidentification rates measured in $\Zee$ events,
including the effect of background contamination in these samples, and accounts for differences observed in the following ``closure'' test:
Simulated events are required to pass all signal selection criteria of the \llss category, except that the two leptons are required to have opposite-sign electric charges.
The selected events are scaled according to the electron charge misidentification probability in simulated events, determined by applying the procedure detailed in Section~\ref{sec:backgroundEstimation_flips} to MC simulation.
The resulting background estimate is compared to the one obtained by applying the nominal signal selection criteria of the \llss category to simulated events.

Uncertainties in the modeling of the trigger and object reconstruction efficiency affect all signal and background processes that are estimated using MC simulation.
Trigger efficiencies for events with at least two \lep are compared between data and MC simulation in control regions enriched in the \ttbar, $\PW\PZ$, and $\PZ\PZ$ background processes, as a function of lepton flavor, \pt, and $\eta$.
This results in a small \pt-dependent uncertainty correlated between the \llss and \lltt categories, and a 1\% normalization uncertainty, which is correlated among the \lllnot, \lllt, and \llll categories.
The data-to-simulation agreement in trigger efficiency for the \noltttt and \lttt categories is computed using an independent set of data, as a function of the \pt and $\eta$ of the \lep and all \tauh, and the reconstructed decay modes of all \tauh.
The trigger uncertainties for these two categories are treated as uncorrelated.
All systematic uncertainties related to trigger modeling are correlated across different physics processes, but uncorrelated among the three data-taking years.

The uncertainties in the reconstruction and identification efficiencies for \Pe, \PGm, and \tauh candidates have been measured in \PZ boson enriched regions in data for each level of identification criteria (tight, medium, and loose), and are applied to each event as a function of \pt and $\eta$ for leptons and of \pt and the reconstructed hadronic decay mode for \tauh.
The reconstructed \tauh energy has an uncertainty of around 1\%, depending on the data-taking year and reconstructed \tauh decay mode.
These uncertainties affect the predicted rate and BDT output shape for signal and background, and are correlated among the different physics processes, but uncorrelated across different data-taking years.

The jet energy scale and resolution are determined using dijet control regions~\cite{Khachatryan:2016kdb,CMS:2017wyc}.
The jet energy scale is evaluated using 11 separate components, accounting for partial correlations between the data recorded in different years.
The jet energy resolution uncertainty is uncorrelated among the three data-taking years.
Jet energy uncertainties are also propagated to the \ptmiss calculation.
An additional uncertainty in \ptvecmiss comes from uncertainty in the energy of ``unclustered'' PF hadrons (PF hadrons not clustered into either small- or large-radius jets), which is uncorrelated across different years.
The probability for true \Pbottom jets to fail the multivariate \Pbottom jet identification criteria, or for jets from gluons or light flavored quarks to be misidentified as \Pbottom jets, is compared in data and MC simulation in event regions that are enriched in light-flavor quark or gluon, or heavy-flavor jets.
The resulting uncertainty in the data-to-simulation agreement affects the yields and BDT output shapes of multiple physics processes.
The statistical component of this uncertainty is treated as uncorrelated across different data-taking years, while other experimental sources are correlated.

The integrated luminosities for data collected in 2016, 2017, and 2018 have 1.2--2.5\% individual uncertainties~\cite{CMS-LUM-17-003,CMS-PAS-LUM-17-004,CMS-PAS-LUM-18-002}, while the overall uncertainty for the 2016--2018 period is 1.6\%.
The uncertainty in the measured cross section for inelastic $\Pp\Pp$ collisions, amounting to 5\%~\cite{CMS:2018mlc}, is taken into account by varying the number of pileup interactions in MC simulation, which impacts the jet reconstruction and the isolation of \lep and \tauh.

The sources of systematic uncertainty which create the largest uncertainties in the measured ratio of the \HH production cross section to its SM prediction are the theoretical uncertainties in the \HH production cross section and decay branching fractions (25\%), the uncertainties in the rate and shape of backgrounds from misidentified \lep or \tauh (22\%), and in the rates of backgrounds modeled using MC simulation (13\%).
These uncertainties in the signal measurement are determined by removing uncertainties that correspond to a given systematic source from an ML fit to pseudodata, as described in Section~\ref{sec:results}, and subtracting the obtained uncertainty in the signal measurement in quadrature from the total uncertainty.
The impacts of systematic uncertainties are small compared to the effect of the statistical uncertainty in the data (79\%), and are comparable to the statistical uncertainties in the distributions in the BDT classifier output for background processes (33\%).
The latter includes the effect of statistical uncertainties in the MC simulation and in the \lep/\tauh misidentification and electron charge misidentification backgrounds obtained from data.
All other sources of uncertainty have an impact of 5\% or less.

\section{Results}
\label{sec:results}

The data selected in the seven search categories are tested against multiple \HH production hypotheses: the SM prediction; variations of the SM coupling strength modifiers \kappal, \kappat, \kappaV, and \kappaVV; the effective couplings \cg, \cgg, and \ctwo in the EFT approach; and resonant production of \PH boson pairs originating from the decay of heavy particles with spins of 0 or 2 and masses $m_{\X}$ ranging from 250 to 1000\GeV.
In each case, the data observed in the seven search categories is fit simultaneously to a model composed of the background prediction (with uncertainties) and the \HH signal hypothesis under consideration.
The distributions in \mT in the \threeLeptonCR CR and in $m_{4\lep}$ in the \fourLeptonCR CR shown in Fig.~\ref{fig:postfitPlotsCR} are included in these fits, 
in order to obtain in-situ constraints on the systematic uncertainties described in Section~\ref{sec:systematicUncertainties}.
This in turn reduces the uncertainties in the signal and background predictions.

The SM ``signal strength'' parameter $\mu$ is defined as the ratio of the measured \HH production cross section to its predicted value in the SM.
This parameter modifies the expected signal yield by the same proportion in each category.
By contrast, variations in the $\kappa$ modifiers may affect the signal yields in each category differently, and also change the BDT classifier output shape for \HH events.
The twenty benchmark scenarios spanning combinations of \kappal, \kappat, \cg, \cgg, and \ctwo values in the coupling parameter space each correspond to different kinematic distributions, so the \HH production cross section for each point is measured separately.
Similarly, signal efficiency and BDT classifier output shapes vary dramatically for different resonant masses, and thus a separate measurement is performed for each mass and spin hypothesis.
The SM signal strength measurement is performed using the output of the BDT classifier
that has been trained for SM nonresonant \HH production,
while the \kappal measurement uses the BDT trained for benchmark scenario JHEP04 BM7.
In the scenario JHEP04 BM7, the \mHH value tends to be close to the lower limit of 250\GeV,
which matches the event kinematics for nonresonant \HH production in the \kappal range of the expected limit.
When setting limits on the twenty different benchmark scenarios, the binary BDT inputs correspond to the given scenario, or in case of the benchmarks from Ref.~\cite{Capozi:2019xsi} the kinematically closest scenario.
In case of resonant \HH production, the BDT input for the resonance mass is set to the $m_{\X}$ value for which the limit is computed.

The SM signal strength is measured using a profile likelihood test statistic~\cite{Cowan:2010js}, with systematic uncertainties treated as nuisance parameters $\theta$ in a frequentist approach~\cite{ATL-PHYS-PUB-2011-011}.
The effect of variations in $\theta$ on the shape of the BDT classifier output distribution for the \HH signal and for background processes is incorporated into the ML fit using the technique described in Ref.~\cite{Conway:2011in}.
Statistical uncertainties in these distributions are also taken into account using the approach detailed in Ref.~\cite{Conway:2011in}.
The likelihood ratio $q_{\mu}$ for a fixed ``test'' signal strength value $\mu$ is
\begin{linenomath}
    \begin{equation*}
        \begin{aligned}
            q_{\mu} & = -2 \Delta \ln \mathcal{L} = -2 \ln \frac{\mathcal{L}(\text{data}|\mu,\hat{\theta}_{\mu})}{\mathcal{L}(\text{data}|\hat{\mu},\hat{\theta})},
        \end{aligned}
    \end{equation*}
\end{linenomath}
where $\hat{\mu}$ and $\hat{\theta}$ are the signal strength and nuisance parameter values that give the maximum value of the likelihood function $\mathcal{L}$ for the given set of data (requiring $\hat{\mu} \geq 0$), and $\hat{\theta}_{\mu}$ is the set of $\theta$ values which maximize $\mathcal{L}$ for the fixed $\mu$.
The 95\% confidence level (\CL) upper limit for $\mu$ is obtained using the \CLs criterion~\cite{Junk:1999kv,Read:2002hq}, with $q_{\mu}$ set to $0$ when $\mu < \hat{\mu}$.
The probabilities to observe a given value of the likelihood ratio $q_{\mu}$ under the signal-plus-background and background-only hypotheses are computed using the asymptotic approximation from Ref.~\cite{Cowan:2010js}.
The limits on $\mu$ obtained using the asymptotic approximation, match the limits obtained with toy MC experiments~\cite{ATL-PHYS-PUB-2011-011} within 10\%.
The SM coupling strength modifiers and the cross sections for the various \HH production hypotheses are measured by scanning the likelihood ratio $q_{\mu}$ as a function of $\mu$.
Theoretical and experimental uncertainties affecting the signal and background yields or the shape of the BDT classifier output distributions may be correlated or uncorrelated across different years, search categories, and BDT output bins, as described in Section~\ref{sec:systematicUncertainties}.

\begin{figure}[!ht]
    \centering\includegraphics[width=0.46\textwidth]{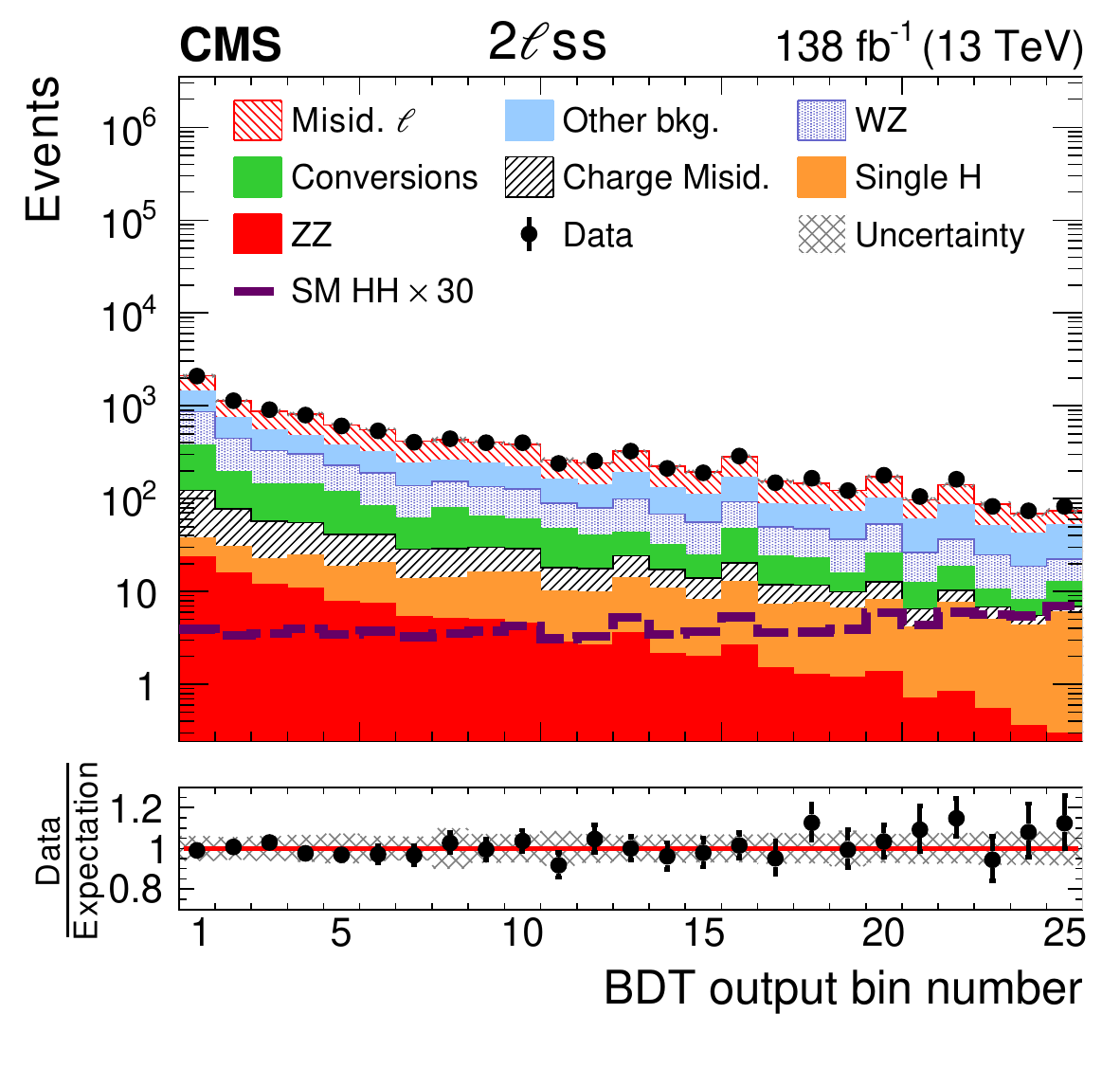}\hfill
    \centering\includegraphics[width=0.46\textwidth]{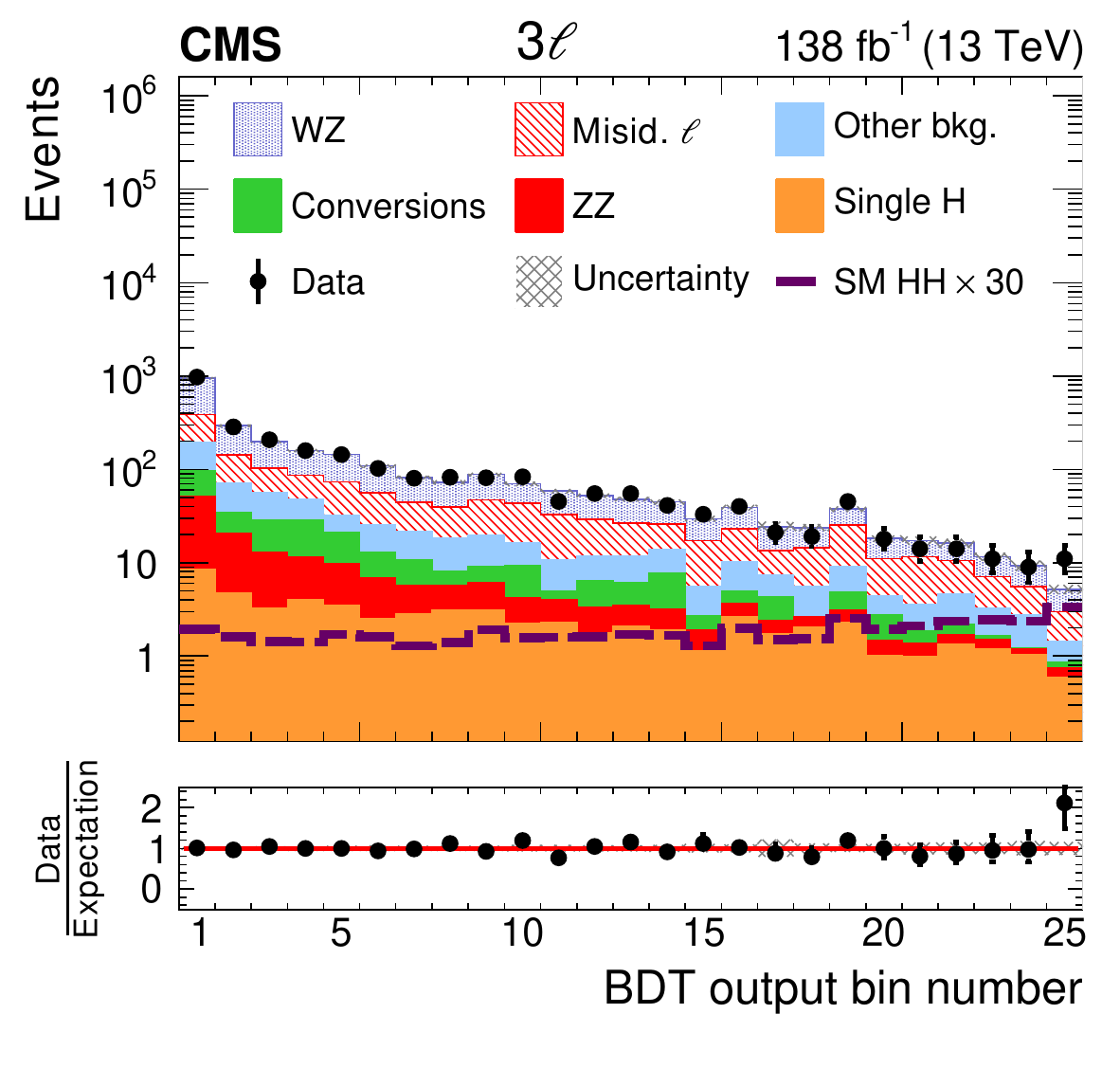}\\
    \centering\includegraphics[width=0.46\textwidth]{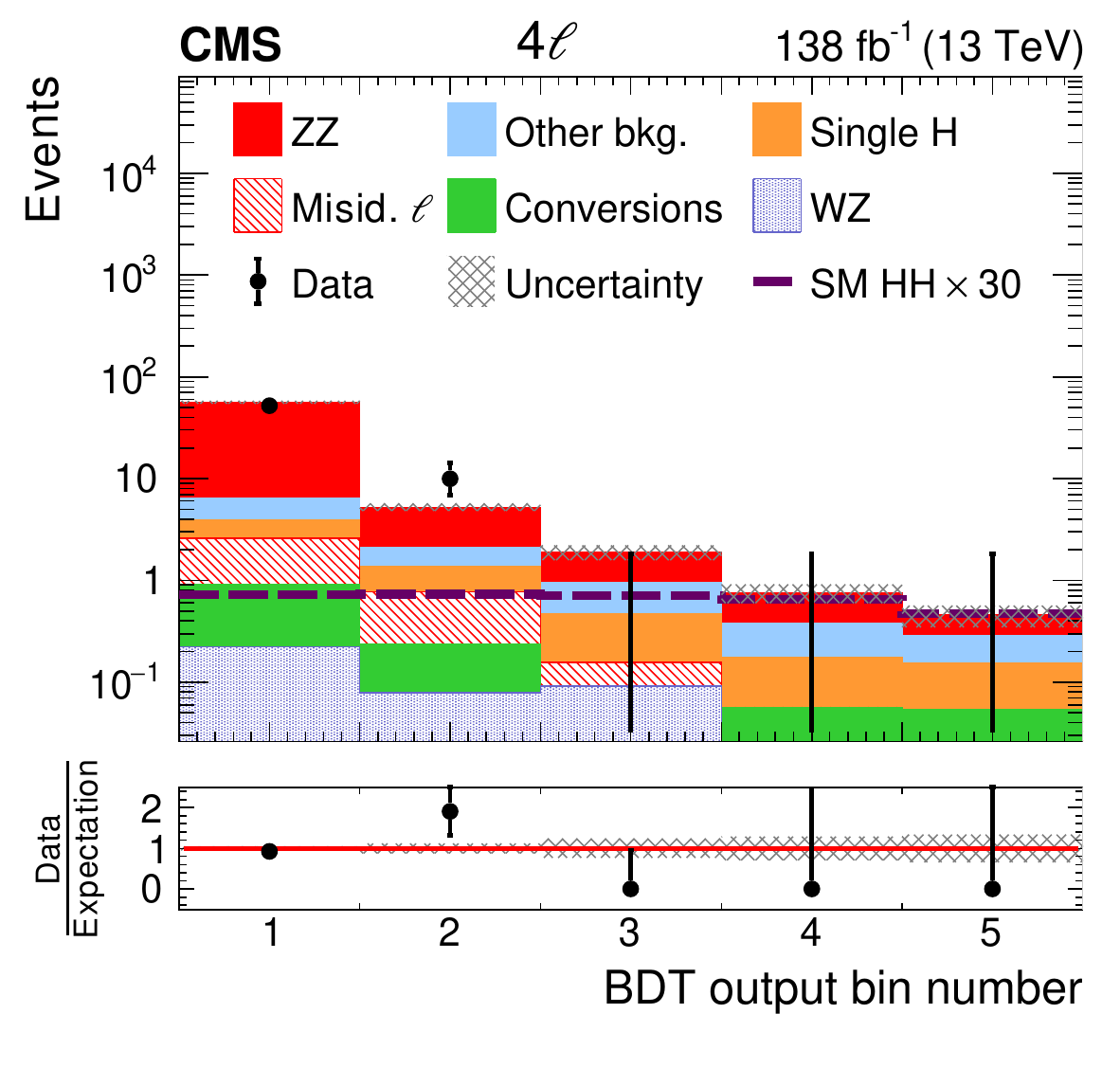}
    \caption{
        Distribution in the output of the BDT trained for nonresonant \HH production and evaluated for the benchmark scenario JHEP04 BM7 for the \llss (\cmsTop \cmsLeft), \lllnot (\cmsTop \cmsRight), and \llll (\cmsBottom) categories.
        The SM \HH signal is shown for a cross section amounting to 30 times the value predicted in the SM.
        The normalization and shape of the distributions expected for the background processes
        are shown for the values of nuisance parameters obtained from the ML fit of the signal+background hypothesis to the data.
        The gray shaded area indicates the sum of statistical and systematic uncertainties on the background prediction obtained from the ML fit.
        No data events are observed in the three rightmost bins of the BDT output distribution in the \llll category.
    }
    \label{fig:postfitPlots1}
\end{figure}

\begin{figure}[!ht]
    \centering\includegraphics[width=0.46\textwidth]{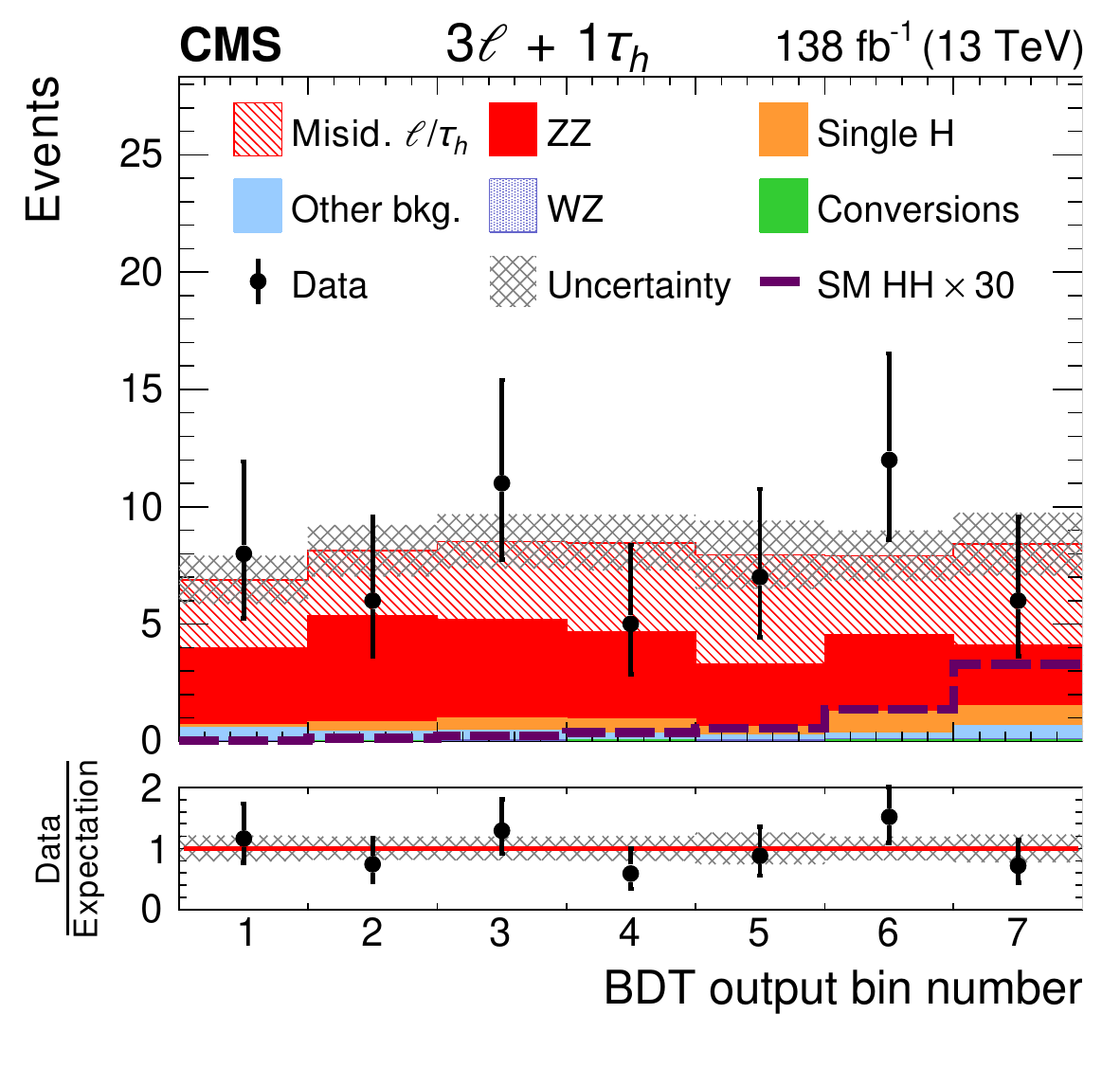}\hfill
    \centering\includegraphics[width=0.46\textwidth]{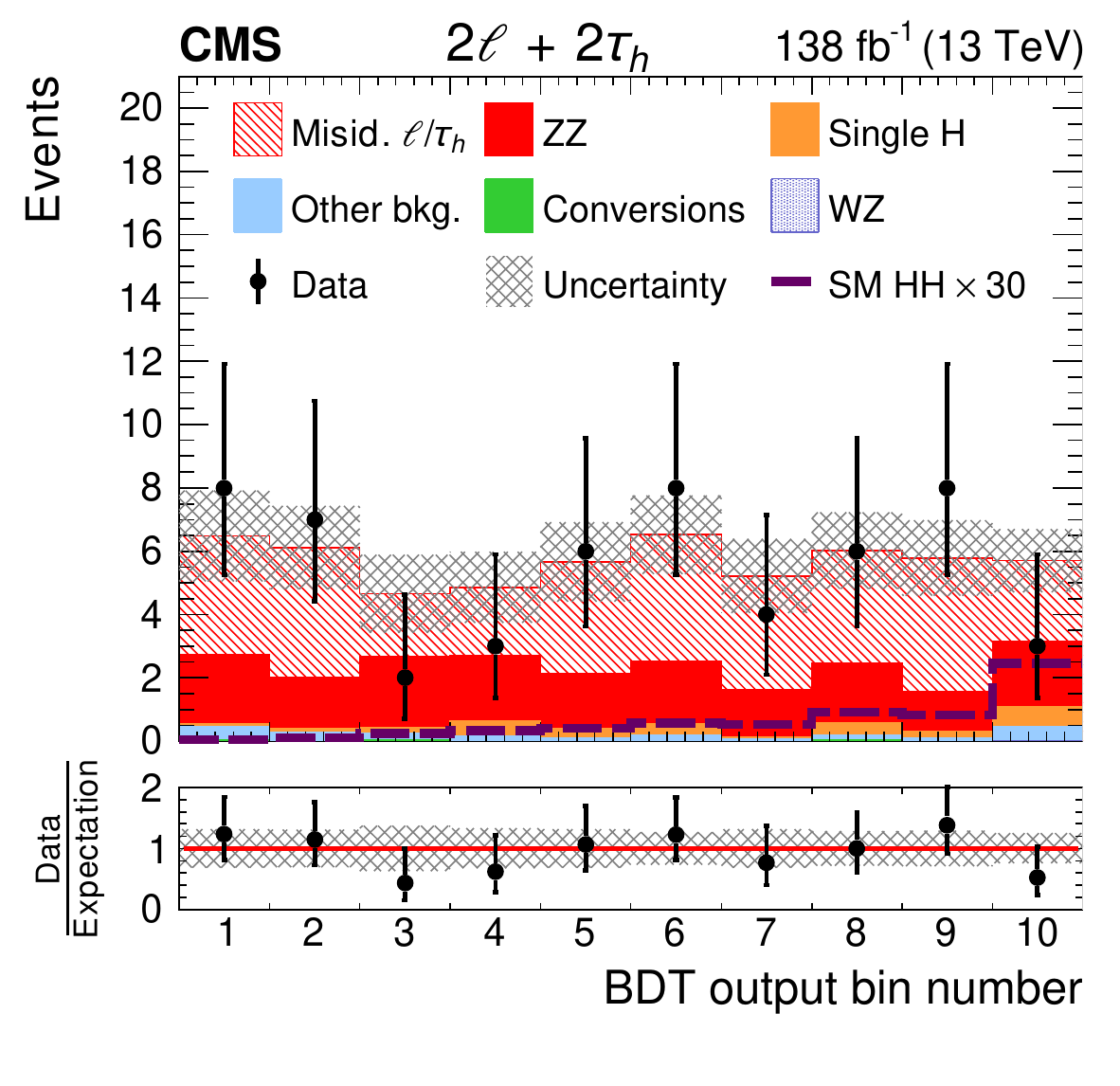}\\
    \centering\includegraphics[width=0.46\textwidth]{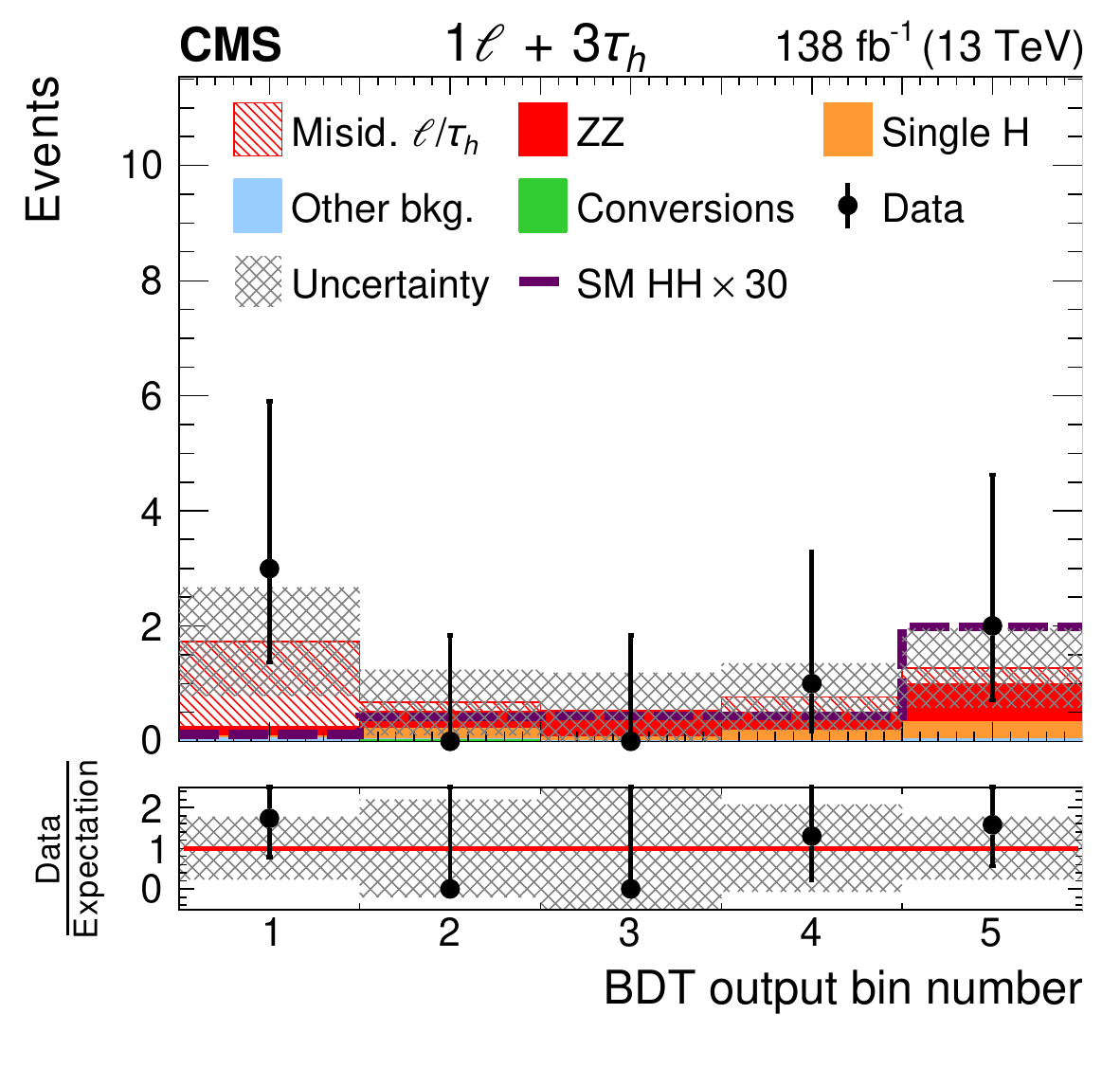}\hfill
    \centering\includegraphics[width=0.46\textwidth]{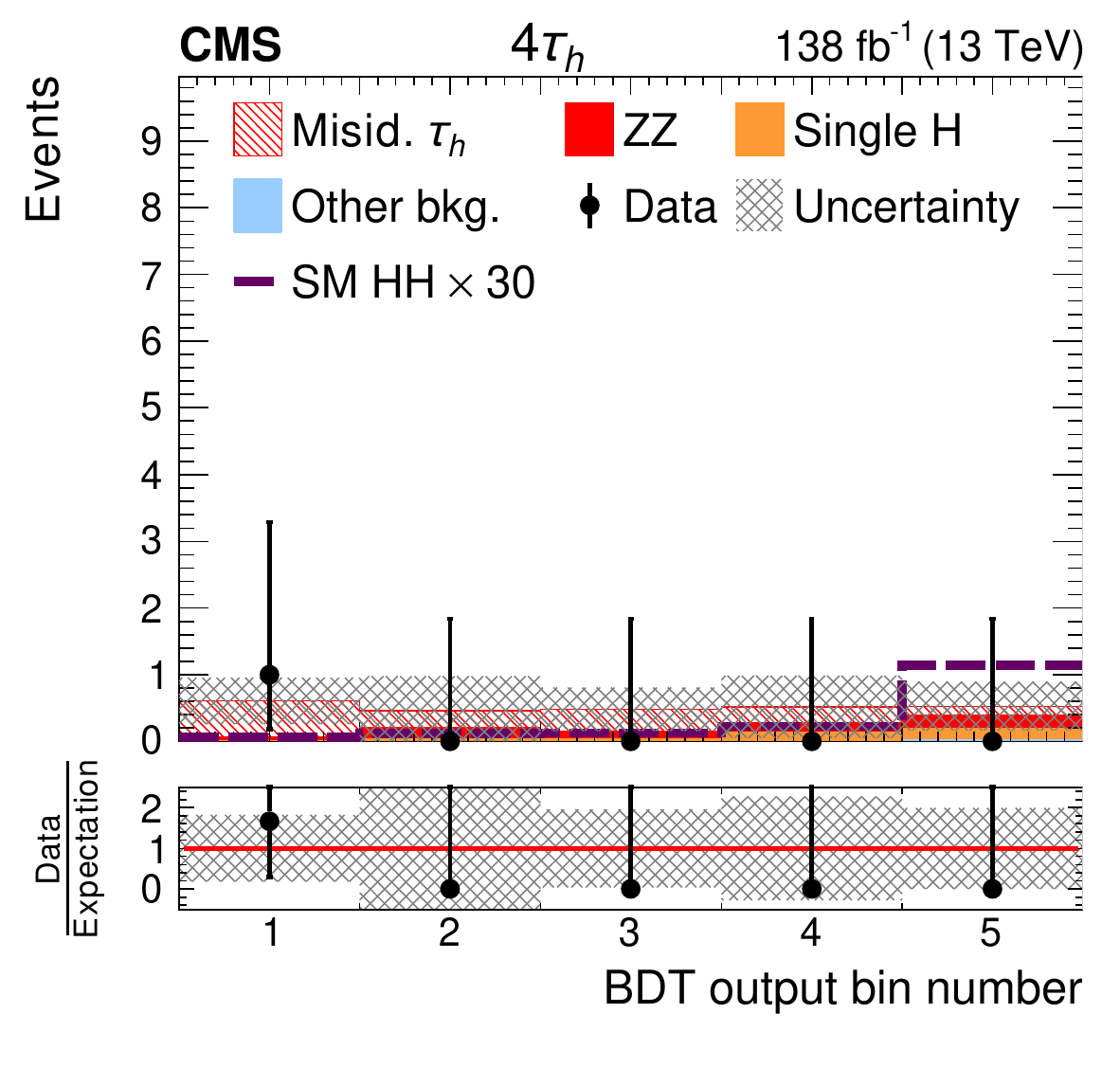}
    \caption{
        Distribution in the output of the BDT trained for nonresonant \HH production and evaluated for the benchmark scenario JHEP04 BM7
        for the \lllt (\cmsTop \cmsLeft), \lltt (\cmsTop \cmsRight), \lttt (\cmsBottom \cmsLeft), and \noltttt (\cmsBottom \cmsRight) categories.
        The SM \HH signal is shown for a cross section amounting to 30 times the value predicted in the SM.
        The normalization and shape of the distributions expected for the background processes
        are shown for the values of nuisance parameters obtained from the ML fit of the signal+background hypothesis to the data.
        The gray shaded area indicates the sum of statistical and systematic uncertainties on the background prediction obtained from the ML fit.
    }
    \label{fig:postfitPlots2}
\end{figure}

\begin{figure}[!ht]
    \centering
    \includegraphics[width=0.52\textwidth]{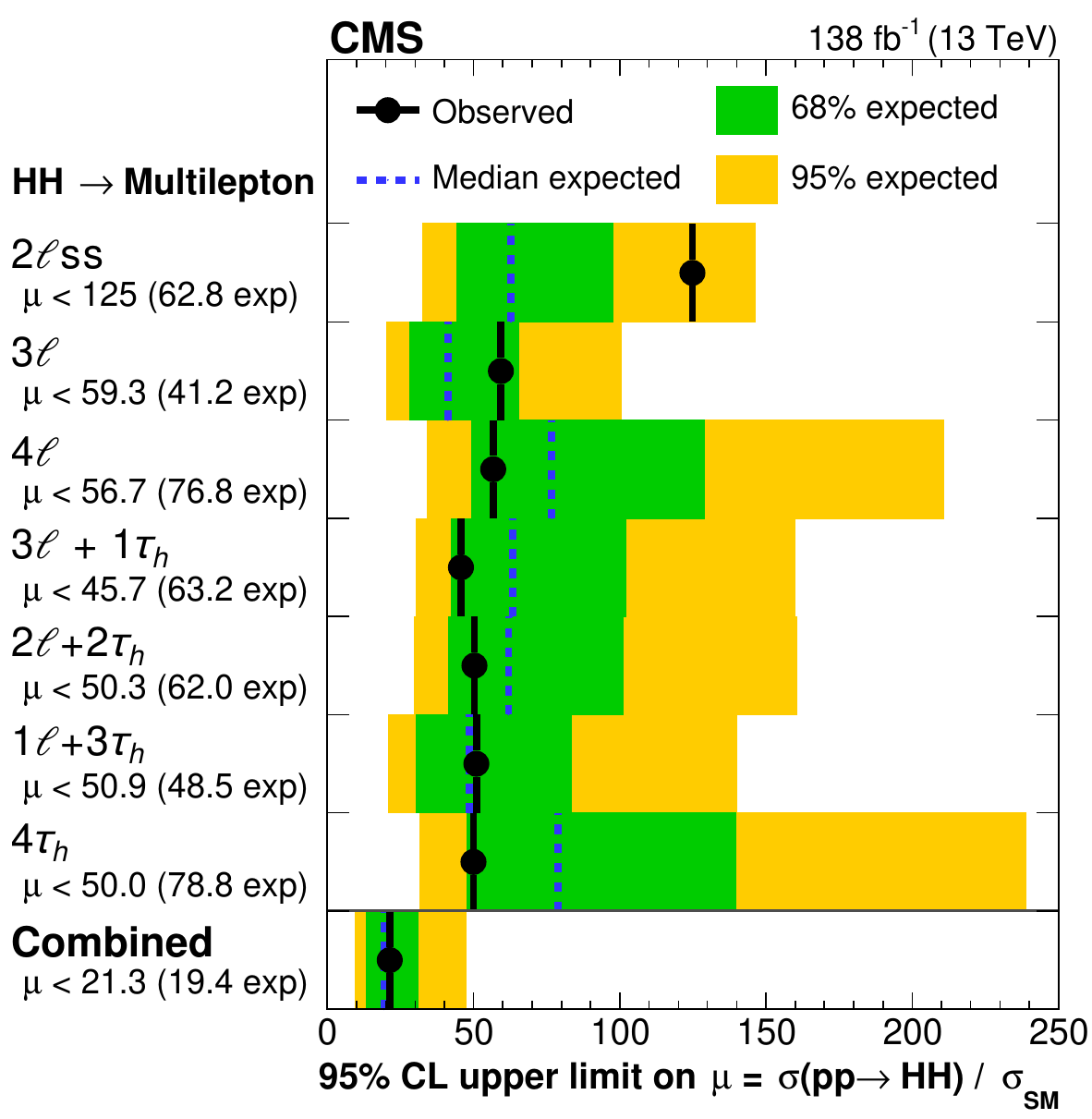}
    \caption{
        Observed and expected 95\% \CL upper limits on the SM \HH production cross section, obtained for both individual search categories and from a simultaneous fit of all seven categories combined.
    }
    \label{fig:HH_limits_SM}
\end{figure}

For the case of nonresonant \HH production with event kinematics as predicted by the SM, the best-fit value of the \HH production rate, 
obtained from the simultaneous fit of all seven search categories, amounts to $\hat{\mu} = 2\pm 8\,$(stat.)$\,\pm 6\,$(syst.) times the SM expectation.
The measured value of the signal strength refers to the sum of \ggHH and \qqHH production and is compatible with both the SM and background-only hypotheses,
within statistical and systematic uncertainties.
Distributions in the output of the BDT classifier for SM nonresonant \HH production in the seven search categories are shown in Figs.~\ref{fig:postfitPlots1} and~\ref{fig:postfitPlots2},
and the corresponding expected event yields are given in Table~\ref{tab:event_yields_postfit}.
The data excess in the rightmost bin of the BDT classifier output distribution for the \threeLeptonZeroTau category is not statistically significant: 11 events are observed in this bin, while $5.2 \pm 0.7\,$(stat.)$\,\pm 0.2\,$(syst.) are expected from background processes, amounting to a local significance of about 1.7 standard deviations.
The observed (expected) 95\% \CL upper limit on the cross section for nonresonant \HH production is 651 (592)\fb.
Taking into account the theoretical uncertainties in the SM \HH production cross section, this corresponds to an observed (expected) limit on the nonresonant \HH production rate of 21.3 (19.4) times the SM expectation.
These limits are shown in Fig.~\ref{fig:HH_limits_SM} for individual categories and for the combination of all seven search categories,
which is referred to as the ``$\HH \to \text{multilepton}$'' result.
The \lllnot and \lttt categories are the most sensitive to SM \HH production, followed closely by the other categories.

\begin{table}[!ht]
    \topcaption{
        The number of expected and observed events in each of the seven search categories, and in two CRs, which validate the modeling of the $\PW\PZ$ and $\PZ\PZ$ backgrounds.
        The \lep/\tauh misidentification and electron charge misidentification backgrounds are determined from data, as described in Section~\ref{sec:backgroundEstimation},
        while the \HH signal and all other backgrounds are modeled using MC simulation.
        The symbol ``\NA'' indicates that the background is not relevant for the category.
        The \HH signal represents the sum of the \ggHH and \qqHH production processes and is normalized to 30 times the event yield expected in the SM, corresponding to a cross section of about 1\pb.
        The expected event yields are computed for the values of nuisance parameters obtained from the ML fit described in Section~\ref{sec:results}.
        Quoted uncertainties represent the sum of statistical and systematic components.
        Uncertainties that are smaller than half the value of the least significant digit have been rounded to zero.
    }
    \label{tab:event_yields_postfit}
    \centering
    \begin{tabular}{lr@{ $\pm$ }lr@{ $\pm$ }lr@{ $\pm$ }l}
        Process                          & \multicolumn{2}{c}{\llss}   & \multicolumn{2}{c}{\lllnot} & \multicolumn{2}{c}{\llll}                                           \\
        \hline
        SM $\HH\to\WWWW$  ($\times\,30$) & 73                          & 6                           & 33                        & 3                          & 2.2  & 0.2 \\
        SM $\HH\to\WWtt$  ($\times\,30$) & 31                          & 3                           & 12                        & 1                          & 0.9  & 0.1 \\
        SM $\HH\to\tttt$  ($\times\,30$) & 3                           & 0                           & 1                         & 0                          & 0.1  & 0.0 \\
        [\cmsTabSkip]
        $\PW\PZ$                         & 2003                        & 58                          & 1321                      & 27                         & 0.4  & 0.1 \\
        $\PZ\PZ$                         & 121                         & 2                           & 109                       & 2                          & 54.7 & 1.8 \\
        Misidentified \lep               & 3939                        & 267                         & 670                       & 55                         & 2.3  & 1.0 \\
        Conversion electrons             & 1009                        & 170                         & 146                       & 24                         & 0.9  & 0.4 \\
        Electron charge misid.           & 366                         & 52                          & \multicolumn{2}{c}{\,\, \NA} & \multicolumn{2}{c}{\, \NA}             \\
        Single Higgs boson               & 216                         & 4                           & 62                        & 1                          & 2.4  & 0.3 \\
        Other backgrounds                & 2690                        & 224                         & 293                       & 20                         & 4.1  & 0.4 \\
        [\cmsTabSkip]
        Total expected background        & 10\,346                     & 396                         & 2601                      & 68                         & 64.8 & 2.1 \\
        \hline
        Data                             & \multicolumn{2}{c}{10\,344} & \multicolumn{2}{c}{2621}    & \multicolumn{2}{c}{62}                                              \\
    \end{tabular}

    \vspace*{0.4 cm}

    \begin{tabular}{lr@{ $\pm$ }lr@{ $\pm$ }lr@{ $\pm$ }lr@{ $\pm$ }l}
        Process                          & \multicolumn{2}{c}{\lllt} & \multicolumn{2}{c}{\lltt} & \multicolumn{2}{c}{\lttt}  & \multicolumn{2}{c}{\noltttt}                                                                       \\
        \hline
        SM $\HH\to\WWWW$  ($\times\,30$) & 0.9                       & 0.1                       & 0.2                        & 0.0                          & 0.2                        & 0.0                        & 0.3 & 0.0 \\
        SM $\HH\to\WWtt$  ($\times\,30$) & 4.1                       & 0.3                       & 3.9                        & 0.4                          & 0.6                        & 0.1                        & 0.1 & 0.0 \\
        SM $\HH\to\tttt$  ($\times\,30$) & 0.9                       & 0.1                       & 2.3                        & 0.3                          & 2.6                        & 0.4                        & 1.3 & 0.2 \\
        [\cmsTabSkip]
        $\PW\PZ$                         & 0.2                       & 0.0                       & \multicolumn{2}{c}{$<$0.1} & \multicolumn{2}{c}{$<$0.1}   & \multicolumn{2}{c}{$<$0.1}                                          \\
        $\PZ\PZ$                         & 24.3                      & 0.8                       & 18.5                       & 1.0                          & 1.9                        & 0.2                        & 0.7 & 0.1 \\
        Misidentified \lep and \tauh     & 25.1                      & 4.4                       & 33.5                       & 4.6                          & 2.1                        & 1.7                        & 1.5 & 0.9 \\
        Conversion electrons             & 0.1                       & 0.0                       & 0.1                        & 0.1                          & \multicolumn{2}{c}{$<$0.1} & \multicolumn{2}{c}{$<$0.1}             \\
        Single Higgs boson               & 3.8                       & 0.2                       & 2.9                        & 0.5                          & 0.8                        & 0.4                        & 0.4 & 0.1 \\
        Other backgrounds                & 2.7                       & 0.3                       & 2.1                        & 0.4                          & 0.1 & 0.0 & \multicolumn{2}{c}{$<$0.1}             \\
        [\cmsTabSkip]
        Total expected background        & 56.2                      & 4.5                       & 57.0                       & 4.8                          & 4.9                        & 1.7                        & 2.6 & 0.9 \\
        \hline
        Data                             & \multicolumn{2}{c}{55}    & \multicolumn{2}{c}{55}    & \multicolumn{2}{c}{$6$}    & \multicolumn{2}{c}{$1$}                                                                            \\
    \end{tabular}

    \vspace*{0.4 cm}

    \begin{tabular}{lr@{ $\pm$ }lr@{ $\pm$ }l}
        Process                   & \multicolumn{2}{c}{\threeLeptonCR CR} & \multicolumn{2}{c}{\fourLeptonCR CR}                                 \\
        \hline
        $\PW\PZ$                  & 12\,546                               & 148                                  & \multicolumn{2}{c}{$<$1}      \\
        $\PZ\PZ$                  & 799                                   & 24                                   & 2032                     & 60 \\
        Misidentified \lep        & 908                                   & 122                                  & 13                       & 4  \\
        Conversion electrons      & 134                                   & 22                                   & 3                        & 0  \\
        Other backgrounds         & 620                                   & 54                                   & 59                       & 6  \\
        [\cmsTabSkip]
        Total expected background & 15\,006                               & 202                                  & 2108                     & 60 \\
        \hline
        Data                      & \multicolumn{2}{c}{14\,994}           & \multicolumn{2}{c}{2096}                                             \\
    \end{tabular}
\end{table}

\begin{figure}[!ht]
    \centering\includegraphics[width=0.46\textwidth]{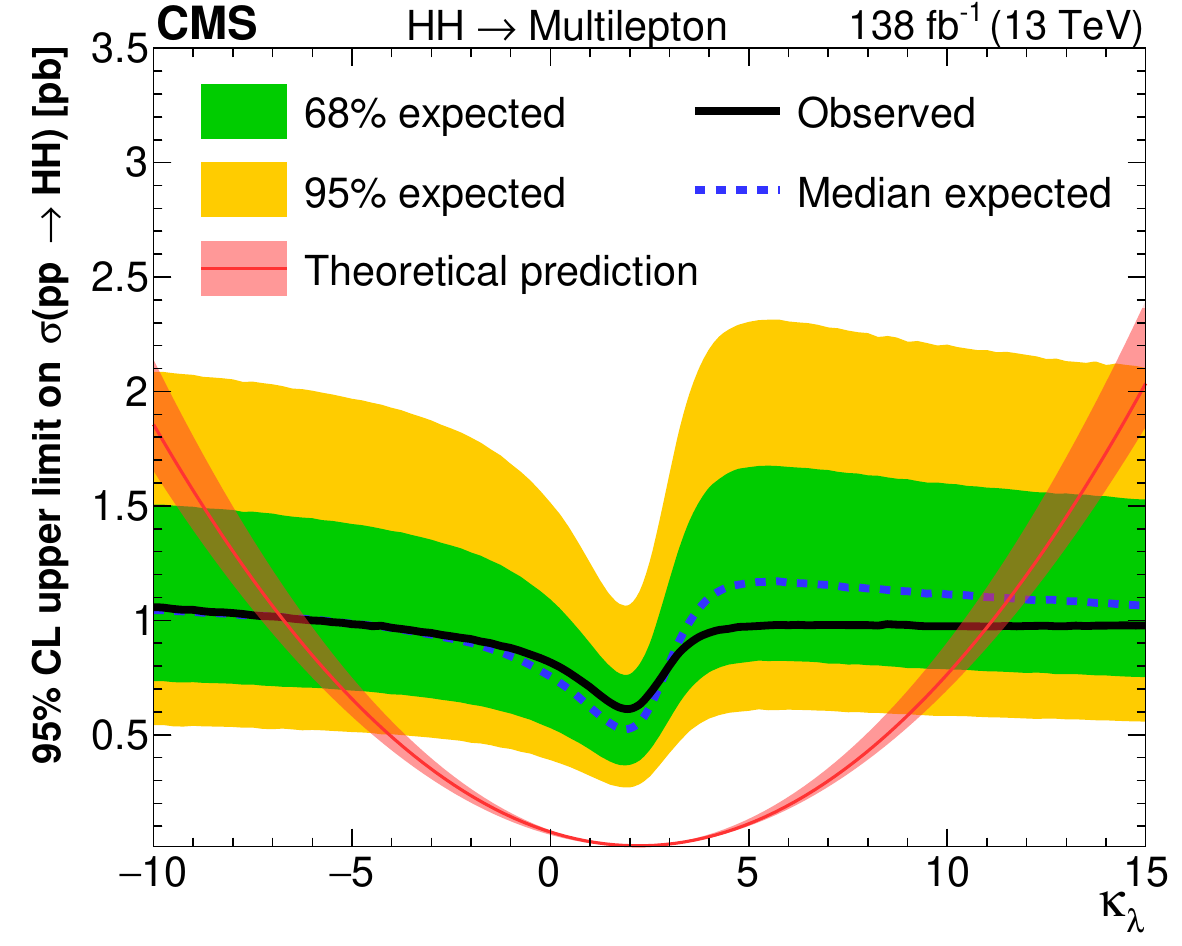}\hfill
    \centering\includegraphics[width=0.46\textwidth]{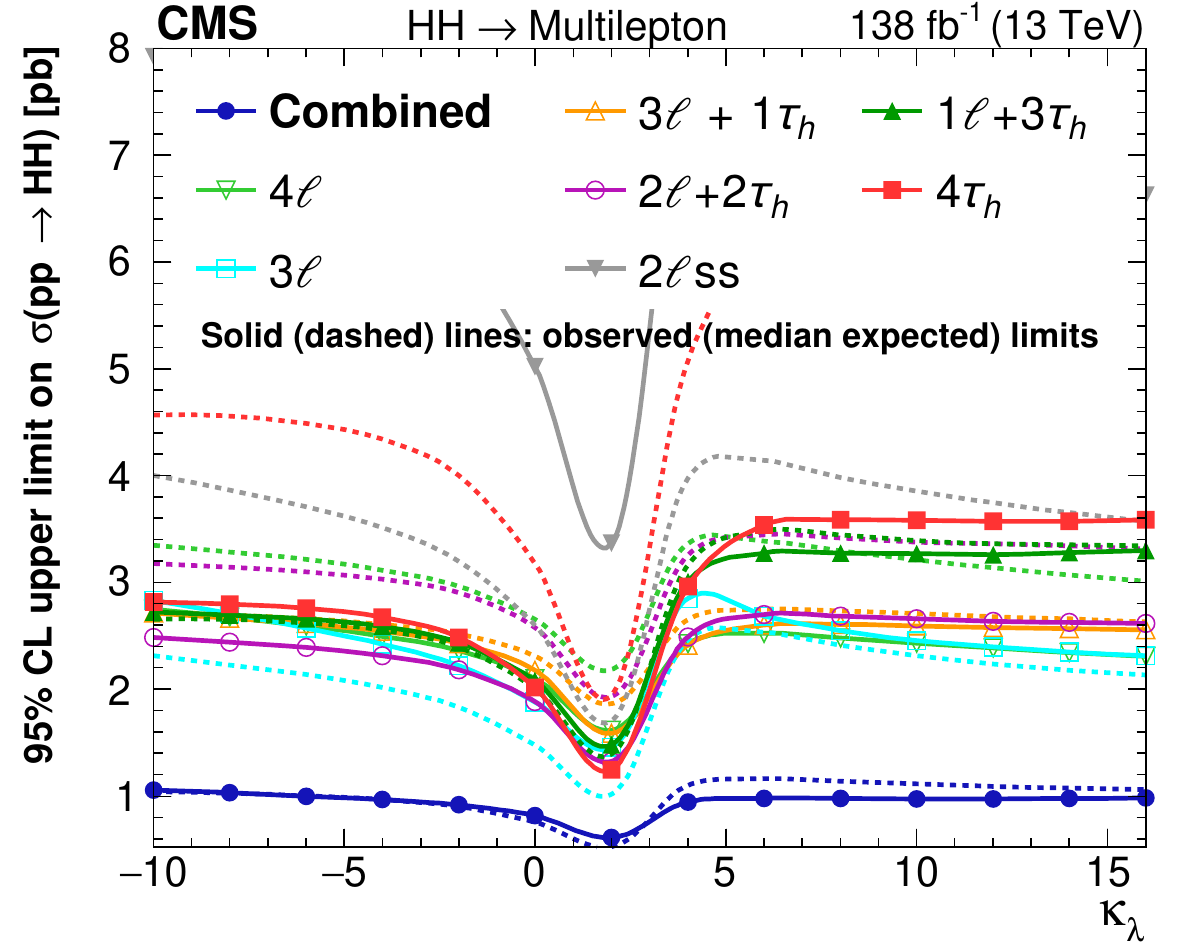}
    \caption{
        Observed and expected 95\% \CL upper limits on the \HH production cross section as a function of the \PH boson self-coupling strength modifier \kappal.
        All \PH boson couplings other than $\lambda$ are assumed to have the values predicted in the SM.
        The left plot shows the result obtained by combining all seven search categories, while the right plot shows the limits obtained for each category separately.
        The red curve in the left plot represents the SM prediction for the \HH production cross section as a function of \kappal, and the red shaded band the theoretical uncertainty in this prediction.
    }
    \label{fig:HH_limits_kLambda}
\end{figure}

The observed (expected) 95\% \CL interval for the \PH boson trilinear self-coupling strength modifier is measured to be $-6.9 < \kappal < 11.1$ ($-6.9 < \kappal < 11.7$).
The upper limit on \kappal is one of the strongest constraints on this fundamental SM parameter to date, with only \HH searches in the $\Pbottom\Pbottom\PGg\PGg$~\cite{CMS:2020tkr,ATLAS-HDBS-2018-34} and $\Pbottom\Pbottom\Pbottom\Pbottom$~\cite{CMS:2022cpr} decay modes providing tighter bounds.
The observed and expected upper limits on the \HH production cross section as a function of \kappal, obtained from the simultaneous fit of all seven search categories, are shown in Fig.~\ref{fig:HH_limits_kLambda}, along with the limits obtained for each category individually.

\begin{figure}[!ht]
    \centering\includegraphics[width=0.80\textwidth]{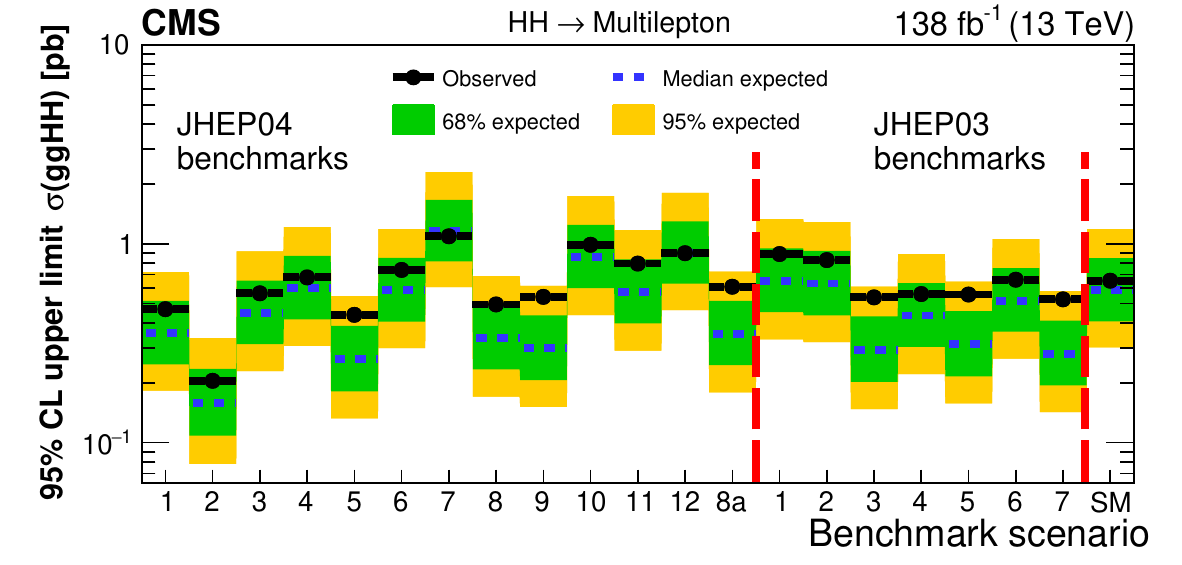}\\
    \centering\includegraphics[width=0.80\textwidth]{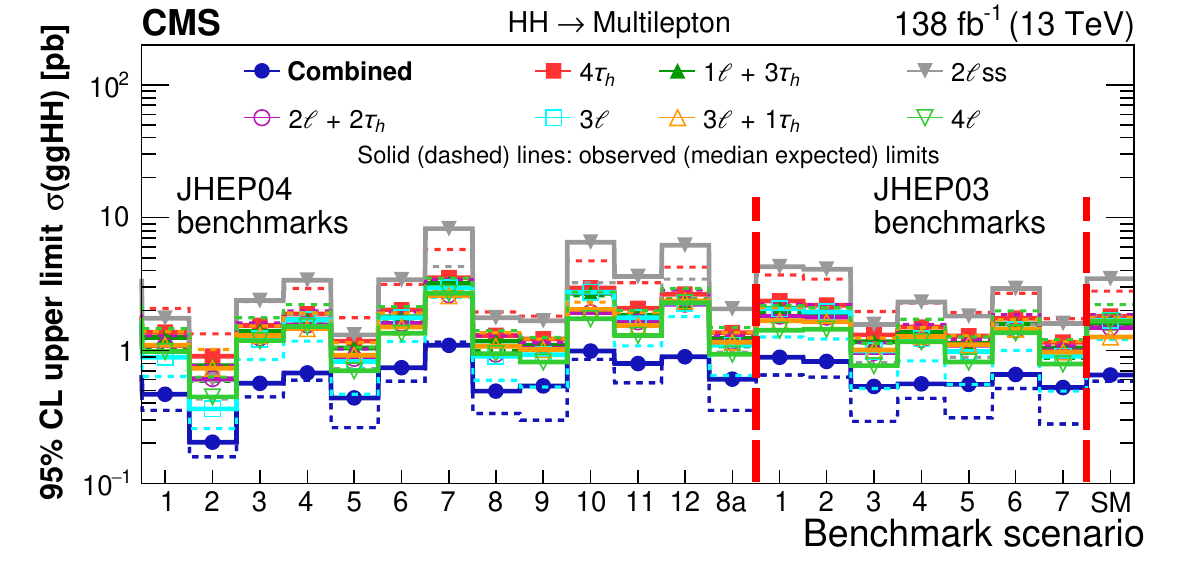}
    \caption{
        Observed and expected 95\% \CL upper limits on the \HH production cross section for the twelve benchmark scenarios from Ref.~\cite{Carvalho:2015ttv}, the additional benchmark scenario 8a from Ref.~\cite{Buchalla:2018yce}, the seven benchmark scenarios from Ref.~\cite{Capozi:2019xsi}, and for the SM.
        The upper plot shows the result obtained by combining all seven search categories, while the lower plot shows the limits obtained for each category separately, and the combined limit.
    }
    \label{fig:HH_limits_EFT}
\end{figure}

The observed and expected limits on the \ggHH production cross section for the twenty benchmark scenarios are shown in Fig.~\ref{fig:HH_limits_EFT} and summarized in Table~\ref{tab:eftres}.
Signal contributions from the \qqHH process, at the rate expected in the SM, are about two orders of magnitude lower than the limits that we set on the rate of the \ggHH signal
in these measurements and are therefore neglected.
The observed (expected) limits on nonresonant \HH production in the different benchmark scenarios range from 0.21 to 1.09 (0.16 to 1.16)\pb, depending on the scenario.
These limits are a factor of 2--3 higher than those obtained by the CMS measurement in the $\Pbottom\Pbottom\PGg\PGg$ final state~\cite{CMS:2020tkr}.
The variation in expected limits reflects differences in the \mHH distribution among the benchmark scenarios,
which in turn affect the \pt and angles between the particles produced in the \PH boson decays.
As a consequence, the signal acceptance can change, along with the separation of the \HH signal from backgrounds through the BDT classifiers described in Section~\ref{sec:analysisStrategy}.
The most and least stringent limits on the cross section are expected for the benchmark scenarios JHEP04 BM2 and BM7, respectively.
The former has a pronounced tail of the \mHH distribution extending to high values, while the latter is characterized by low \mHH values, as seen in Fig.~5 of Ref.~\cite{Carvalho:2015ttv}.

\begin{table}[!ht]
    \topcaption{
        Observed (expected) 95\% \CL upper limits on the \ggHH production cross section for the twelve benchmark scenarios from Ref.~\cite{Carvalho:2015ttv}, the additional benchmark scenario 8a from Ref.~\cite{Buchalla:2018yce} and the seven benchmark scenarios from Ref.~\cite{Capozi:2019xsi}.
        The corresponding observed (expected) upper limit for the SM is 652 (583)\fb.
        The limits correspond to the combination of all seven search categories.}
    \centering
    \hspace{0.05\textwidth}
    \begin{minipage}[]{0.35\textwidth}
        \begin{tabular}{c|c}
            JHEP04    & Observed (expected) \\
            benchmark & limit [fb]          \\
            \hline
            BM1       & 469 (354)           \\
            BM2       & 205 (159)           \\
            BM3       & 563 (447)           \\
            BM4       & 677 (600)           \\
            BM5       & 439 (263)           \\
            BM6       & 739 (584)           \\
            BM7       & 1090 (1156)         \\
            BM8       & 495 (336)           \\
            BM9       & 541 (298)           \\
            BM10      & 988 (855)           \\
            BM11      & 795 (572)           \\
            BM12      & 897 (898)           \\
            BM8a      & 608 (353)           \\
        \end{tabular}
    \end{minipage}\hspace{0.15\textwidth}
    \begin{minipage}[]{0.35\textwidth}
        \begin{tabular}{c|c}
            JHEP03    & Observed (expected) \\
            benchmark & limit [fb]          \\
            \hline
            BM1       & 888 (650)           \\
            BM2       & 828 (632)           \\
            BM3       & 538 (293)           \\
            BM4       & 559 (436)           \\
            BM5       & 556 (313)           \\
            BM6       & 660 (518)           \\
            BM7       & 525 (280)           \\
        \end{tabular}
    \end{minipage}
    \hspace{0.10\textwidth}
    \label{tab:eftres}
\end{table}

\begin{figure}[!ht]
    \centering\includegraphics[width=0.46\textwidth]{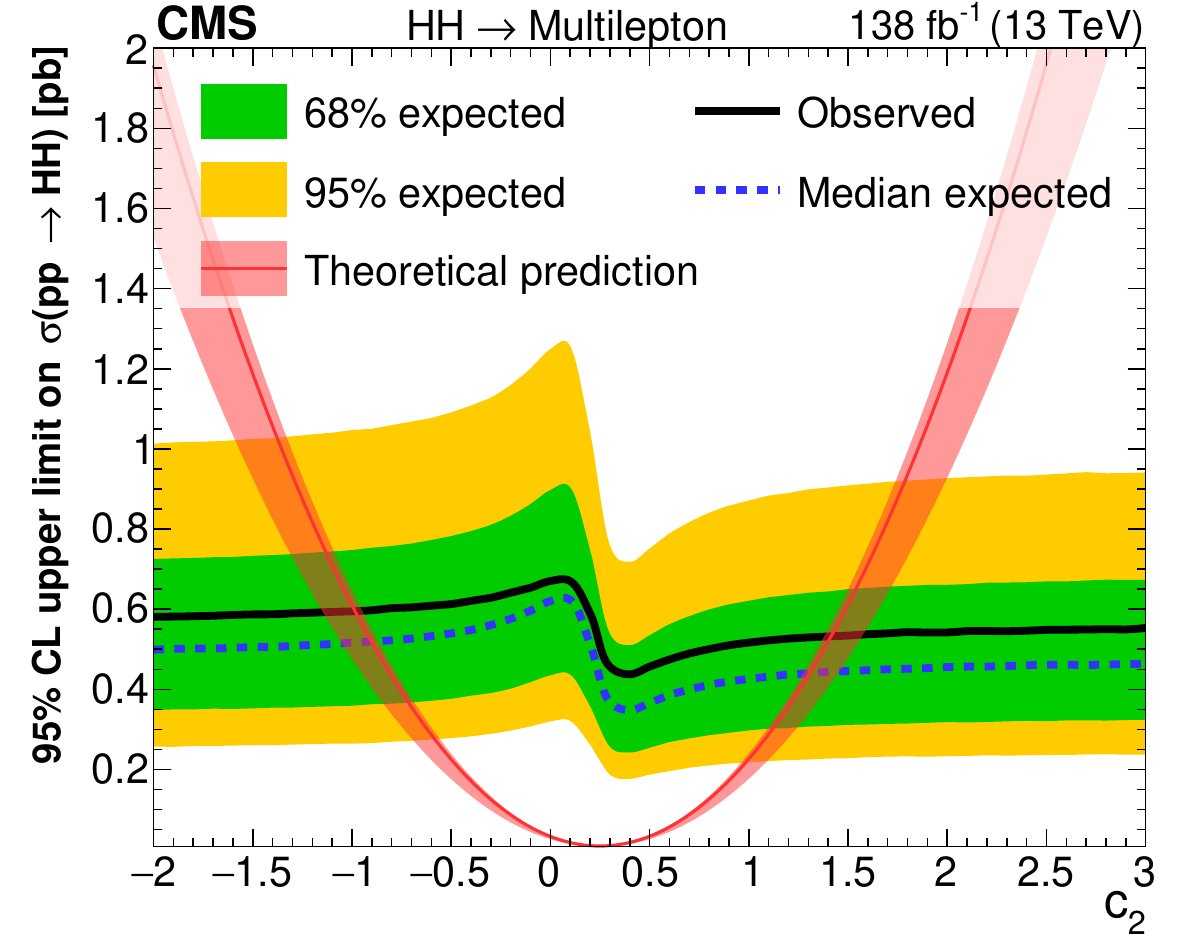}\hfill
    \centering\includegraphics[width=0.46\textwidth]{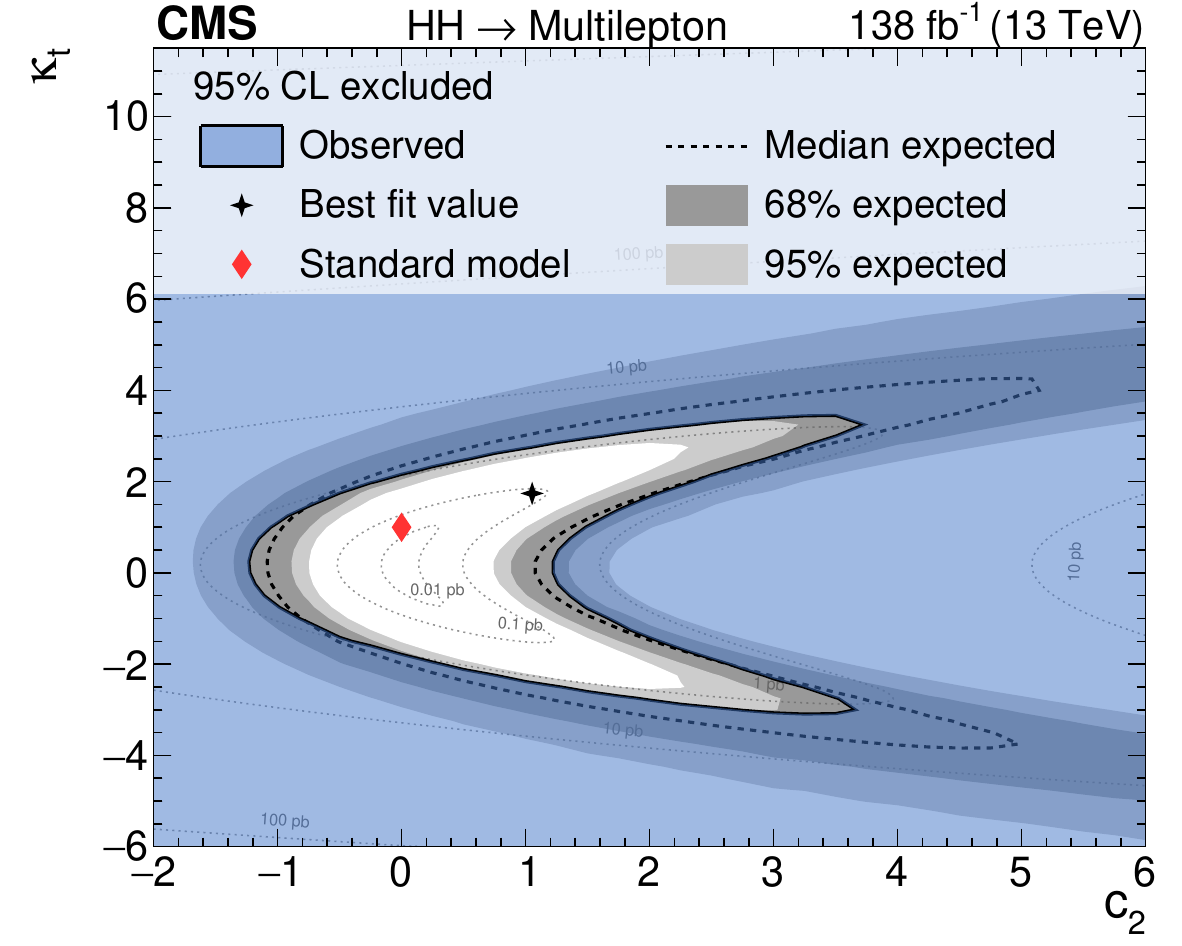}
    \caption{
        Observed and expected limits on the \HH production cross section as a function of the effective coupling \ctwo (\cmsLeft), and the region excluded in the \kappat--\,\ctwo plane (\cmsRight).
        All limits are computed at 95\% \CL.
        \PH boson couplings other than the ones shown in the plots (\ctwo in the left plot and \ctwo and \kappat in the right plot) are assumed to have the values predicted by the SM.
    }
    \label{fig:HH_limits_c2}
\end{figure}

Figure~\ref{fig:HH_limits_c2} shows the observed and expected upper limits on the \HH production cross section as a function of the coupling \ctwo, and the region excluded in the \kappat--\,\ctwo plane.
The effects of variations in \kappal and \kappat on the rate of the SM single \PH boson background~\cite{Maltoni:2017ims} and on the \PH boson decay branching fractions~\cite{Degrassi:2016wml} are taken into account when computing these limits and those shown in Fig.~\ref{fig:HH_limits_kLambda}.
The magnitude of these effects is typically 5 to 10\% within the scanned range of \kappal and \kappat.
Assuming \kappat and \kappal are both equal to 1, the coupling \ctwo is observed (expected) to be constrained to the interval $-1.05 < \ctwo < 1.48$ ($-0.96 < \ctwo < 1.37$) at 95\% \CL.

\begin{figure}[!ht]
    \centering\includegraphics[width=0.46\textwidth]{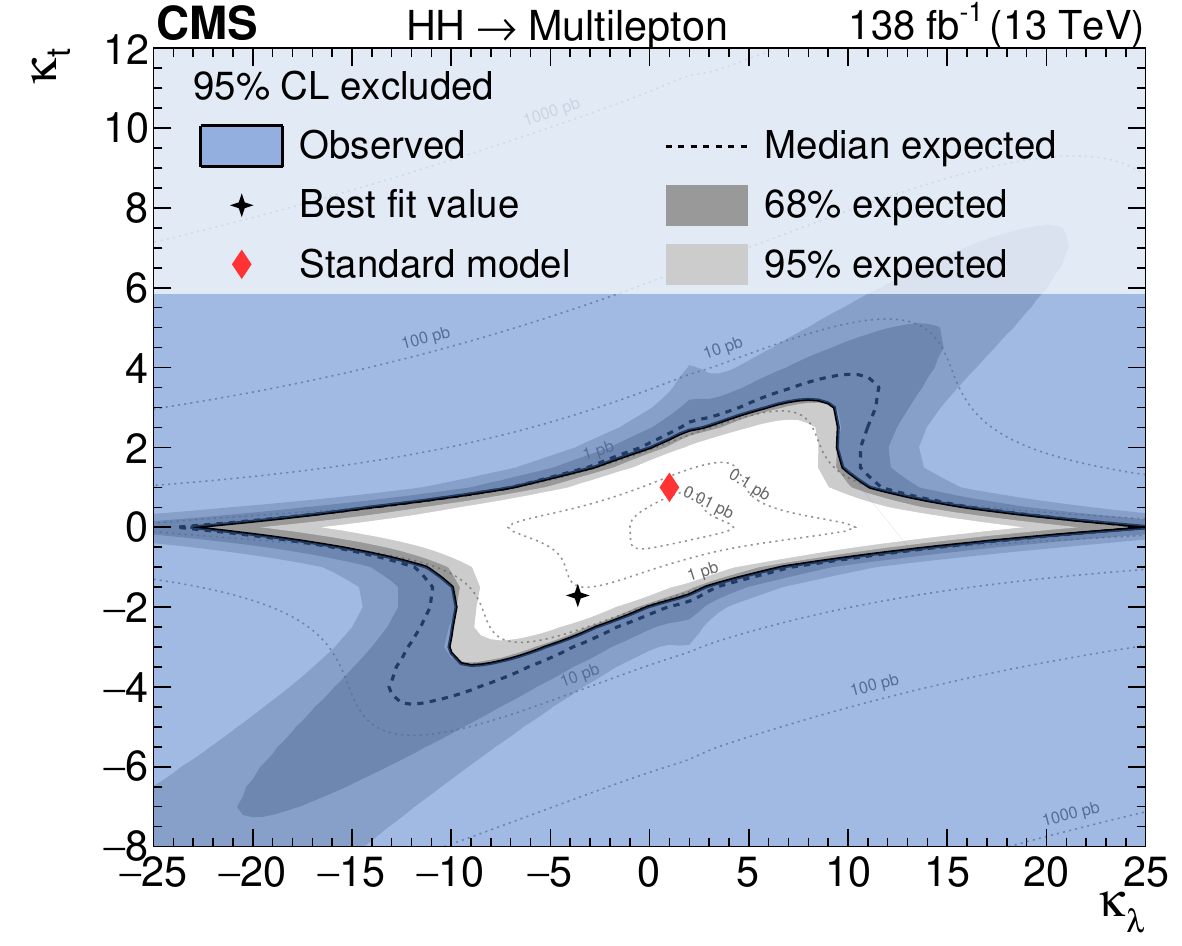}\hfill
    \centering\includegraphics[width=0.46\textwidth]{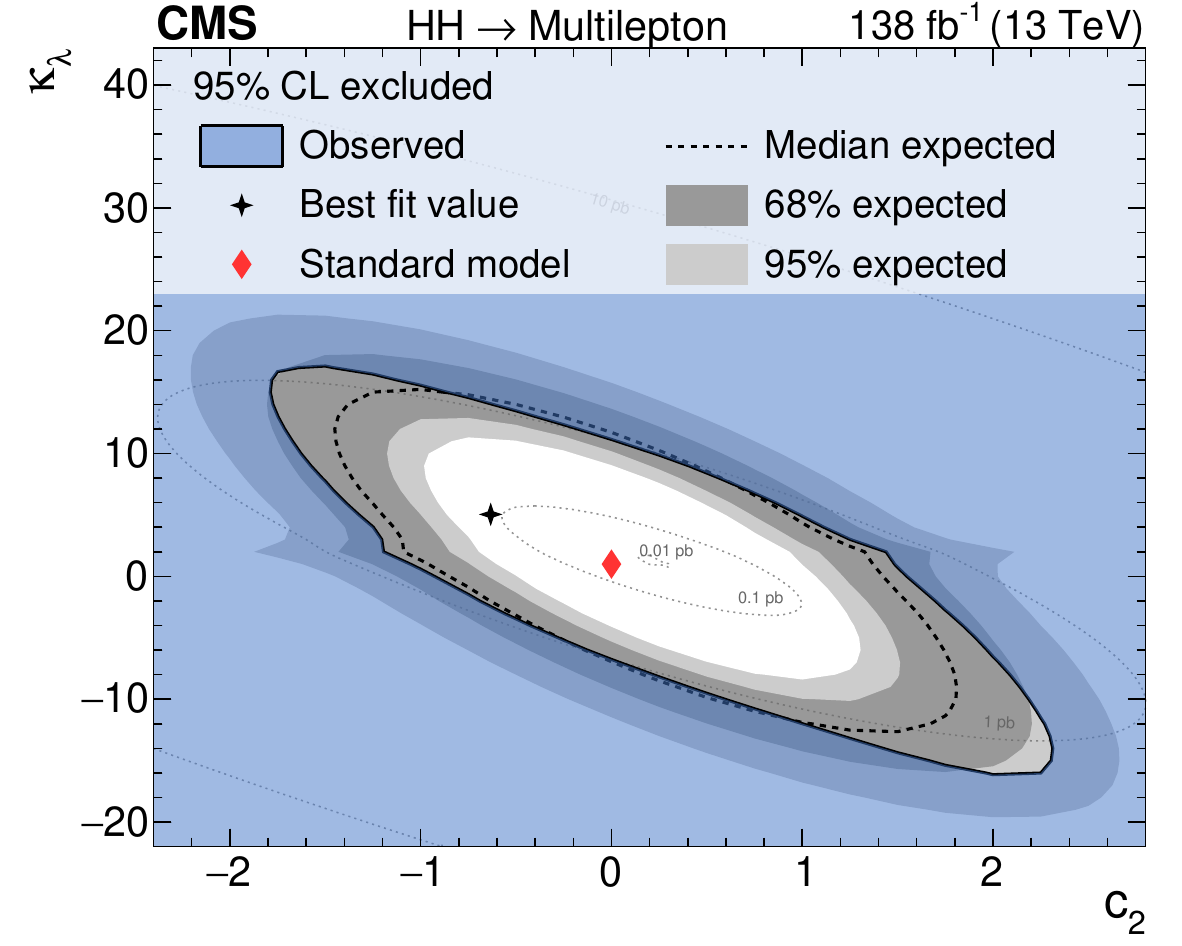}
    \caption{
        Observed and expected regions excluded in the \kappat--\,\kappal (\cmsLeft) and \kappal--\,\ctwo (\cmsRight) planes.
        \PH boson couplings other than the ones shown in the plots (\kappal and \kappat in the left plot, and \ctwo and \kappal in the right plot) are assumed to have the values predicted by the SM.
    }
    \label{fig:HH_limits_scans}
\end{figure}

Similar to the right part of Figure~\ref{fig:HH_limits_c2}, Figure~\ref{fig:HH_limits_scans} shows the observed and expected regions excluded in the \kappat--\,\kappal and \kappal--\,\ctwo planes.

\begin{figure}[!ht]
    \centering\includegraphics[width=0.46\textwidth]{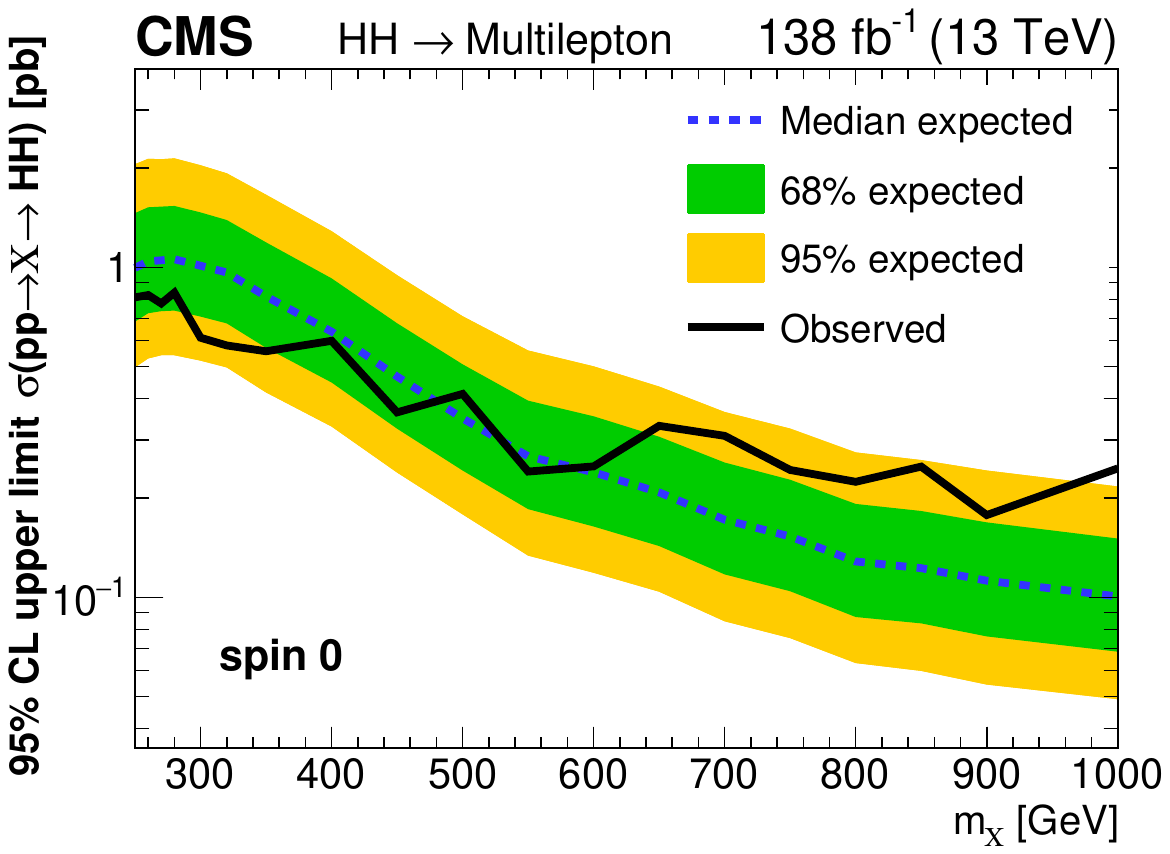}\hfill
    \centering\includegraphics[width=0.46\textwidth]{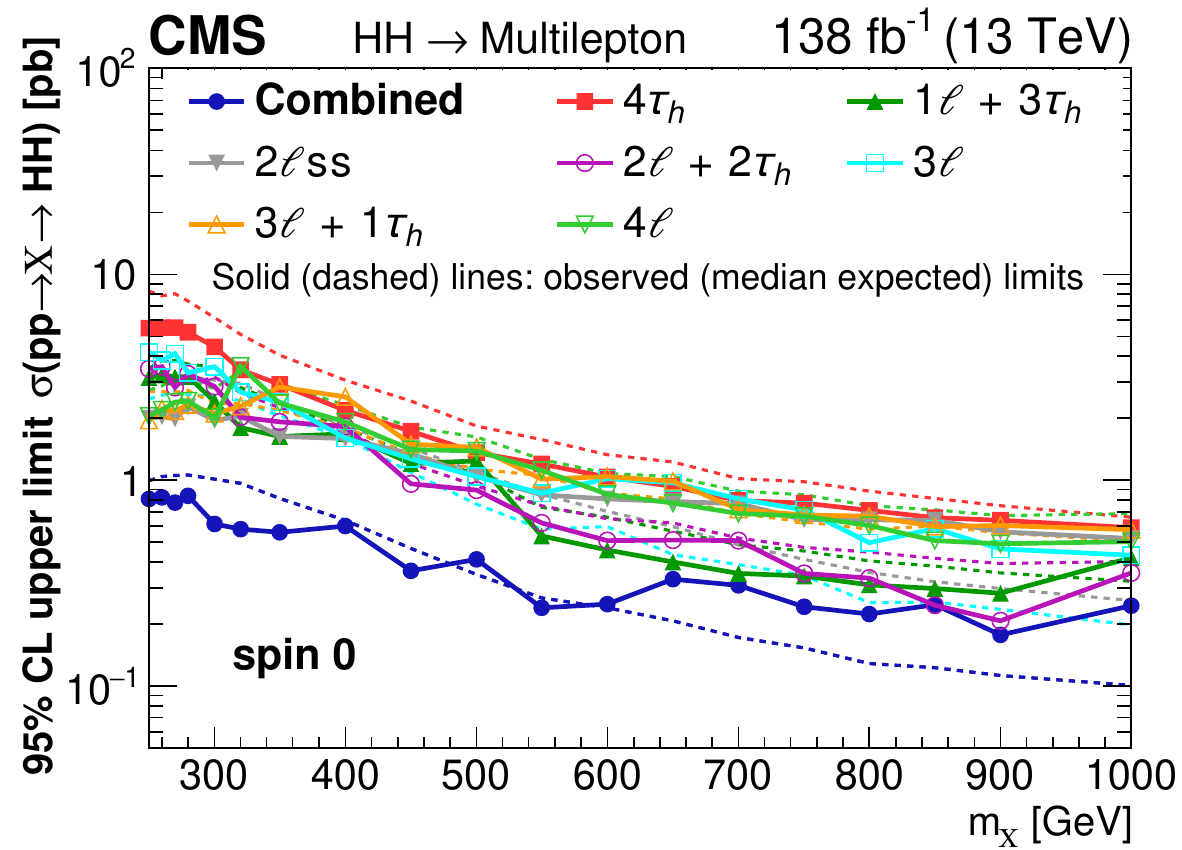}\\
    \centering\includegraphics[width=0.46\textwidth]{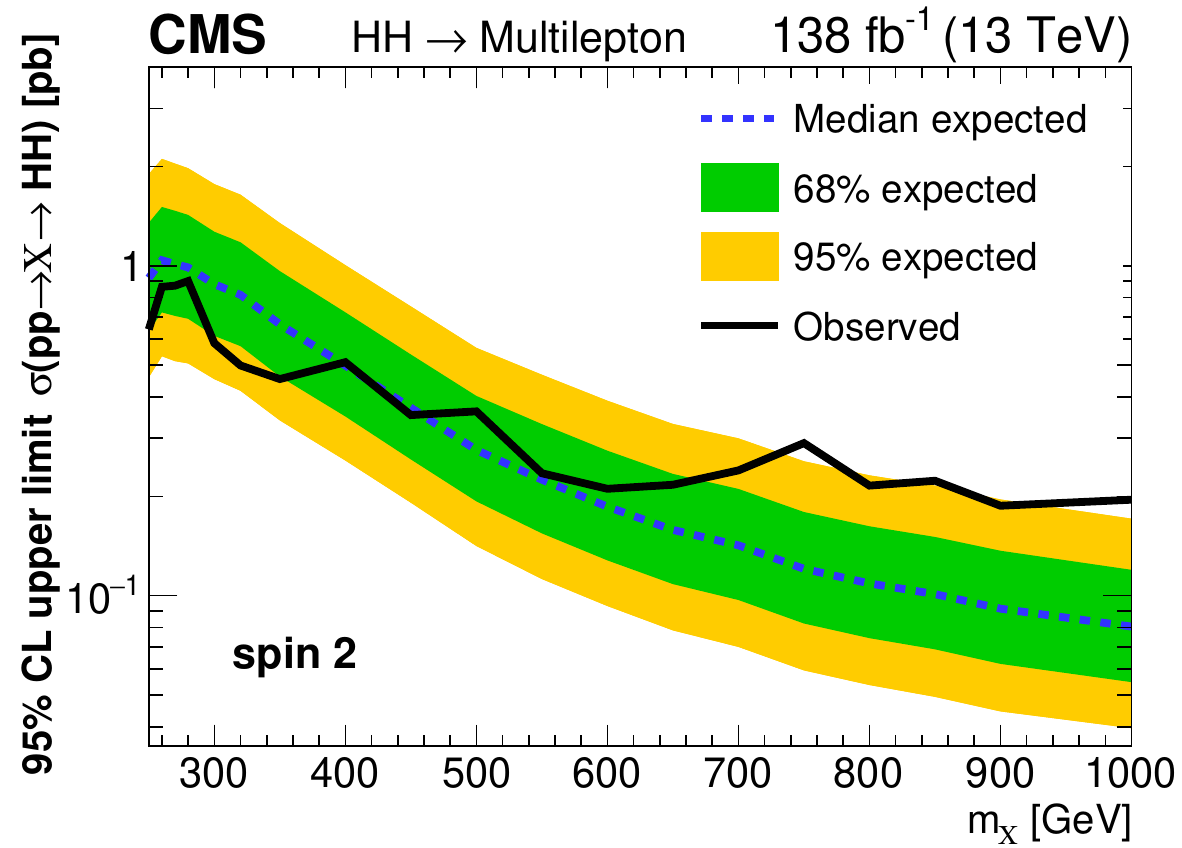}\hfill
    \centering\includegraphics[width=0.46\textwidth]{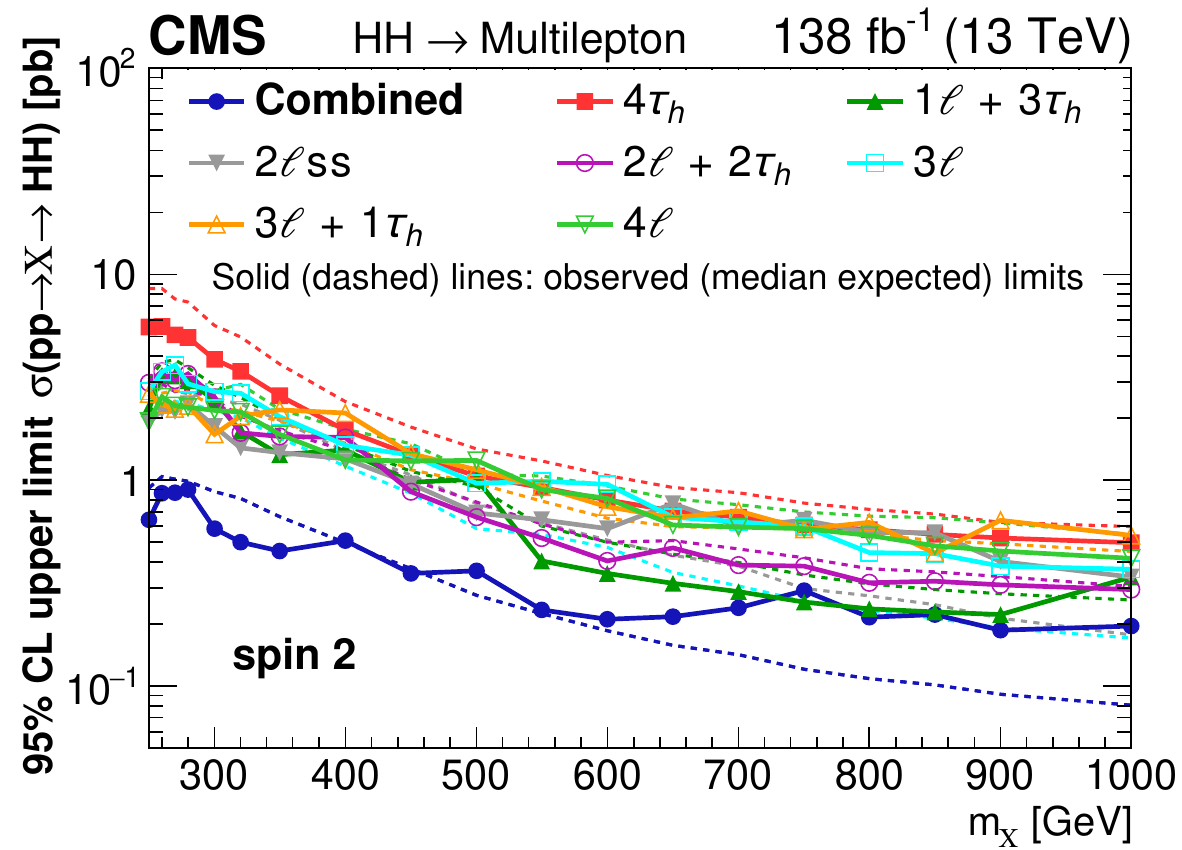}
    \caption{
        Observed and expected 95\% \CL upper limits on the production of new particles \X of spin 0 (upper) and spin 2 (lower) and mass $m_{\X}$ in the range 250--1000\GeV, which decay to \PH boson pairs.
        The plot on the left shows the result obtained by combining all seven search categories, while the plot on the right shows the limits obtained for each category separately, and the combined limit.
    }
    \label{fig:HH_limits_resonant}
\end{figure}

Figure~\ref{fig:HH_limits_resonant} shows the observed and expected limits on the resonant \HH production cross section as a function of $m_{\X}$ for a spin-0 or spin-2 particle \X decaying to \HH.
The mass points probed are listed in the fourth paragraph of Section~\ref{sec:datasets}.
The limits are expected to become more stringent as $m_{\X}$ increases, as the acceptance for the \HH signal increases and the signal can be more easily distinguished from backgrounds.
The observed (expected) 95\% \CL upper limits on the resonant \HH production cross section range from 0.18 to 0.90 (0.08 to 1.06)\pb, depending on the mass and spin.
Tabulated results are provided in the HEPData record for this analysis~\cite{hepdata}.
Only the ATLAS search in the $\Pbottom\Pbottom\PGg\PGg$ final state achieves more stringent limits at low masses (close to 250\GeV)~\cite{ATLAS-HDBS-2018-34},
while the low-mass limits from ATLAS in the $\Pbottom\Pbottom\PGt\PGt$ decay mode are roughly the same~\cite{ATLAS-HDBS-2018-40}.
Both these analyses, along with the ATLAS search for $\Pbottom\Pbottom\Pbottom\Pbottom$ decays~\cite{ATLAS:2022hwc},
set much more stringent limits at higher masses.

\begin{figure}[!ht]
    \centering\includegraphics[width=0.46\textwidth]{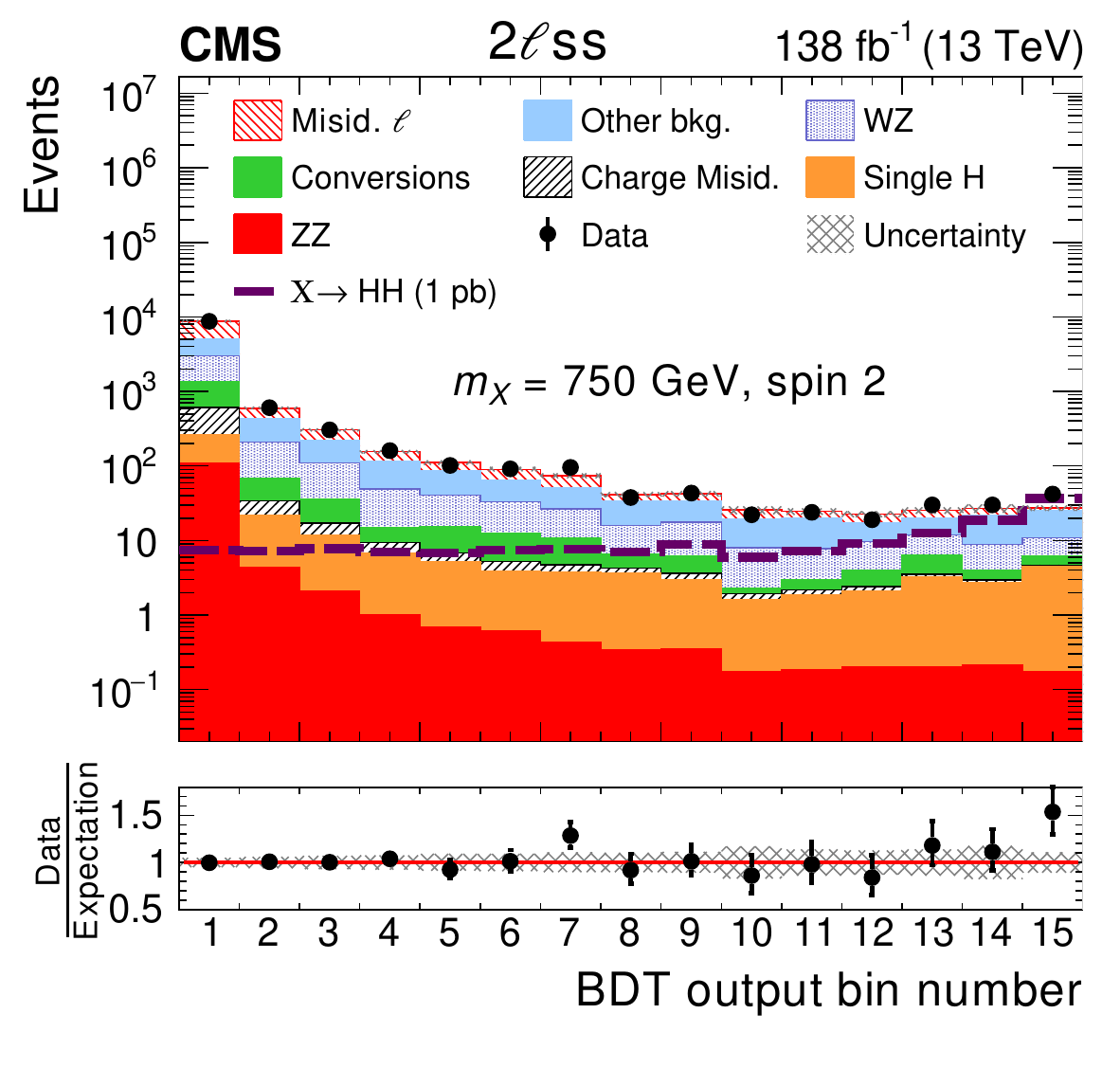}\hfill
    \centering\includegraphics[width=0.46\textwidth]{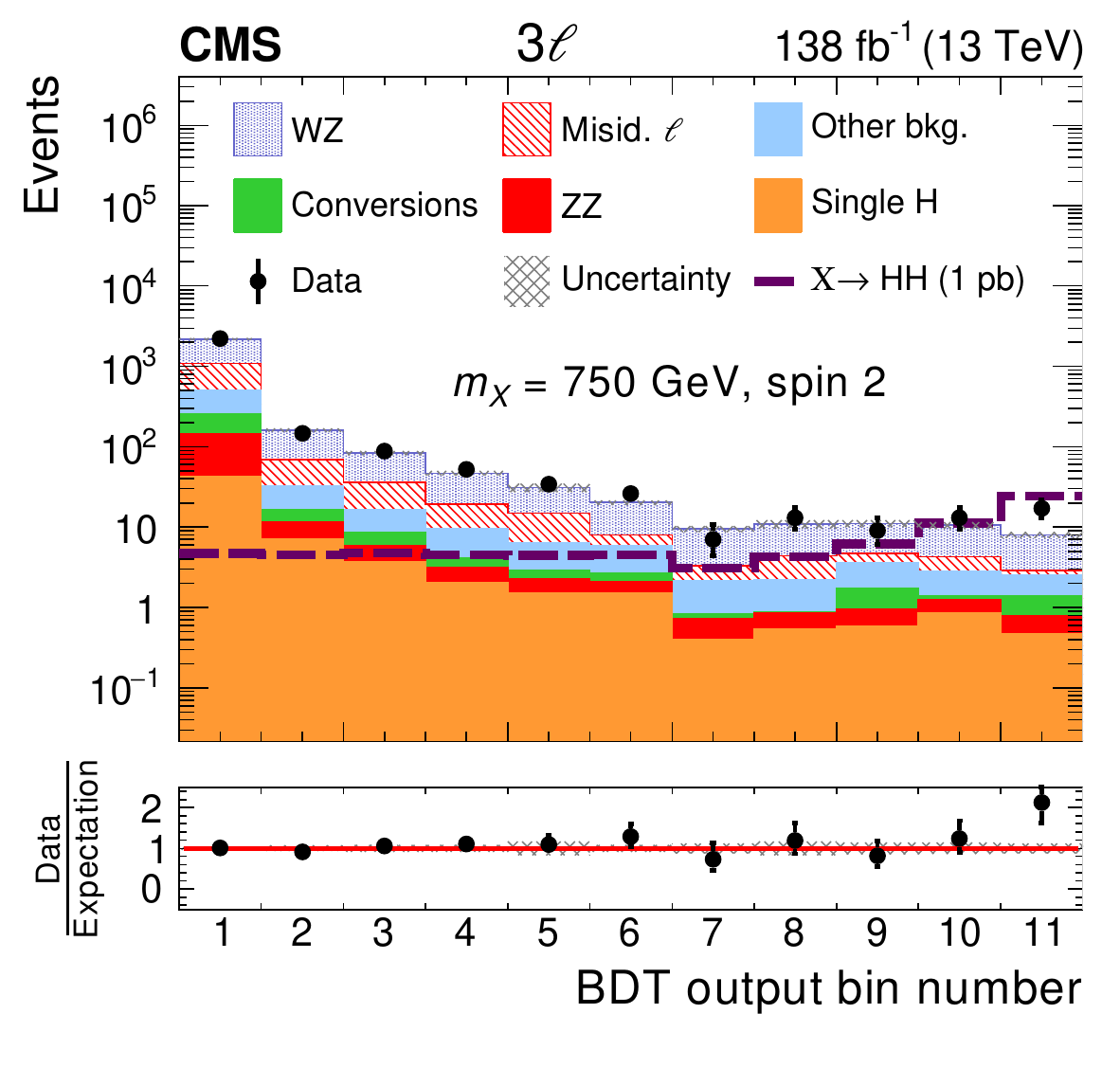}
    \caption{
        Distribution in BDT classifier output for resonances of spin 2 and mass 750\GeV in the \twoLeptonssZeroTau (\cmsLeft) and \threeLeptonZeroTau (\cmsRight) categories.
        The resonant \HH signal is shown for a cross section amounting to 1\pb.
        The distributions expected for the background processes are shown for the values of nuisance parameters obtained from the ML fit of the signal+background hypothesis to the data.
    }
    \label{fig:HH_BDT_resonant}
\end{figure}

For $m_{\X} \gtrsim 600$\GeV, the observed limit is less stringent than the expected limit, due to a small excess of events in the data that is concentrated near $m_{\X} = 750$\GeV in the \twoLeptonssZeroTau and \threeLeptonZeroTau categories.
The distributions in the output of the BDT classifier targeting resonances with spin 2 and mass 750\GeV in the \twoLeptonssZeroTau and \threeLeptonZeroTau categories are shown in Fig.~\ref{fig:HH_BDT_resonant}.
A small excess of events can be seen in the rightmost bin of both distributions.
In the \twoLeptonssZeroTau (\threeLeptonZeroTau) category, 42 (17) events are observed in this bin in the data, while $27.3 \pm 2.8\,$(stat.)$\,\pm 0.7\,$(syst.) ($8.0 \pm 0.8\,$(stat.)$\,\pm 0.5\,$(syst.)) are expected from background processes, 
amounting to a local significance of about 2.1 (2.1) standard deviations.
The excess affects the observed limits in a broad mass range from 600 to 1000\GeV.
No measurement is made for masses above 1000\GeV, as limits on \HH decays producing at least
one bottom quark pair are much more stringent in this phase space~\cite{ATLAS:2022hwc,CMS:2021roc}.
The presence of multiple neutrinos in \HH signal events in these categories, coming from \PW boson or \Pgt lepton decays, limits the experimental resolution on $m_{\X}$ and causes the BDT classifier output distributions to be highly correlated for resonances of similar mass.
No significant excess is observed in any of the other five search categories.
The significance for the combination of all seven search categories at 750\GeV amounts to 1.9 standard deviations,
without accounting for the ``look elsewhere effect''~\cite{Gross:2010qma}.

\section{Summary}
\label{sec:summary}

The results of a search for nonresonant and resonant Higgs boson pair (\HH) production in final states with multiple reconstructed leptons, including electrons and muons (\lep) and hadronically decaying tau leptons (\tauh), has been presented.
The search targets the \HH decay modes \WWWW, \WWtt, and \tttt, using proton-proton collision data recorded by the CMS experiment at a center-of-mass energy of 13\TeV and corresponding to an integrated luminosity of 138\fbinv.
Seven search categories, distinguished by \lep and \tauh multiplicity, are included in the analysis: \llss, \lllnot, \llll, \lllt, \lltt, \lttt, and \noltttt, where ``ss'' indicates an \leplep pair with the same charge.
No evidence for a signal is found in the data.
Upper limits on the cross sections for both nonresonant and resonant \HH production are set.
The observed (expected) limits on the nonresonant \HH production cross section in twenty EFT benchmark scenarios range from 0.21 to 1.09 (0.16 to 1.16)\pb at 95\% confidence level (\CL), depending on the scenario.
For nonresonant \HH production with event kinematics as predicted by the standard model (SM), the observed (expected) 95\% \CL upper limit on the \HH production rate is 21.3 (19.4) times the rate expected in the SM.
The results of the search for nonresonant \HH production are used to exclude regions in the plane of the \PH boson coupling to the top quark, \yt, and of the trilinear Higgs boson self-coupling, $\lambda$.
Assuming \yt has the value expected in the SM, the observed (expected) 95\% \CL interval for $\lambda$ is between $-6.9$ and 11.1 ($-6.9$ and 11.7) times the value expected in the SM.
The resonant production of \PH boson pairs, resulting from decays of new heavy particles \X with mass $m_{\X}$, is probed within the mass range 250--1000\GeV.
The corresponding observed (expected) 95\% \CL upper limits on the cross section for resonant \HH production range from 0.18 to 0.90 (0.08 to 1.06)\pb, depending on the mass and spin of the resonance.

\begin{acknowledgments}
We congratulate our colleagues in the CERN accelerator departments for the excellent performance of the LHC and thank the technical and administrative staffs at CERN and at other CMS institutes for their contributions to the success of the CMS effort. In addition, we gratefully acknowledge the computing centers and personnel of the Worldwide LHC Computing Grid and other centers for delivering so effectively the computing infrastructure essential to our analyses. Finally, we acknowledge the enduring support for the construction and operation of the LHC, the CMS detector, and the supporting computing infrastructure provided by the following funding agencies: BMBWF and FWF (Austria); FNRS and FWO (Belgium); CNPq, CAPES, FAPERJ, FAPERGS, and FAPESP (Brazil); MES and BNSF (Bulgaria); CERN; CAS, MoST, and NSFC (China); MINCIENCIAS (Colombia); MSES and CSF (Croatia); RIF (Cyprus); SENESCYT (Ecuador); MoER, ERC PUT and ERDF (Estonia); Academy of Finland, MEC, and HIP (Finland); CEA and CNRS/IN2P3 (France); BMBF, DFG, and HGF (Germany); GSRI (Greece); NKFIH (Hungary); DAE and DST (India); IPM (Iran); SFI (Ireland); INFN (Italy); MSIP and NRF (Republic of Korea); MES (Latvia); LAS (Lithuania); MOE and UM (Malaysia); BUAP, CINVESTAV, CONACYT, LNS, SEP, and UASLP-FAI (Mexico); MOS (Montenegro); MBIE (New Zealand); PAEC (Pakistan); MES and NSC (Poland); FCT (Portugal); MESTD (Serbia); MCIN/AEI and PCTI (Spain); MOSTR (Sri Lanka); Swiss Funding Agencies (Switzerland); MST (Taipei); MHESI and NSTDA (Thailand); TUBITAK and TENMAK (Turkey); NASU (Ukraine); STFC (United Kingdom); DOE and NSF (USA).
    
\hyphenation{Rachada-pisek} Individuals have received support from the Marie-Curie program and the European Research Council and Horizon 2020 Grant, contract Nos.\ 675440, 724704, 752730, 758316, 765710, 824093, 884104, and COST Action CA16108 (European Union); the Leventis Foundation; the Alfred P.\ Sloan Foundation; the Alexander von Humboldt Foundation; the Belgian Federal Science Policy Office; the Fonds pour la Formation \`a la Recherche dans l'Industrie et dans l'Agriculture (FRIA-Belgium); the Agentschap voor Innovatie door Wetenschap en Technologie (IWT-Belgium); the F.R.S.-FNRS and FWO (Belgium) under the ``Excellence of Science -- EOS" -- be.h project n.\ 30820817; the Beijing Municipal Science \& Technology Commission, No. Z191100007219010; the Ministry of Education, Youth and Sports (MEYS) of the Czech Republic; the Hellenic Foundation for Research and Innovation (HFRI), Project Number 2288 (Greece); the Deutsche Forschungsgemeinschaft (DFG), under Germany's Excellence Strategy -- EXC 2121 ``Quantum Universe" -- 390833306, and under project number 400140256 - GRK2497; the Hungarian Academy of Sciences, the New National Excellence Program - \'UNKP, the NKFIH research grants K 124845, K 124850, K 128713, K 128786, K 129058, K 131991, K 133046, K 138136, K 143460, K 143477, 2020-2.2.1-ED-2021-00181, and TKP2021-NKTA-64 (Hungary); the Council of Science and Industrial Research, India; the Latvian Council of Science; the Ministry of Education and Science, project no. 2022/WK/14, and the National Science Center, contracts Opus 2021/41/B/ST2/01369 and 2021/43/B/ST2/01552 (Poland); the Funda\c{c}\~ao para a Ci\^encia e a Tecnologia, grant CEECIND/01334/2018 (Portugal); the National Priorities Research Program by Qatar National Research Fund; MCIN/AEI/10.13039/501100011033, ERDF ``a way of making Europe", and the Programa Estatal de Fomento de la Investigaci{\'o}n Cient{\'i}fica y T{\'e}cnica de Excelencia Mar\'{\i}a de Maeztu, grant MDM-2017-0765 and Programa Severo Ochoa del Principado de Asturias (Spain); the Chulalongkorn Academic into Its 2nd Century Project Advancement Project, and the National Science, Research and Innovation Fund via the Program Management Unit for Human Resources \& Institutional Development, Research and Innovation, grant B05F650021 (Thailand); the Kavli Foundation; the Nvidia Corporation; the SuperMicro Corporation; the Welch Foundation, contract C-1845; and the Weston Havens Foundation (USA).    
\end{acknowledgments}

\bibliography{auto_generated}

\providecommand{\href}[2]{#2}\begingroup\raggedright\begin{thebibliography}{100}%
\makeatletter
\providecommand{\hrefCMSnoop }[0]{\@secondoftwo}%
\makeatother
\providecommand{\doi}{\texttt{doi:}\begingroup \urlstyle{tt}\Url}

\bibitem{Higgs-Discovery_ATLAS}
\hrefCMSnoop {}{{ATLAS Collaboration}, ``{Observation of a new particle in the
  search for the standard model {Higgs} boson with the {ATLAS} detector at the
  {LHC}}'',} \textit{ Phys. Lett. B} \textbf{ 716} (2012) 1,
  \href{http://dx.doi.org/10.1016/j.physletb.2012.08.020}{\doi{10.1016/j.physletb.2012.08.020}},
\href{http://www.arXiv.org/abs/1207.7214}{\texttt{arXiv:1207.7214}}.

\bibitem{Higgs-Discovery_CMS}
\hrefCMSnoop {}{{CMS Collaboration}, ``Observation of a new boson at a mass of
  {125\GeV} with the {CMS} experiment at the {LHC}'',} \textit{ Phys. Lett. B}
  \textbf{ 716} (2012) 30,
  \href{http://dx.doi.org/10.1016/j.physletb.2012.08.021}{\doi{10.1016/j.physletb.2012.08.021}},
\href{http://www.arXiv.org/abs/1207.7235}{\texttt{arXiv:1207.7235}}.

\bibitem{Higgs-Discovery_CMS_long}
\hrefCMSnoop {}{{CMS Collaboration}, ``Observation of a new boson with mass
  near {125\GeV} in {$\Pp\Pp$} collisions at {$\sqrt{s} = 7$} and {8\TeV}'',}
  \textit{ JHEP} \textbf{ 06} (2013) 081,
  \href{http://dx.doi.org/10.1007/JHEP06(2013)081}{\doi{10.1007/JHEP06(2013)081}},
\href{http://www.arXiv.org/abs/1303.4571}{\texttt{arXiv:1303.4571}}.

\bibitem{CMS-HIG-17-031}
\hrefCMSnoop {}{{CMS Collaboration}, ``{Combined measurements of Higgs boson
  couplings in proton-proton collisions at $\sqrt{s} = 13\TeV$}'',} \textit{
  Eur. Phys. J. C} \textbf{ 79} (2019) 421,
  \href{http://dx.doi.org/10.1140/epjc/s10052-019-6909-y}{\doi{10.1140/epjc/s10052-019-6909-y}},
  \href{http://www.arXiv.org/abs/1809.10733}{\texttt{arXiv:1809.10733}}.

\bibitem{ATLAS-HIGG-2018-57}
\hrefCMSnoop {}{{ATLAS Collaboration}, ``{Combined measurements of Higgs boson
  production and decay using up to $80\fbinv$ of proton-proton collision data
  at $\sqrt{s} = 13\TeV$ collected with the ATLAS experiment}'',} \textit{
  Phys. Rev. D} \textbf{ 101} (2020) 012002,
  \href{http://dx.doi.org/10.1103/PhysRevD.101.012002}{\doi{10.1103/PhysRevD.101.012002}},
  \href{http://www.arXiv.org/abs/1909.02845}{\texttt{arXiv:1909.02845}}.

\bibitem{CMS-HIG-21-013}
\hrefCMSnoop {}{{CMS Collaboration}, ``{Measurement of the Higgs boson width
  and evidence of its off-shell contributions to ZZ production}'',} \textit{
  Nat. Phys.} (2022)
  \href{http://dx.doi.org/10.1038/s41567-022-01682-0}{\doi{10.1038/s41567-022-01682-0}},
  \href{http://www.arXiv.org/abs/2202.06923}{\texttt{arXiv:2202.06923}}.

\bibitem{Englert:1964et}
\hrefCMSnoop {}{F.~Englert and R.~Brout, ``{Broken symmetry and the mass of
  gauge vector mesons}'',} \textit{ Phys. Rev. Lett.} \textbf{ 13} (1964) 321,
  \href{http://dx.doi.org/10.1103/PhysRevLett.13.321}{\doi{10.1103/PhysRevLett.13.321}}.

\bibitem{Higgs:1964ia}
\hrefCMSnoop {}{P.~W. Higgs, ``{Broken symmetries, massless particles and gauge
  fields}'',} \textit{ Phys. Lett.} \textbf{ 12} (1964) 132,
  \href{http://dx.doi.org/10.1016/0031-9163(64)91136-9}{\doi{10.1016/0031-9163(64)91136-9}}.

\bibitem{Higgs:1964pj}
\hrefCMSnoop {}{P.~W. Higgs, ``{Broken symmetries and the masses of gauge
  bosons}'',} \textit{ Phys. Rev. Lett.} \textbf{ 13} (1964) 508,
  \href{http://dx.doi.org/10.1103/PhysRevLett.13.508}{\doi{10.1103/PhysRevLett.13.508}}.

\bibitem{Guralnik:1964eu}
\hrefCMSnoop {}{G.~S. Guralnik, C.~R. Hagen, and T.~W.~B. Kibble, ``{Global
  conservation laws and massless particles}'',} \textit{ Phys. Rev. Lett.}
  \textbf{ 13} (1964) 585,
  \href{http://dx.doi.org/10.1103/PhysRevLett.13.585}{\doi{10.1103/PhysRevLett.13.585}}.

\bibitem{Higgs:1966ev}
\hrefCMSnoop {}{P.~W. Higgs, ``{Spontaneous symmetry breakdown without massless
  bosons}'',} \textit{ Phys. Rev.} \textbf{ 145} (1966) 1156,
  \href{http://dx.doi.org/10.1103/PhysRev.145.1156}{\doi{10.1103/PhysRev.145.1156}}.

\bibitem{Kibble:1967sv}
\hrefCMSnoop {}{T.~W.~B. Kibble, ``{Symmetry breaking in non-Abelian gauge
  theories}'',} \textit{ Phys. Rev.} \textbf{ 155} (1967) 1554,
  \href{http://dx.doi.org/10.1103/PhysRev.155.1554}{\doi{10.1103/PhysRev.155.1554}}.

\bibitem{Glashow:1961tr}
\hrefCMSnoop {}{S.~L. Glashow, ``{Partial symmetries of weak interactions}'',}
  \textit{ Nucl. Phys.} \textbf{ 22} (1961) 579,
  \href{http://dx.doi.org/10.1016/0029-5582(61)90469-2}{\doi{10.1016/0029-5582(61)90469-2}}.

\bibitem{Weinberg:1967tq}
\hrefCMSnoop {}{S.~Weinberg, ``{A model of leptons}'',} \textit{ Phys. Rev.
  Lett.} \textbf{ 19} (1967) 1264,
  \href{http://dx.doi.org/10.1103/PhysRevLett.19.1264}{\doi{10.1103/PhysRevLett.19.1264}}.

\bibitem{Salam:1968rm}
\hrefCMSnoop {}{A.~Salam, ``{Weak and electromagnetic interactions}'',}
  \textit{ Conf. Proc. C} \textbf{ 680519} (1968) 367,
  \href{http://dx.doi.org/10.1142/9789812795915_0034}{\doi{10.1142/9789812795915_0034}}.

\bibitem{deFlorian:2019app}
\hrefCMSnoop {}{D.~de~Florian, I.~Fabre, and J.~Mazzitelli, ``{Triple Higgs
  production at hadron colliders at NNLO in QCD}'',} \textit{ JHEP} \textbf{
  03} (2020) 155,
  \href{http://dx.doi.org/10.1007/JHEP03(2020)155}{\doi{10.1007/JHEP03(2020)155}},
  \href{http://www.arXiv.org/abs/1912.02760}{\texttt{arXiv:1912.02760}}.

\bibitem{HL-LHC-TDR}
\hrefCMSnoop {}{O.~Aberle {et~al.}, ``{High-Luminosity Large Hadron Collider
  (HL-LHC)}: {Technical design report V. 0.1}'',} \textit{ CERN Yellow Rep.
  Monogr.} \textbf{ 4} (2017)
  \href{http://dx.doi.org/10.23731/CYRM-2017-004}{\doi{10.23731/CYRM-2017-004}}.

\bibitem{Grazzini:2018hh}
M.~Grazzini\hrefCMSnoop {}{ {et~al.}, ``Higgs boson pair production at {NNLO}
  with top quark mass effects'',} \textit{ JHEP} \textbf{ 05} (2018) 059,
  \href{http://dx.doi.org/10.1007/JHEP05(2018)059}{\doi{10.1007/JHEP05(2018)059}},
  \href{http://www.arXiv.org/abs/1803.02463}{\texttt{arXiv:1803.02463}}.

\bibitem{Dreyer:2018qbw}
\hrefCMSnoop {}{F.~A. Dreyer and A.~Karlberg, ``{Vector-boson fusion Higgs pair
  production at N$^3$LO}'',} \textit{ Phys. Rev. D} \textbf{ 98} (2018) 114016,
  \href{http://dx.doi.org/10.1103/PhysRevD.98.114016}{\doi{10.1103/PhysRevD.98.114016}},
  \href{http://www.arXiv.org/abs/1811.07906}{\texttt{arXiv:1811.07906}}.

\bibitem{Degrassi:2016wml}
\hrefCMSnoop {}{G.~Degrassi, P.~P. Giardino, F.~Maltoni, and D.~Pagani,
  ``{Probing the Higgs self coupling via single Higgs production at the
  LHC}'',} \textit{ JHEP} \textbf{ 12} (2016) 080,
  \href{http://dx.doi.org/10.1007/JHEP12(2016)080}{\doi{10.1007/JHEP12(2016)080}},
  \href{http://www.arXiv.org/abs/1607.04251}{\texttt{arXiv:1607.04251}}.

\bibitem{Maltoni:2017ims}
\hrefCMSnoop {}{F.~Maltoni, D.~Pagani, A.~Shivaji, and X.~Zhao, ``{Trilinear
  Higgs coupling determination via single-Higgs differential measurements at
  the LHC}'',} \textit{ Eur. Phys. J. C} \textbf{ 77} (2017) 887,
  \href{http://dx.doi.org/10.1140/epjc/s10052-017-5410-8}{\doi{10.1140/epjc/s10052-017-5410-8}},
  \href{http://www.arXiv.org/abs/1709.08649}{\texttt{arXiv:1709.08649}}.

\bibitem{Buchmuller:1985jz}
\hrefCMSnoop {}{W.~Buchmuller and D.~Wyler, ``{Effective Lagrangian analysis of
  new interactions and flavor conservation}'',} \textit{ Nucl. Phys. B}
  \textbf{ 268} (1986) 621,
  \href{http://dx.doi.org/10.1016/0550-3213(86)90262-2}{\doi{10.1016/0550-3213(86)90262-2}}.

\bibitem{Grzadkowski:2010es}
\hrefCMSnoop {}{B.~Grzadkowski, M.~Iskrzynski, M.~Misiak, and J.~Rosiek,
  ``{Dimension-six terms in the standard model Lagrangian}'',} \textit{ JHEP}
  \textbf{ 10} (2010) 085,
  \href{http://dx.doi.org/10.1007/JHEP10(2010)085}{\doi{10.1007/JHEP10(2010)085}},
  \href{http://www.arXiv.org/abs/1008.4884}{\texttt{arXiv:1008.4884}}.

\bibitem{Carvalho:2015ttv}
A.~Carvalho\hrefCMSnoop {}{ {et~al.}, ``{Higgs pair production: choosing
  benchmarks with cluster analysis}'',} \textit{ JHEP} \textbf{ 04} (2016) 126,
  \href{http://dx.doi.org/10.1007/JHEP04(2016)126}{\doi{10.1007/JHEP04(2016)126}},
  \href{http://www.arXiv.org/abs/1507.02245}{\texttt{arXiv:1507.02245}}.

\bibitem{Craig:2013hca}
\hrefCMSnoop {}{N.~Craig, J.~Galloway, and S.~Thomas, ``{Searching for signs of
  the second Higgs doublet}'',} 2013.
\href{http://www.arXiv.org/abs/1305.2424}{\texttt{arXiv:1305.2424}}.

\bibitem{Nhung:2013lpa}
\hrefCMSnoop {}{D.~T. Nhung, M.~M{\"u}hlleitner, J.~Streicher, and K.~Walz,
  ``{Higher order corrections to the trilinear Higgs self-couplings in the real
  NMSSM}'',} \textit{ JHEP} \textbf{ 11} (2013) 181,
  \href{http://dx.doi.org/10.1007/JHEP11(2013)181}{\doi{10.1007/JHEP11(2013)181}},
\href{http://www.arXiv.org/abs/1306.3926}{\texttt{arXiv:1306.3926}}.

\bibitem{Grober:2010yv}
\hrefCMSnoop {}{R.~Grober and M.~M{\"{u}}hlleitner, ``{Composite Higgs boson
  pair production at the LHC}'',} \textit{ JHEP} \textbf{ 06} (2011) 020,
  \href{http://dx.doi.org/10.1007/JHEP06(2011)020}{\doi{10.1007/JHEP06(2011)020}},
\href{http://www.arXiv.org/abs/1012.1562}{\texttt{arXiv:1012.1562}}.

\bibitem{Contino:2010mh}
R.~Contino\hrefCMSnoop {}{ {et~al.}, ``{Strong double Higgs production at the
  LHC}'',} \textit{ JHEP} \textbf{ 05} (2010) 089,
  \href{http://dx.doi.org/10.1007/JHEP05(2010)089}{\doi{10.1007/JHEP05(2010)089}},
\href{http://www.arXiv.org/abs/1002.1011}{\texttt{arXiv:1002.1011}}.

\bibitem{Englert:2011yb}
\hrefCMSnoop {}{C.~Englert, T.~Plehn, D.~Zerwas, and P.~M. Zerwas, ``{Exploring
  the Higgs portal}'',} \textit{ Phys. Lett. B} \textbf{ 703} (2011) 298,
  \href{http://dx.doi.org/10.1016/j.physletb.2011.08.002}{\doi{10.1016/j.physletb.2011.08.002}},
\href{http://www.arXiv.org/abs/1106.3097}{\texttt{arXiv:1106.3097}}.

\bibitem{No:2013wsa}
\hrefCMSnoop {}{J.~M. No and M.~Ramsey-Musolf, ``{Probing the Higgs portal at
  the LHC through resonant di-Higgs production}'',} \textit{ Phys. Rev. D}
  \textbf{ 89} (2014) 095031,
  \href{http://dx.doi.org/10.1103/PhysRevD.89.095031}{\doi{10.1103/PhysRevD.89.095031}},
\href{http://www.arXiv.org/abs/1310.6035}{\texttt{arXiv:1310.6035}}.

\bibitem{Randall:1999ee}
\hrefCMSnoop {}{L.~Randall and R.~Sundrum, ``{A large mass hierarchy from a
  small extra dimension}'',} \textit{ Phys. Rev. Lett.} \textbf{ 83} (1999)
  3370,
  \href{http://dx.doi.org/10.1103/PhysRevLett.83.3370}{\doi{10.1103/PhysRevLett.83.3370}},
\href{http://www.arXiv.org/abs/hep-ph/9905221}{\texttt{arXiv:hep-ph/9905221}}.

\bibitem{Cheung:2000rw}
\hrefCMSnoop {}{K.~Cheung, ``{Phenomenology of radion in Randall-Sundrum
  scenario}'',} \textit{ Phys. Rev. D} \textbf{ 63} (2001) 056007,
  \href{http://dx.doi.org/10.1103/PhysRevD.63.056007}{\doi{10.1103/PhysRevD.63.056007}},
\href{http://www.arXiv.org/abs/hep-ph/0009232}{\texttt{arXiv:hep-ph/0009232}}.

\bibitem{ATLAS:2022hwc}
\hrefCMSnoop {}{{ATLAS Collaboration}, ``{Search for resonant pair production
  of Higgs bosons in the $\Pbottom\APbottom\Pbottom\APbottom$ final state using
  $\Pp\Pp$ collisions at $\sqrt{s} = 13\TeV$ with the ATLAS detector}'',}
  \textit{ Phys. Rev. D} \textbf{ 105} (2022) 092002,
  \href{http://dx.doi.org/10.1103/PhysRevD.105.092002}{\doi{10.1103/PhysRevD.105.092002}},
  \href{http://www.arXiv.org/abs/2202.07288}{\texttt{arXiv:2202.07288}}.

\bibitem{CMS:2021roc}
\hrefCMSnoop {}{{CMS Collaboration}, ``{Search for heavy resonances decaying to
  a pair of Lorentz-boosted Higgs bosons in final states with leptons and a
  bottom quark pair at $\sqrt{s} = 13\TeV$}'',} \textit{ JHEP} \textbf{ 05}
  (2022) 005,
  \href{http://dx.doi.org/10.1007/JHEP05(2022)005}{\doi{10.1007/JHEP05(2022)005}},
  \href{http://www.arXiv.org/abs/2112.03161}{\texttt{arXiv:2112.03161}}.

\bibitem{ATLAS-HDBS-2018-40}
\hrefCMSnoop {}{{ATLAS Collaboration}, ``{Search for resonant and non-resonant
  Higgs boson pair production in the $\Pbottom\APbottom\PGt^{+}\PGt^{-}$ decay
  channel using $13\TeV$ $\Pp\Pp$ collision data from the ATLAS detector}'',}
  2022. \href{http://www.arXiv.org/abs/2209.10910}{\texttt{arXiv:2209.10910}}.
  Submitted to \textit{JHEP}.

\bibitem{Baur:2002rb}
\hrefCMSnoop {}{U.~Baur, T.~Plehn, and D.~L. Rainwater, ``{Measuring the Higgs
  boson self-coupling at the LHC and finite top mass matrix elements}'',}
  \textit{ Phys. Rev. Lett.} \textbf{ 89} (2002) 151801,
  \href{http://dx.doi.org/10.1103/PhysRevLett.89.151801}{\doi{10.1103/PhysRevLett.89.151801}},
  \href{http://www.arXiv.org/abs/hep-ph/0206024}{\texttt{arXiv:hep-ph/0206024}}.

\bibitem{Baur:2002qd}
\hrefCMSnoop {}{U.~Baur, T.~Plehn, and D.~L. Rainwater, ``{Determining the
  Higgs boson self-coupling at hadron colliders}'',} \textit{ Phys. Rev. D}
  \textbf{ 67} (2003) 033003,
  \href{http://dx.doi.org/10.1103/PhysRevD.67.033003}{\doi{10.1103/PhysRevD.67.033003}},
  \href{http://www.arXiv.org/abs/hep-ph/0211224}{\texttt{arXiv:hep-ph/0211224}}.

\bibitem{Li:2015yia}
\hrefCMSnoop {}{Q.~Li, Z.~Li, Q.-S. Yan, and X.~Zhao, ``{Probe Higgs boson pair
  production via the $3\Plepton \, 2\jet + \ptmiss$ mode}'',} \textit{ Phys.
  Rev. D} \textbf{ 92} (2015) 014015,
  \href{http://dx.doi.org/10.1103/PhysRevD.92.014015}{\doi{10.1103/PhysRevD.92.014015}},
  \href{http://www.arXiv.org/abs/1503.07611}{\texttt{arXiv:1503.07611}}.

\bibitem{Adhikary:2017jtu}
A.~Adhikary\hrefCMSnoop {}{ {et~al.}, ``{Revisiting the non-resonant Higgs pair
  production at the HL-LHC}'',} \textit{ JHEP} \textbf{ 07} (2018) 116,
  \href{http://dx.doi.org/10.1007/JHEP07(2018)116}{\doi{10.1007/JHEP07(2018)116}},
  \href{http://www.arXiv.org/abs/1712.05346}{\texttt{arXiv:1712.05346}}.

\bibitem{Ren:2017jbg}
J.~Ren\hrefCMSnoop {}{ {et~al.}, ``{LHC search of new Higgs boson via resonant
  di-Higgs production with decays into $4\PW$}'',} \textit{ JHEP} \textbf{ 06}
  (2018) 090,
  \href{http://dx.doi.org/10.1007/JHEP06(2018)090}{\doi{10.1007/JHEP06(2018)090}},
  \href{http://www.arXiv.org/abs/1706.05980}{\texttt{arXiv:1706.05980}}.

\bibitem{Aaboud:2018ksn}
\hrefCMSnoop {}{{ATLAS Collaboration}, ``{Search for Higgs boson pair
  production in the $\PW\PW^{\ast}\PW\PW^{\ast}$ decay channel using ATLAS data
  recorded at $\sqrt{s} = 13\TeV$}'',} \textit{ JHEP} \textbf{ 05} (2019) 124,
  \href{http://dx.doi.org/10.1007/JHEP05(2019)124}{\doi{10.1007/JHEP05(2019)124}},
  \href{http://www.arXiv.org/abs/1811.11028}{\texttt{arXiv:1811.11028}}.

\bibitem{CMS:2020tkr}
\hrefCMSnoop {}{{CMS Collaboration}, ``{Search for nonresonant Higgs boson pair
  production in final states with two bottom quarks and two photons in
  proton-proton collisions at $\sqrt{s} = 13\TeV$}'',} \textit{ JHEP} \textbf{
  03} (2021) 257,
  \href{http://dx.doi.org/10.1007/JHEP03(2021)257}{\doi{10.1007/JHEP03(2021)257}},
  \href{http://www.arXiv.org/abs/2011.12373}{\texttt{arXiv:2011.12373}}.

\bibitem{ATLAS-HDBS-2018-34}
\hrefCMSnoop {}{{ATLAS Collaboration}, ``{Search for Higgs boson pair
  production in the two bottom quarks plus two photons final state in $\Pp\Pp$
  collisions at $\sqrt{s} = 13\TeV$ with the ATLAS detector}'',} \textit{ Phys.
  Rev. D} \textbf{ 106} (2022) 052001,
  \href{http://dx.doi.org/10.1103/PhysRevD.106.052001}{\doi{10.1103/PhysRevD.106.052001}},
  \href{http://www.arXiv.org/abs/2112.11876}{\texttt{arXiv:2112.11876}}.

\bibitem{ATLAS:2020jgy}
\hrefCMSnoop {}{{ATLAS Collaboration}, ``{Search for the $\HH \rightarrow
  \Pbottom\APbottom\Pbottom\APbottom$ process via vector-boson fusion
  production using proton-proton collisions at $\sqrt{s} = 13\TeV$ with the
  ATLAS detector}'',} \textit{ JHEP} \textbf{ 07} (2020) 108,
  \href{http://dx.doi.org/10.1007/JHEP07(2020)108}{\doi{10.1007/JHEP07(2020)108}},
  \href{http://www.arXiv.org/abs/2001.05178}{\texttt{arXiv:2001.05178}}.
  [Errata: \DOI{10.1007/JHEP01(2021)145}, \DOI{10.1007/JHEP05(2021)207}].

\bibitem{CMS:2022cpr}
\hrefCMSnoop {}{{CMS Collaboration}, ``{Search for Higgs boson pair production
  in the four $\Pbottom$ quark final state in proton-proton collisions at
  $\sqrt{s} = 13\TeV$}'',} \textit{ Phys. Rev. Lett.} \textbf{ 129} (2022)
  081802,
  \href{http://dx.doi.org/10.1103/PhysRevLett.129.081802}{\doi{10.1103/PhysRevLett.129.081802}},
  \href{http://www.arXiv.org/abs/2202.09617}{\texttt{arXiv:2202.09617}}.

\bibitem{CMS-B2G-22-003}
\hrefCMSnoop {}{{CMS Collaboration}, ``{Search for nonresonant pair production
  of highly energetic Higgs bosons decaying to bottom quarks}'',} 2022.
  \href{http://www.arXiv.org/abs/2205.06667}{\texttt{arXiv:2205.06667}}.
  Submitted to \textit{Phys. Rev. Lett.}

\bibitem{Sirunyan:2018zkk}
\hrefCMSnoop {}{{CMS Collaboration}, ``{Search for resonant pair production of
  Higgs bosons decaying to bottom quark-antiquark pairs in proton-proton
  collisions at $13\TeV$}'',} \textit{ JHEP} \textbf{ 08} (2018) 152,
  \href{http://dx.doi.org/10.1007/JHEP08(2018)152}{\doi{10.1007/JHEP08(2018)152}},
  \href{http://www.arXiv.org/abs/1806.03548}{\texttt{arXiv:1806.03548}}.

\bibitem{CMS-HIG-20-010}
\hrefCMSnoop {}{{CMS Collaboration}, ``{Search for nonresonant Higgs boson pair
  production in final state with two bottom quarks and two tau leptons in
  proton-proton collisions at $\sqrt{s} = 13\TeV$}'',} 2022.
  \href{http://www.arXiv.org/abs/2206.09401}{\texttt{arXiv:2206.09401}}.
  Submitted to \textit{Phys. Lett. B}.

\bibitem{ATLAS:2020azv}
\hrefCMSnoop {}{{ATLAS Collaboration}, ``{Reconstruction and identification of
  boosted di-$\PGt$ systems in a search for Higgs boson pairs using $13\TeV$
  proton-proton collision data in ATLAS}'',} \textit{ JHEP} \textbf{ 11} (2020)
  163,
  \href{http://dx.doi.org/10.1007/JHEP11(2020)163}{\doi{10.1007/JHEP11(2020)163}},
  \href{http://www.arXiv.org/abs/2007.14811}{\texttt{arXiv:2007.14811}}.

\bibitem{Sirunyan:2017guj}
\hrefCMSnoop {}{{CMS Collaboration}, ``{Search for resonant and non-resonant
  Higgs boson pair production in the
  $\Pbottom\APbottom\Plepton\Pnu\Plepton\Pnu$ final state in proton-proton
  collisions at $\sqrt{s} = 13\TeV$}'',} \textit{ JHEP} \textbf{ 01} (2018)
  054,
  \href{http://dx.doi.org/10.1007/JHEP01(2018)054}{\doi{10.1007/JHEP01(2018)054}},
  \href{http://www.arXiv.org/abs/1708.04188}{\texttt{arXiv:1708.04188}}.

\bibitem{ATLAS-HDBS-2018-33}
\hrefCMSnoop {}{{ATLAS Collaboration}, ``{Search for non-resonant Higgs boson
  pair production in the $\Pbottom\APbottom\Plepton\Pnu\Plepton\Pnu$ final
  state with the ATLAS detector in $\Pp\Pp$ collisions at $\sqrt{s} =
  13\TeV$}'',} \textit{ Phys. Lett. B} \textbf{ 801} (2020) 135145,
  \href{http://dx.doi.org/10.1016/j.physletb.2019.135145}{\doi{10.1016/j.physletb.2019.135145}},
  \href{http://www.arXiv.org/abs/1908.06765}{\texttt{arXiv:1908.06765}}.

\bibitem{ATLAS:2018fpd}
\hrefCMSnoop {}{{ATLAS Collaboration}, ``{Search for Higgs boson pair
  production in the $\Pbottom\APbottom\PW\PWst$ decay mode at $\sqrt{s} =
  13\TeV$ with the ATLAS detector}'',} \textit{ JHEP} \textbf{ 04} (2019) 092,
  \href{http://dx.doi.org/10.1007/JHEP04(2019)092}{\doi{10.1007/JHEP04(2019)092}},
  \href{http://www.arXiv.org/abs/1811.04671}{\texttt{arXiv:1811.04671}}.

\bibitem{Aaboud:2018ewm}
\hrefCMSnoop {}{{ATLAS Collaboration}, ``{Search for Higgs boson pair
  production in the $\PGg\PGg\PW\PW^{\ast}$ channel using $\Pp\Pp$ collision
  data recorded at $\sqrt{s} = 13\TeV$ with the ATLAS detector}'',} \textit{
  Eur. Phys. J. C} \textbf{ 78} (2018) 1007,
  \href{http://dx.doi.org/10.1140/epjc/s10052-018-6457-x}{\doi{10.1140/epjc/s10052-018-6457-x}},
  \href{http://www.arXiv.org/abs/1807.08567}{\texttt{arXiv:1807.08567}}.

\bibitem{Sirunyan:2018ayu}
\hrefCMSnoop {}{{CMS Collaboration}, ``{Combination of searches for Higgs boson
  pair production in proton-proton collisions at $\sqrt{s} = 13\TeV$}'',}
  \textit{ Phys. Rev. Lett.} \textbf{ 122} (2019) 121803,
  \href{http://dx.doi.org/10.1103/PhysRevLett.122.121803}{\doi{10.1103/PhysRevLett.122.121803}},
  \href{http://www.arXiv.org/abs/1811.09689}{\texttt{arXiv:1811.09689}}.

\bibitem{2020135103}
\hrefCMSnoop {}{{ATLAS Collaboration}, ``{Combination of searches for Higgs
  boson pairs in $\Pp\Pp$ collisions at $\sqrt{s} = 13\TeV$ with the ATLAS
  detector}'',} \textit{ Phys. Lett. B} \textbf{ 800} (2020) 135103,
  \href{http://dx.doi.org/10.1016/j.physletb.2019.135103}{\doi{10.1016/j.physletb.2019.135103}},
  \href{http://www.arXiv.org/abs/1906.02025}{\texttt{arXiv:1906.02025}}.

\bibitem{Sirunyan:2020zal}
\hrefCMSnoop {}{{CMS Collaboration}, ``{Performance of the CMS level-1 trigger
  in proton-proton collisions at $\sqrt{s} = 13\TeV$}'',} \textit{ JINST}
  \textbf{ 15} (2020) P10017,
  \href{http://dx.doi.org/10.1088/1748-0221/15/10/P10017}{\doi{10.1088/1748-0221/15/10/P10017}},
  \href{http://www.arXiv.org/abs/2006.10165}{\texttt{arXiv:2006.10165}}.

\bibitem{Khachatryan:2016bia}
\hrefCMSnoop {}{{CMS Collaboration}, ``{The CMS trigger system}'',} \textit{
  JINST} \textbf{ 12} (2017) P01020,
  \href{http://dx.doi.org/10.1088/1748-0221/12/01/P01020}{\doi{10.1088/1748-0221/12/01/P01020}},
\href{http://www.arXiv.org/abs/1609.02366}{\texttt{arXiv:1609.02366}}.

\bibitem{Chatrchyan:2008zzk}
\hrefCMSnoop {}{{CMS Collaboration}, ``The {CMS} experiment at the {CERN}
  {LHC}'',} \textit{ JINST} \textbf{ 3} (2008) S08004,
  \href{http://dx.doi.org/10.1088/1748-0221/3/08/S08004}{\doi{10.1088/1748-0221/3/08/S08004}}.

\bibitem{CMS-LUM-17-003}
\hrefCMSnoop {}{{CMS Collaboration}, ``{Precision luminosity measurement in
  proton-proton collisions at $\sqrt{s} = 13\TeV$ in 2015 and 2016 at CMS}'',}
  \textit{ Eur. Phys. J. C} \textbf{ 81} (2021) 800,
  \href{http://dx.doi.org/10.1140/epjc/s10052-021-09538-2}{\doi{10.1140/epjc/s10052-021-09538-2}},
  \href{http://www.arXiv.org/abs/2104.01927}{\texttt{arXiv:2104.01927}}.

\bibitem{CMS-PAS-LUM-17-004}
\href {https://cds.cern.ch/record/2621960/}{{CMS Collaboration}, ``{CMS
  luminosity measurement for the 2017 data-taking period at $\sqrt{s} =
  13\TeV$}'',} CMS Physics Analysis Summary CMS-PAS-LUM-17-004, 2018.

\bibitem{CMS-PAS-LUM-18-002}
\href {https://cds.cern.ch/record/2676164/}{{CMS Collaboration}, ``{CMS
  luminosity measurement for the 2018 data-taking period at $\sqrt{s} =
  13\TeV$}'',} CMS Physics Analysis Summary CMS-PAS-LUM-18-002, 2019.

\bibitem{Alwall:2014hca}
J.~Alwall\hrefCMSnoop {}{ {et~al.}, ``The automated computation of tree-level
  and next-to-leading order differential cross sections, and their matching to
  parton shower simulations'',} \textit{ JHEP} \textbf{ 07} (2014) 079,
  \href{http://dx.doi.org/10.1007/JHEP07(2014)079}{\doi{10.1007/JHEP07(2014)079}},
\href{http://www.arXiv.org/abs/1405.0301}{\texttt{arXiv:1405.0301}}.

\bibitem{Kalogeropoulos:2018cke}
\hrefCMSnoop {}{A.~Kalogeropoulos and J.~Alwall, ``{The SysCalc code: A tool to
  derive theoretical systematic uncertainties}'',} 1, 2018.
  \href{http://www.arXiv.org/abs/1801.08401}{\texttt{arXiv:1801.08401}}.

\bibitem{Nason:2004rx}
\hrefCMSnoop {}{P.~Nason, ``A new method for combining {NLO} {QCD} with shower
  {Monte} {Carlo} algorithms'',} \textit{ JHEP} \textbf{ 11} (2004) 040,
  \href{http://dx.doi.org/10.1088/1126-6708/2004/11/040}{\doi{10.1088/1126-6708/2004/11/040}},
\href{http://www.arXiv.org/abs/hep-ph/0409146}{\texttt{arXiv:hep-ph/0409146}}.

\bibitem{Frixione:2007vw}
\hrefCMSnoop {}{S.~Frixione, P.~Nason, and C.~Oleari, ``Matching {NLO} {QCD}
  computations with parton shower simulations: The powheg method'',} \textit{
  JHEP} \textbf{ 11} (2007) 070,
  \href{http://dx.doi.org/10.1088/1126-6708/2007/11/070}{\doi{10.1088/1126-6708/2007/11/070}},
\href{http://www.arXiv.org/abs/0709.2092}{\texttt{arXiv:0709.2092}}.

\bibitem{Alioli:2010xd}
\hrefCMSnoop {}{S.~Alioli, P.~Nason, C.~Oleari, and E.~Re, ``A general
  framework for implementing {NLO} calculations in shower {Monte} {Carlo}
  programs: The {POWHEG} {BOX}'',} \textit{ JHEP} \textbf{ 06} (2010) 043,
  \href{http://dx.doi.org/10.1007/JHEP06(2010)043}{\doi{10.1007/JHEP06(2010)043}},
\href{http://www.arXiv.org/abs/1002.2581}{\texttt{arXiv:1002.2581}}.

\bibitem{MCFM1}
\hrefCMSnoop {}{J.~M. Campbell and R.~K. Ellis, ``{An Update on vector boson
  pair production at hadron colliders}'',} \textit{ Phys. Rev. D} \textbf{ 60}
  (1999) 113006,
  \href{http://dx.doi.org/10.1103/PhysRevD.60.113006}{\doi{10.1103/PhysRevD.60.113006}},
  \href{http://www.arXiv.org/abs/hep-ph/9905386}{\texttt{arXiv:hep-ph/9905386}}.

\bibitem{MCFM2}
\hrefCMSnoop {}{J.~M. Campbell, R.~K. Ellis, and C.~Williams, ``{Vector boson
  pair production at the LHC}'',} \textit{ JHEP} \textbf{ 07} (2011) 018,
  \href{http://dx.doi.org/10.1007/JHEP07(2011)018}{\doi{10.1007/JHEP07(2011)018}},
  \href{http://www.arXiv.org/abs/1105.0020}{\texttt{arXiv:1105.0020}}.

\bibitem{MCFM3}
\hrefCMSnoop {}{J.~M. Campbell, R.~K. Ellis, and W.~T. Giele, ``{A
  Multi-Threaded Version of MCFM}'',} \textit{ Eur. Phys. J. C} \textbf{ 75}
  (2015) 246,
  \href{http://dx.doi.org/10.1140/epjc/s10052-015-3461-2}{\doi{10.1140/epjc/s10052-015-3461-2}},
  \href{http://www.arXiv.org/abs/1503.06182}{\texttt{arXiv:1503.06182}}.

\bibitem{Sjostrand:2014zea}
T.~Sj{\"o}strand\hrefCMSnoop {}{ {et~al.}, ``An introduction to {PYTHIA}
  $8.2$'',} \textit{ Comput. Phys. Commun.} \textbf{ 191} (2015) 159,
  \href{http://dx.doi.org/10.1016/j.cpc.2015.01.024}{\doi{10.1016/j.cpc.2015.01.024}},
\href{http://www.arXiv.org/abs/1410.3012}{\texttt{arXiv:1410.3012}}.

\bibitem{Hespel:2014sla}
\hrefCMSnoop {}{B.~Hespel, D.~Lopez-Val, and E.~Vryonidou, ``{Higgs pair
  production via gluon fusion in the two-Higgs-doublet model}'',} \textit{
  JHEP} \textbf{ 09} (2014) 124,
  \href{http://dx.doi.org/10.1007/JHEP09(2014)124}{\doi{10.1007/JHEP09(2014)124}},
  \href{http://www.arXiv.org/abs/1407.0281}{\texttt{arXiv:1407.0281}}.

\bibitem{Oliveira:2014kla}
\hrefCMSnoop {}{A.~Carvalho, ``{Gravity particles from warped extra dimensions,
  predictions for LHC}'',} 3, 2014.
  \href{http://www.arXiv.org/abs/1404.0102}{\texttt{arXiv:1404.0102}}.

\bibitem{POWHEGHH1}
G.~Heinrich\hrefCMSnoop {}{ {et~al.}, ``{NLO predictions for Higgs boson pair
  production with full top quark mass dependence matched to parton showers}'',}
  \textit{ JHEP} \textbf{ 08} (2017) 088,
  \href{http://dx.doi.org/10.1007/JHEP08(2017)088}{\doi{10.1007/JHEP08(2017)088}},
  \href{http://www.arXiv.org/abs/1703.09252}{\texttt{arXiv:1703.09252}}.

\bibitem{POWHEGHH2}
G.~Heinrich\hrefCMSnoop {}{ {et~al.}, ``{Probing the trilinear Higgs boson
  coupling in di-Higgs production at NLO QCD including parton shower
  effects}'',} \textit{ JHEP} \textbf{ 06} (2019) 066,
  \href{http://dx.doi.org/10.1007/JHEP06(2019)066}{\doi{10.1007/JHEP06(2019)066}},
  \href{http://www.arXiv.org/abs/1903.08137}{\texttt{arXiv:1903.08137}}.

\bibitem{POWHEGHH3}
\hrefCMSnoop {}{G.~Heinrich, S.~P. Jones, M.~Kerner, and L.~Scyboz, ``{A
  non-linear EFT description of $\Pg\Pg \to \PH\PH$ at NLO interfaced to
  POWHEG}'',} \textit{ JHEP} \textbf{ 10} (2020) 021,
  \href{http://dx.doi.org/10.1007/JHEP10(2020)021}{\doi{10.1007/JHEP10(2020)021}},
  \href{http://www.arXiv.org/abs/2006.16877}{\texttt{arXiv:2006.16877}}.

\bibitem{POWHEGGGFH}
\hrefCMSnoop {}{E.~Bagnaschi, G.~Degrassi, P.~Slavich, and A.~Vicini, ``{Higgs
  production via gluon fusion in the POWHEG approach in the SM and in the
  MSSM}'',} \textit{ JHEP} \textbf{ 02} (2012) 088,
  \href{http://dx.doi.org/10.1007/JHEP02(2012)088}{\doi{10.1007/JHEP02(2012)088}},
  \href{http://www.arXiv.org/abs/1111.2854}{\texttt{arXiv:1111.2854}}.

\bibitem{POWHEGVBFH}
\hrefCMSnoop {}{P.~Nason and C.~Oleari, ``{NLO Higgs boson production via
  vector-boson fusion matched with shower in POWHEG}'',} \textit{ JHEP}
  \textbf{ 02} (2010) 037,
  \href{http://dx.doi.org/10.1007/JHEP02(2010)037}{\doi{10.1007/JHEP02(2010)037}},
  \href{http://www.arXiv.org/abs/0911.5299}{\texttt{arXiv:0911.5299}}.

\bibitem{POWHEGVH}
\hrefCMSnoop {}{G.~Luisoni, P.~Nason, C.~Oleari, and F.~Tramontano,
  ``{$\PH\PW^{\pm}$/$\PH\PZ$ + $0$ and $1$ jet at NLO with the POWHEG BOX
  interfaced to GoSam and their merging within MiNLO}'',} \textit{ JHEP}
  \textbf{ 10} (2013) 083,
  \href{http://dx.doi.org/10.1007/JHEP10(2013)083}{\doi{10.1007/JHEP10(2013)083}},
  \href{http://www.arXiv.org/abs/1306.2542}{\texttt{arXiv:1306.2542}}.

\bibitem{POWHEGVV1}
\hrefCMSnoop {}{T.~Melia, P.~Nason, R.~Rontsch, and G.~Zanderighi,
  ``{$\PW^{+}\PW^{-}$, $\PW\PZ$ and $\PZ\PZ$ production in the POWHEG BOX}'',}
  \textit{ JHEP} \textbf{ 11} (2011) 078,
  \href{http://dx.doi.org/10.1007/JHEP11(2011)078}{\doi{10.1007/JHEP11(2011)078}},
  \href{http://www.arXiv.org/abs/1107.5051}{\texttt{arXiv:1107.5051}}.

\bibitem{POWHEGVV2}
\hrefCMSnoop {}{P.~Nason and G.~Zanderighi, ``{$\PW^{+}\PW^{-}$, $\PW\PZ$ and
  $\PZ\PZ$ production in the POWHEG-BOX-V2}'',} \textit{ Eur. Phys. J. C}
  \textbf{ 74} (2014) 2702,
  \href{http://dx.doi.org/10.1140/epjc/s10052-013-2702-5}{\doi{10.1140/epjc/s10052-013-2702-5}},
  \href{http://www.arXiv.org/abs/1311.1365}{\texttt{arXiv:1311.1365}}.

\bibitem{MCFMGGZZ}
\hrefCMSnoop {}{J.~M. Campbell, R.~K. Ellis, and C.~Williams, ``{Bounding the
  Higgs Width at the LHC Using Full Analytic Results for $gg \to e^- e^+ \mu^-
  \mu^+$}'',} \textit{ JHEP} \textbf{ 04} (2014) 060,
  \href{http://dx.doi.org/10.1007/JHEP04(2014)060}{\doi{10.1007/JHEP04(2014)060}},
  \href{http://www.arXiv.org/abs/1311.3589}{\texttt{arXiv:1311.3589}}.

\bibitem{POWHEGST2}
\hrefCMSnoop {}{R.~Frederix, E.~Re, and P.~Torrielli, ``{Single-top $t$-channel
  hadroproduction in the four-flavour scheme with POWHEG and aMC@NLO}'',}
  \textit{ JHEP} \textbf{ 09} (2012) 130,
  \href{http://dx.doi.org/10.1007/JHEP09(2012)130}{\doi{10.1007/JHEP09(2012)130}},
  \href{http://www.arXiv.org/abs/1207.5391}{\texttt{arXiv:1207.5391}}.

\bibitem{POWHEGST1}
\hrefCMSnoop {}{E.~Re, ``{Single-top $Wt$ production matched with parton
  showers using the POWHEG method}'',} \textit{ Eur. Phys. J. C} \textbf{ 71}
  (2011) 1547,
  \href{http://dx.doi.org/10.1140/epjc/s10052-011-1547-z}{\doi{10.1140/epjc/s10052-011-1547-z}},
  \href{http://www.arXiv.org/abs/1009.2450}{\texttt{arXiv:1009.2450}}.

\bibitem{POWHEGTTBAR}
\hrefCMSnoop {}{S.~Frixione, P.~Nason, and G.~Ridolfi, ``{A positive-weight
  next-to-leading-order Monte Carlo for heavy flavour hadroproduction}'',}
  \textit{ JHEP} \textbf{ 09} (2007) 126,
  \href{http://dx.doi.org/10.1088/1126-6708/2007/09/126}{\doi{10.1088/1126-6708/2007/09/126}},
  \href{http://www.arXiv.org/abs/0707.3088}{\texttt{arXiv:0707.3088}}.

\bibitem{NNPDF:2014otw}
\hrefCMSnoop {}{{NNPDF} Collaboration, ``{Parton distributions for the LHC Run
  II}'',} \textit{ JHEP} \textbf{ 04} (2015) 040,
  \href{http://dx.doi.org/10.1007/JHEP04(2015)040}{\doi{10.1007/JHEP04(2015)040}},
  \href{http://www.arXiv.org/abs/1410.8849}{\texttt{arXiv:1410.8849}}.

\bibitem{NNPDF:2017mvq}
\hrefCMSnoop {}{{NNPDF} Collaboration, ``{Parton distributions from
  high-precision collider data}'',} \textit{ Eur. Phys. J. C} \textbf{ 77}
  (2017) 663,
  \href{http://dx.doi.org/10.1140/epjc/s10052-017-5199-5}{\doi{10.1140/epjc/s10052-017-5199-5}},
  \href{http://www.arXiv.org/abs/1706.00428}{\texttt{arXiv:1706.00428}}.

\bibitem{Butterworth:2015oua}
\hrefCMSnoop {}{J.~Butterworth {et~al.}, ``{PDF4LHC recommendations for {LHC}
  Run II}'',} \textit{ J. Phys. G} \textbf{ 43} (2016) 023001,
  \href{http://dx.doi.org/10.1088/0954-3899/43/2/023001}{\doi{10.1088/0954-3899/43/2/023001}},
\href{http://www.arXiv.org/abs/1510.03865}{\texttt{arXiv:1510.03865}}.

\bibitem{Rojo:2015acz}
\hrefCMSnoop {}{J.~Rojo {et~al.}, ``{The PDF4LHC report on PDFs and LHC data:
  Results from Run I and preparation for Run II}'',} \textit{ J. Phys. G}
  \textbf{ 42} (2015) 103103,
  \href{http://dx.doi.org/10.1088/0954-3899/42/10/103103}{\doi{10.1088/0954-3899/42/10/103103}},
  \href{http://www.arXiv.org/abs/1507.00556}{\texttt{arXiv:1507.00556}}.

\bibitem{Accardi:2016ndt}
A.~Accardi\hrefCMSnoop {}{ {et~al.}, ``{A critical appraisal and evaluation of
  modern PDFs}'',} \textit{ Eur. Phys. J. C} \textbf{ 76} (2016) 471,
  \href{http://dx.doi.org/10.1140/epjc/s10052-016-4285-4}{\doi{10.1140/epjc/s10052-016-4285-4}},
  \href{http://www.arXiv.org/abs/1603.08906}{\texttt{arXiv:1603.08906}}.

\bibitem{PYTHIA_CUETP8M1tune_CMS}
\hrefCMSnoop {}{{CMS Collaboration}, ``{Event generator tunes obtained from
  underlying event and multiparton scattering measurements}'',} \textit{ Eur.
  Phys. J. C} \textbf{ 76} (2016) 155,
  \href{http://dx.doi.org/10.1140/epjc/s10052-016-3988-x}{\doi{10.1140/epjc/s10052-016-3988-x}},
  \href{http://www.arXiv.org/abs/1512.00815}{\texttt{arXiv:1512.00815}}.

\bibitem{Sirunyan:2019dfx}
\hrefCMSnoop {}{{CMS Collaboration}, ``Extraction and validation of a new set
  of {CMS} {PYTHIA}$8$ tunes from underlying-event measurements'',} \textit{
  Eur. Phys. J. C.} \textbf{ 80} (2020) 4,
  \href{http://dx.doi.org/10.1140/epjc/s10052-019-7499-4}{\doi{10.1140/epjc/s10052-019-7499-4}},
  \href{http://www.arXiv.org/abs/1903.12179}{\texttt{arXiv:1903.12179}}.

\bibitem{PYTHIA_MonashTune}
\hrefCMSnoop {}{P.~Skands, S.~Carrazza, and J.~Rojo, ``{Tuning PYTHIA $8.1$:
  The Monash 2013 tune}'',} \textit{ Eur. Phys. J. C} \textbf{ 74} (2014) 3024,
  \href{http://dx.doi.org/10.1140/epjc/s10052-014-3024-y}{\doi{10.1140/epjc/s10052-014-3024-y}},
\href{http://www.arXiv.org/abs/1404.5630}{\texttt{arXiv:1404.5630}}.

\bibitem{Alwall:2007fs}
J.~Alwall\hrefCMSnoop {}{ {et~al.}, ``Comparative study of various algorithms
  for the merging of parton showers and matrix elements in hadronic
  collisions'',} \textit{ Eur. Phys. J. C} \textbf{ 53} (2008) 473,
  \href{http://dx.doi.org/10.1140/epjc/s10052-007-0490-5}{\doi{10.1140/epjc/s10052-007-0490-5}},
\href{http://www.arXiv.org/abs/0706.2569}{\texttt{arXiv:0706.2569}}.

\bibitem{Frederix:2012ps}
\hrefCMSnoop {}{R.~Frederix and S.~Frixione, ``Merging meets matching in
  {MC@NLO}'',} \textit{ JHEP} \textbf{ 12} (2012) 061,
  \href{http://dx.doi.org/10.1007/JHEP12(2012)061}{\doi{10.1007/JHEP12(2012)061}},
\href{http://www.arXiv.org/abs/1209.6215}{\texttt{arXiv:1209.6215}}.

\bibitem{Agostinelli:2002hh}
\hrefCMSnoop {}{{GEANT4} Collaboration, ``{\GEANTfour}: A simulation
  toolkit'',} \textit{ Nucl. Instrum. Meth. A} \textbf{ 506} (2003) 250,
\href{http://dx.doi.org/10.1016/S0168-9002(03)01368-8}{\doi{10.1016/S0168-9002(03)01368-8}}.

\bibitem{Capozi:2019xsi}
\hrefCMSnoop {}{M.~Capozi and G.~Heinrich, ``{Exploring anomalous couplings in
  Higgs boson pair production through shape analysis}'',} \textit{ JHEP}
  \textbf{ 03} (2020) 091,
  \href{http://dx.doi.org/10.1007/JHEP03(2020)091}{\doi{10.1007/JHEP03(2020)091}},
  \href{http://www.arXiv.org/abs/1908.08923}{\texttt{arXiv:1908.08923}}.

\bibitem{Buchalla:2018yce}
G.~Buchalla\hrefCMSnoop {}{ {et~al.}, ``{Higgs boson pair production in
  non-linear effective field theory with full $m_{\Ptop}$-dependence at NLO
  QCD}'',} \textit{ JHEP} \textbf{ 09} (2018) 057,
  \href{http://dx.doi.org/10.1007/JHEP09(2018)057}{\doi{10.1007/JHEP09(2018)057}},
  \href{http://www.arXiv.org/abs/1806.05162}{\texttt{arXiv:1806.05162}}.

\bibitem{Carvalho:2017vnu}
A.~Carvalho\hrefCMSnoop {}{ {et~al.}, ``{On the reinterpretation of
  non-resonant searches for Higgs boson pairs}'',} \textit{ JHEP} \textbf{ 02}
  (2021) 049,
  \href{http://dx.doi.org/10.1007/JHEP02(2021)049}{\doi{10.1007/JHEP02(2021)049}},
  \href{http://www.arXiv.org/abs/1710.08261}{\texttt{arXiv:1710.08261}}.

\bibitem{Sirunyan:2017ulk}
\hrefCMSnoop {}{{CMS Collaboration}, ``Particle-flow reconstruction and global
  event description with the {CMS} detector'',} \textit{ JINST} \textbf{ 12}
  (2017) P10003,
  \href{http://dx.doi.org/10.1088/1748-0221/12/10/P10003}{\doi{10.1088/1748-0221/12/10/P10003}},
\href{http://www.arXiv.org/abs/1706.04965}{\texttt{arXiv:1706.04965}}.

\bibitem{Cacciari:2008gp}
\hrefCMSnoop {}{M.~Cacciari, G.~P. Salam, and G.~Soyez, ``{The anti-$\kt$ jet
  clustering algorithm}'',} \textit{ JHEP} \textbf{ 04} (2008) 063,
  \href{http://dx.doi.org/10.1088/1126-6708/2008/04/063}{\doi{10.1088/1126-6708/2008/04/063}},
\href{http://www.arXiv.org/abs/0802.1189}{\texttt{arXiv:0802.1189}}.

\bibitem{Cacciari:2011ma}
\hrefCMSnoop {}{M.~Cacciari, G.~P. Salam, and G.~Soyez, ``{\FASTJET user
  manual}'',} \textit{ Eur. Phys. J. C} \textbf{ 72} (2012) 1896,
  \href{http://dx.doi.org/10.1140/epjc/s10052-012-1896-2}{\doi{10.1140/epjc/s10052-012-1896-2}},
\href{http://www.arXiv.org/abs/1111.6097}{\texttt{arXiv:1111.6097}}.

\bibitem{CMS:2020uim}
\hrefCMSnoop {}{{CMS Collaboration}, ``{Electron and photon reconstruction and
  identification with the CMS experiment at the CERN LHC}'',} \textit{ JINST}
  \textbf{ 16} (2021) P05014,
  \href{http://dx.doi.org/10.1088/1748-0221/16/05/P05014}{\doi{10.1088/1748-0221/16/05/P05014}},
  \href{http://www.arXiv.org/abs/2012.06888}{\texttt{arXiv:2012.06888}}.

\bibitem{Sirunyan:2020icl}
\hrefCMSnoop {}{{CMS Collaboration}, ``{Measurement of the Higgs boson
  production rate in association with top quarks in final states with
  electrons, muons, and hadronically decaying tau leptons at $\sqrt{s} =
  13\TeV$}'',} \textit{ Eur. Phys. J. C} \textbf{ 81} (2021) 378,
  \href{http://dx.doi.org/10.1140/epjc/s10052-021-09014-x}{\doi{10.1140/epjc/s10052-021-09014-x}},
  \href{http://www.arXiv.org/abs/2011.03652}{\texttt{arXiv:2011.03652}}.

\bibitem{Khachatryan:2015hwa}
\hrefCMSnoop {}{{CMS Collaboration}, ``{Performance of electron reconstruction
  and selection with the {CMS} detector in $\Pp\Pp$ collisions at $\sqrt{s} =
  8\TeV$}'',} \textit{ JINST} \textbf{ 10} (2015) P06005,
  \href{http://dx.doi.org/10.1088/1748-0221/10/06/P06005}{\doi{10.1088/1748-0221/10/06/P06005}},
\href{http://www.arXiv.org/abs/1502.02701}{\texttt{arXiv:1502.02701}}.

\bibitem{Sirunyan:2018}
\hrefCMSnoop {}{{CMS Collaboration}, ``{Performance of the CMS muon detector
  and muon reconstruction with proton-proton collisions at $\sqrt{s} =
  13\TeV$}'',} \textit{ JINST} \textbf{ 13} (2018) P06015,
  \href{http://dx.doi.org/10.1088/1748-0221/13/06/P06015}{\doi{10.1088/1748-0221/13/06/P06015}},
\href{http://www.arXiv.org/abs/1804.04528}{\texttt{arXiv:1804.04528}}.

\bibitem{Sirunyan:2018pgf}
\hrefCMSnoop {}{{CMS Collaboration}, ``Performance of reconstruction and
  identification of {$\PGt$} leptons decaying to hadrons and {$\Pnu_{\PGt}$} in
  {$\Pp\Pp$} collisions at {$\sqrt{s} = 13\TeV$}'',} \textit{ JINST} \textbf{
  13} (2018) P10005,
  \href{http://dx.doi.org/10.1088/1748-0221/13/10/P10005}{\doi{10.1088/1748-0221/13/10/P10005}},
\href{http://www.arXiv.org/abs/1809.02816}{\texttt{arXiv:1809.02816}}.

\bibitem{CMS:2022prd}
\hrefCMSnoop {}{{CMS Collaboration}, ``{Identification of hadronic tau lepton
  decays using a deep neural network}'',} \textit{ JINST} \textbf{ 17} (2022)
  P07023,
  \href{http://dx.doi.org/10.1088/1748-0221/17/07/P07023}{\doi{10.1088/1748-0221/17/07/P07023}},
  \href{http://www.arXiv.org/abs/2201.08458}{\texttt{arXiv:2201.08458}}.

\bibitem{lecun1989}
\hrefCMSnoop {}{Y.~LeCun, ``Generalization and network design strategies'',} in
  \textit{ Connectionism in perspective}, R.~Pfeifer, Z.~Schreter, F.~Fogelman,
  and L.~Steels, eds.
\newblock Elsevier, Zurich, Switzerland, 1989.
\newblock {An extended version was published as a technical report of the
  University of Toronto}.

\bibitem{Sirunyan:2020foa}
\hrefCMSnoop {}{{CMS Collaboration}, ``{Pileup mitigation at CMS in $13\TeV$
  data}'',} \textit{ JINST} \textbf{ 15} (2020) P09018,
  \href{http://dx.doi.org/10.1088/1748-0221/15/09/p09018}{\doi{10.1088/1748-0221/15/09/p09018}},
  \href{http://www.arXiv.org/abs/2003.00503}{\texttt{arXiv:2003.00503}}.

\bibitem{Bertolini:2014bba}
\hrefCMSnoop {}{D.~Bertolini, P.~Harris, M.~Low, and N.~Tran, ``{Pileup per
  particle identification}'',} \textit{ JHEP} \textbf{ 10} (2014) 059,
  \href{http://dx.doi.org/10.1007/JHEP10(2014)059}{\doi{10.1007/JHEP10(2014)059}},
\href{http://www.arXiv.org/abs/1407.6013}{\texttt{arXiv:1407.6013}}.

\bibitem{Khachatryan:2016kdb}
\hrefCMSnoop {}{{CMS Collaboration}, ``{Jet energy scale and resolution in the
  CMS experiment in $\Pp\Pp$ collisions at $8\TeV$}'',} \textit{ JINST}
  \textbf{ 12} (2017) P02014,
  \href{http://dx.doi.org/10.1088/1748-0221/12/02/P02014}{\doi{10.1088/1748-0221/12/02/P02014}},
\href{http://www.arXiv.org/abs/1607.03663}{\texttt{arXiv:1607.03663}}.

\bibitem{Sirunyan:2019quj}
\hrefCMSnoop {}{{CMS Collaboration}, ``{Search for resonances decaying to a
  pair of Higgs bosons in the $\Pbottom\APbottom\Pquark\APquark'\Plepton\Pnu$
  final state in proton-proton collisions at $\sqrt{s} = 13\TeV$}'',} \textit{
  JHEP} \textbf{ 10} (2019) 125,
  \href{http://dx.doi.org/10.1007/JHEP10(2019)125}{\doi{10.1007/JHEP10(2019)125}},
  \href{http://www.arXiv.org/abs/1904.04193}{\texttt{arXiv:1904.04193}}.

\bibitem{Bols_2020}
E.~Bols\hrefCMSnoop {}{ {et~al.}, ``Jet flavour classification using
  {DeepJet}'',} \textit{ JINST} \textbf{ 15} (2020) P12012,
  \href{http://dx.doi.org/10.1088/1748-0221/15/12/P12012}{\doi{10.1088/1748-0221/15/12/P12012}},
  \href{http://www.arXiv.org/abs/2008.10519}{\texttt{arXiv:2008.10519}}.

\bibitem{Sirunyan:2019kia}
\hrefCMSnoop {}{{CMS Collaboration}, ``{Performance of missing transverse
  momentum reconstruction in proton-proton collisions at $\sqrt{s} = 13\TeV$
  using the CMS detector}'',} \textit{ JINST} \textbf{ 14} (2019) P07004,
  \href{http://dx.doi.org/10.1088/1748-0221/14/07/P07004}{\doi{10.1088/1748-0221/14/07/P07004}},
\href{http://www.arXiv.org/abs/1903.06078}{\texttt{arXiv:1903.06078}}.

\bibitem{Sirunyan:2018shy}
\hrefCMSnoop {}{{CMS Collaboration}, ``Evidence for associated production of a
  {Higgs} boson with a top quark pair in final states with electrons, muons,
  and hadronically decaying {$\Pgt$} leptons at {$\sqrt{s} = 13\TeV$}'',}
  \textit{ JHEP} \textbf{ 08} (2018) 066,
  \href{http://dx.doi.org/10.1007/JHEP08(2018)066}{\doi{10.1007/JHEP08(2018)066}},
\href{http://www.arXiv.org/abs/1803.05485}{\texttt{arXiv:1803.05485}}.

\bibitem{Breiman:1984jka}
L.~Breiman, J.~Friedman, R.~A. Olshen, and C.~J. Stone, ``{Classification and
  regression trees}''.
\newblock Chapman and Hall/CRC, 1984.
\newblock ISBN~978-0-412-04841-8.

\bibitem{Ehataht:2018nql}
\hrefCMSnoop {}{K.~Ehat{\"{a}}ht, L.~Marzola, and C.~Veelken, ``{Reconstruction
  of the mass of Higgs boson pairs in events with Higgs boson pairs decaying
  into four $\PGt$ leptons}'',} 9, 2018.
  \href{http://www.arXiv.org/abs/1809.06140}{\texttt{arXiv:1809.06140}}.

\bibitem{Chen:2016btl}
\hrefCMSnoop {}{T.~Chen and C.~Guestrin, ``{\textsc{XGBoost}: A scalable tree
  boosting system}'',} in \textit{ Proceedings of the 22nd ACM SIGKDD
  International Conference on Knowledge Discovery and Data Mining}, p.~785.
\newblock ACM, New York, NY, USA, 2016.
\newblock
  \href{http://www.arXiv.org/abs/1603.02754}{\texttt{arXiv:1603.02754}}.
\newblock
  \href{http://dx.doi.org/10.1145/2939672.2939785}{\doi{10.1145/2939672.2939785}}.

\bibitem{scikit-learn}
\href {http://jmlr.org/papers/v12/pedregosa11a.html}{F.~Pedregosa {et~al.},
  ``{\textsc{Scikit-learn}: Machine learning in Python}'',} \textit{ J. Mach.
  Learn. Res.} \textbf{ 12} (2011) 2825,
  \href{http://www.arXiv.org/abs/1201.0490}{\texttt{arXiv:1201.0490}}.

\bibitem{Tani:2020dyi}
\hrefCMSnoop {}{L.~Tani, D.~Rand, C.~Veelken, and M.~Kadastik, ``{Evolutionary
  algorithms for hyperparameter optimization in machine learning for
  application in high energy physics}'',} \textit{ Eur. Phys. J. C} \textbf{
  81} (2021) 170,
  \href{http://dx.doi.org/10.1140/epjc/s10052-021-08950-y}{\doi{10.1140/epjc/s10052-021-08950-y}},
  \href{http://www.arXiv.org/abs/2011.04434}{\texttt{arXiv:2011.04434}}.

\bibitem{CMS:2017moi}
\hrefCMSnoop {}{{CMS Collaboration}, ``{Search for electroweak production of
  charginos and neutralinos in multilepton final states in proton-proton
  collisions at $\sqrt{s} = 13\TeV$}'',} \textit{ JHEP} \textbf{ 03} (2018)
  166,
  \href{http://dx.doi.org/10.1007/JHEP03(2018)166}{\doi{10.1007/JHEP03(2018)166}},
  \href{http://www.arXiv.org/abs/1709.05406}{\texttt{arXiv:1709.05406}}.

\bibitem{Baglio:2020wgt}
J.~Baglio\hrefCMSnoop {}{ {et~al.}, ``{$gg\to HH$: Combined uncertainties}'',}
  \textit{ Phys. Rev. D} \textbf{ 103} (2021) 056002,
  \href{http://dx.doi.org/10.1103/PhysRevD.103.056002}{\doi{10.1103/PhysRevD.103.056002}},
  \href{http://www.arXiv.org/abs/2008.11626}{\texttt{arXiv:2008.11626}}.

\bibitem{LHCHiggsCrossSectionWorkingGroup:2016ypw}
\hrefCMSnoop {}{{LHC Higgs Cross Section Working Group}, ``{Handbook of LHC
  Higgs Cross Sections: 4. Deciphering the nature of the Higgs sector}'',}
  \textit{ CERN Yellow Rep. Monogr.} \textbf{ 2} (2017)
  \href{http://dx.doi.org/10.23731/CYRM-2017-002}{\doi{10.23731/CYRM-2017-002}},
  \href{http://www.arXiv.org/abs/1610.07922}{\texttt{arXiv:1610.07922}}.

\bibitem{GRAZZINI2016179}
\hrefCMSnoop {}{M.~Grazzini, S.~Kallweit, D.~Rathlev, and M.~Wiesemann,
  ``{$\PW^{\pm}\PZ$ production at hadron colliders in NNLO QCD}'',} \textit{
  Phys. Lett. B} \textbf{ 761} (2016) 179,
  \href{http://dx.doi.org/10.1016/j.physletb.2016.08.017}{\doi{10.1016/j.physletb.2016.08.017}},
  \href{http://www.arXiv.org/abs/1604.08576}{\texttt{arXiv:1604.08576}}.

\bibitem{CASCIOLI2014311}
F.~Cascioli\hrefCMSnoop {}{ {et~al.}, ``{$\PZ\PZ$ production at hadron
  colliders in NNLO QCD}'',} \textit{ Phys. Lett. B} \textbf{ 735} (2014) 311,
  \href{http://dx.doi.org/10.1016/j.physletb.2014.06.056}{\doi{10.1016/j.physletb.2014.06.056}},
  \href{http://www.arXiv.org/abs/1405.2219}{\texttt{arXiv:1405.2219}}.

\bibitem{PhysRevD.92.094028}
\hrefCMSnoop {}{{Caola, F. and Melnikov, K. and Rontsch, R. and Tancredi, L.},
  ``{QCD corrections to $\PZ\PZ$ production in gluon fusion at the LHC}'',}
  \textit{ Phys. Rev. D} \textbf{ 92} (2015) 094028,
  \href{http://dx.doi.org/10.1103/PhysRevD.92.094028}{\doi{10.1103/PhysRevD.92.094028}},
  \href{http://www.arXiv.org/abs/1509.06734}{\texttt{arXiv:1509.06734}}.

\bibitem{CMS:2017wyc}
\href {http://cds.cern.ch/record/2256875}{{CMS Collaboration}, ``{Jet
  algorithms performance in $13\TeV$ data}'',} {CMS} Physics Analysis Summary
  CMS-PAS-JME-16-003, 2017.

\bibitem{CMS:2018mlc}
\hrefCMSnoop {}{{CMS Collaboration}, ``{Measurement of the inelastic
  proton-proton cross section at $\sqrt{s} = 13\TeV$}'',} \textit{ JHEP}
  \textbf{ 07} (2018) 161,
  \href{http://dx.doi.org/10.1007/JHEP07(2018)161}{\doi{10.1007/JHEP07(2018)161}},
  \href{http://www.arXiv.org/abs/1802.02613}{\texttt{arXiv:1802.02613}}.

\bibitem{Cowan:2010js}
\hrefCMSnoop {}{G.~Cowan, K.~Cranmer, E.~Gross, and O.~Vitells, ``{Asymptotic
  formulae for likelihood-based tests of new physics}'',} \textit{ Eur. Phys.
  J. C} \textbf{ 71} (2011) 1554,
  \href{http://dx.doi.org/10.1140/epjc/s10052-011-1554-0}{\doi{10.1140/epjc/s10052-011-1554-0}},
  \href{http://www.arXiv.org/abs/1007.1727}{\texttt{arXiv:1007.1727}}.
  [Erratum: \DOI{10.1140/epjc/s10052-013-2501-z}].

\bibitem{ATL-PHYS-PUB-2011-011}
\href {https://cds.cern.ch/record/1379837}{{{ATLAS} and {CMS} Collaborations,
  and LHC Higgs Combination Group}, ``{Procedure for the {LHC} {Higgs} boson
  search combination in summer 2011}'',} technical report, 2011.

\bibitem{Conway:2011in}
\hrefCMSnoop {}{J.~S. Conway, ``{Incorporating nuisance parameters in
  likelihoods for multisource spectra}'',} in \textit{ {PHYSTAT 2011}}, p.~115.
\newblock 2011.
\newblock \href{http://www.arXiv.org/abs/1103.0354}{\texttt{arXiv:1103.0354}}.
\newblock
  \href{http://dx.doi.org/10.5170/CERN-2011-006.115}{\doi{10.5170/CERN-2011-006.115}}.

\bibitem{Junk:1999kv}
\hrefCMSnoop {}{T.~Junk, ``{Confidence level computation for combining searches
  with small statistics}'',} \textit{ Nucl. Instrum. Meth. A} \textbf{ 434}
  (1999) 435,
  \href{http://dx.doi.org/10.1016/S0168-9002(99)00498-2}{\doi{10.1016/S0168-9002(99)00498-2}},
\href{http://www.arXiv.org/abs/hep-ex/9902006}{\texttt{arXiv:hep-ex/9902006}}.

\bibitem{Read:2002hq}
\hrefCMSnoop {}{A.~L. Read, ``{Presentation of search results: The
  $\textrm{CL}_{s}$ technique}'',} \textit{ J. Phys. G} \textbf{ 28} (2002)
  2693,
\href{http://dx.doi.org/10.1088/0954-3899/28/10/313}{\doi{10.1088/0954-3899/28/10/313}}.

\bibitem{hepdata}
\hrefCMSnoop {}{}{HEPD}ata record for this analysis, 2022.
\newblock
  \href{http://dx.doi.org/10.17182/hepdata.130795}{\doi{10.17182/hepdata.130795}}.

\bibitem{Gross:2010qma}
\hrefCMSnoop {}{E.~Gross and O.~Vitells, ``{Trial factors for the look
  elsewhere effect in high energy physics}'',} \textit{ Eur. Phys. J. C}
  \textbf{ 70} (2010) 525,
  \href{http://dx.doi.org/10.1140/epjc/s10052-010-1470-8}{\doi{10.1140/epjc/s10052-010-1470-8}},
  \href{http://www.arXiv.org/abs/1005.1891}{\texttt{arXiv:1005.1891}}.

\end{thebibliography}\endgroup

\cleardoublepage \appendix\section{The CMS Collaboration \label{app:collab}}\begin{sloppypar}\hyphenpenalty=5000\widowpenalty=500\clubpenalty=5000
\cmsinstitute{Yerevan Physics Institute, Yerevan, Armenia}
{\tolerance=6000
A.~Tumasyan\cmsAuthorMark{1}\cmsorcid{0009-0000-0684-6742}
\par}
\cmsinstitute{Institut f\"{u}r Hochenergiephysik, Vienna, Austria}
{\tolerance=6000
W.~Adam\cmsorcid{0000-0001-9099-4341}, J.W.~Andrejkovic, T.~Bergauer\cmsorcid{0000-0002-5786-0293}, S.~Chatterjee\cmsorcid{0000-0003-2660-0349}, K.~Damanakis\cmsorcid{0000-0001-5389-2872}, M.~Dragicevic\cmsorcid{0000-0003-1967-6783}, A.~Escalante~Del~Valle\cmsorcid{0000-0002-9702-6359}, M.~Jeitler\cmsAuthorMark{2}\cmsorcid{0000-0002-5141-9560}, N.~Krammer\cmsorcid{0000-0002-0548-0985}, L.~Lechner\cmsorcid{0000-0002-3065-1141}, D.~Liko\cmsorcid{0000-0002-3380-473X}, I.~Mikulec\cmsorcid{0000-0003-0385-2746}, P.~Paulitsch, F.M.~Pitters, J.~Schieck\cmsAuthorMark{2}\cmsorcid{0000-0002-1058-8093}, R.~Sch\"{o}fbeck\cmsorcid{0000-0002-2332-8784}, D.~Schwarz\cmsorcid{0000-0002-3821-7331}, S.~Templ\cmsorcid{0000-0003-3137-5692}, W.~Waltenberger\cmsorcid{0000-0002-6215-7228}, C.-E.~Wulz\cmsAuthorMark{2}\cmsorcid{0000-0001-9226-5812}
% ...
\par}
\vskip\cmsinstskip
\dag:~Deceased\\
$^{1}$Also at Yerevan State University, Yerevan, Armenia\\
$^{2}$Also at TU Wien, Vienna, Austria\\
% ...
\end{sloppypar}
\end{document}